\documentclass[aps,prb,twocolumn,superscriptaddress]{revtex4-2}
\usepackage{bm,color,amsmath,amssymb,mathrsfs,latexsym,graphicx,psfrag}
\usepackage{color}
\usepackage[dvipsnames]{xcolor}
\usepackage[hidelinks,colorlinks,linkcolor=blue,
citecolor=blue,urlcolor=blue]{hyperref}
\usepackage{dsfont}
\usepackage{comment}
\usepackage[normalem]{ulem}
\usepackage{chemformula}

\newcommand{\bsl}[1]{\boldsymbol{#1}}

\newcommand{\bra}[1]{\langle #1|}
\newcommand{\ket}[1]{|#1 \rangle}
\newcommand{\braket}[2]{\left\langle #1 | #2  \right\rangle}

\newcommand{\ii}{\mathrm{i}}

\newcommand{\mat}[1]{\left(\begin{matrix}#1\end{matrix}\right)}

\newcommand{\eq}[1]{\begin{equation} #1 \end{equation}}

\newcommand{\eqa}[1]{\begin{align}\begin{split} #1 \end{split}\end{align}}

\newcommand{\vrr}{\mathbf{r}}
\newcommand{\vk}{\mathbf{k}}
\newcommand{\vG}{\mathbf{G}}
\newcommand{\vR}{\mathbf{R}}

\renewcommand{\eqref}[1]{Eq.\,(\ref{#1})}
\newcommand{\eqnref}{\eqref}
\newcommand{\figref}[1]{Fig.\,\ref{#1}}
\newcommand{\tabref}[1]{Tab.\,\ref{#1}}
\newcommand{\secref}[1]{Sec.\,\ref{#1}}
\newcommand{\appref}[1]{Appendix.\,\ref{#1}}

\let\oldAA\AA
\renewcommand{\AA}{\text{\normalfont\oldAA}}

\newcommand{\ie}{{\emph{i.e.}}}

\newcommand{\TR}{\mathcal{T}}

\newcommand{\eV}{\mathrm{eV}}

\newcommand{\E}{\mathcal{E}}

\newcommand{\diag}{\text{diag}}

\newcommand{\Nb}{\text{Nb}}
\newcommand{\Se}{\text{Se}}
\newcommand{\nbse}{\text{NbSe$_2$}}

\newcommand{\BZ}{\text{1BZ}}

\usepackage{cancel}

\begin{document}
\title{Quantum Geometry in the NbSe$_2$ Family I: Obstructed Compact Wannier Function and New Perturbation Theory }

\author{Jiabin Yu}
\affiliation{Department of Physics, University of Florida, Gainesville, FL, USA}
\affiliation{Department of Physics, Princeton University, Princeton, New Jersey 08544, USA}

\author{Yi Jiang}
\affiliation{Donostia International Physics Center (DIPC), Paseo Manuel de Lardizábal. 20018, San Sebastián, Spain}

\author{Yuanfeng Xu}
\affiliation{Center for Correlated Matter and School of Physics, Zhejiang University, Hangzhou 310058, China}

\author{Dumitru C\u{a}lug\u{a}ru}
\affiliation{Department of Physics, Princeton University, Princeton, New Jersey 08544, USA}
\affiliation{Rudolf Peierls Centre for Theoretical Physics, University of Oxford, Oxford OX1 3PU, United Kingdom}

\author{Haoyu Hu}
\affiliation{Department of Physics, Princeton University, Princeton, New Jersey 08544, USA}
\affiliation{Donostia International Physics Center (DIPC), Paseo Manuel de Lardizábal. 20018, San Sebastián, Spain}

\author{Haojie Guo}
\affiliation{Donostia International Physics Center (DIPC), Paseo Manuel de Lardizábal. 20018, San Sebastián, Spain}
\author{Sandra Sajan}
\affiliation{Donostia International Physics Center (DIPC), Paseo Manuel de Lardizábal. 20018, San Sebastián, Spain}
\author{Yongsong Wang}
\affiliation{Donostia International Physics Center (DIPC), Paseo Manuel de Lardizábal. 20018, San Sebastián, Spain}

\author{Miguel M.~Ugeda}
\affiliation{Donostia International Physics Center, P. Manuel de Lardizabal 4, 20018 Donostia-San Sebastian, Spain}
\affiliation{Centro de Física de Materiales, Paseo Manuel de Lardizábal 5, 20018 San Sebastián, Spain.}
\affiliation{IKERBASQUE, Basque Foundation for Science, Maria Diaz de Haro 3, 48013 Bilbao, Spain}

\author{Fernando De Juan}
\affiliation{Donostia International Physics Center (DIPC), Paseo Manuel de Lardizábal. 20018, San Sebastián, Spain}
\affiliation{IKERBASQUE, Basque Foundation for Science, Maria Diaz de Haro 3, 48013 Bilbao, Spain}

\author{B. Andrei Bernevig}
\affiliation{Department of Physics, Princeton University, Princeton, New Jersey 08544, USA}
\affiliation{Donostia International Physics Center (DIPC), Paseo Manuel de Lardizábal. 20018, San Sebastián, Spain}
\affiliation{IKERBASQUE, Basque Foundation for Science, Maria Diaz de Haro 3, 48013 Bilbao, Spain}

\begin{abstract}
We revisit the electronic structure and band topology of monolayer 1H-NbSe$_2$, which hosts both superconductivity and charge density wave, and its related compounds 1H-MoS$_2$, NbS$_2$, TaS$_2$, TaSe$_2$ and WS$_2$. 
We construct a 6-band, a 3-band, and --- simplest of all --- a single-band model for this material family, by directly Wannierizing the \emph{ab initio} bands.
All host obstructed atomic isolated bands away from the atomic positions near the Fermi energy.
We find that in the 3-band model, the obstructed atomic Wannier function can be well approximated by an optimally compact Wannier function with more than 90\% accuracy for all the compounds, rising to a remarkable 94\% accuracy in NbSe$_2$.
Interestingly, the simplest single-band model has next nearest-neighboring hopping larger than the nearest-neighboring hopping (by nearly an order of magnitude for MoS$_2$, NbSe$_2$, TaSe$_2$ and WS$_2$), which comes from the cancellation between the atomic onsite terms and the atomic nearest-neighboring hopping after projecting to the obstructed atomic Wannier functions.
Furthermore for NbSe$_2$, we employ a novel approximation scheme to obtain an effective Hamiltonian that captures the 3 bands originating mainly from the Nb atom. We also use conventional perturbation theory to derive the \emph{ab initio} obstructed Wannier function with 95\% accuracy. 
Our results pave the way for future study of the effect of quantum geometry on the correlated phases in this family of materials.

\end{abstract}

\maketitle

\begin{figure}[t]
    \centering
    \includegraphics[width=\columnwidth]{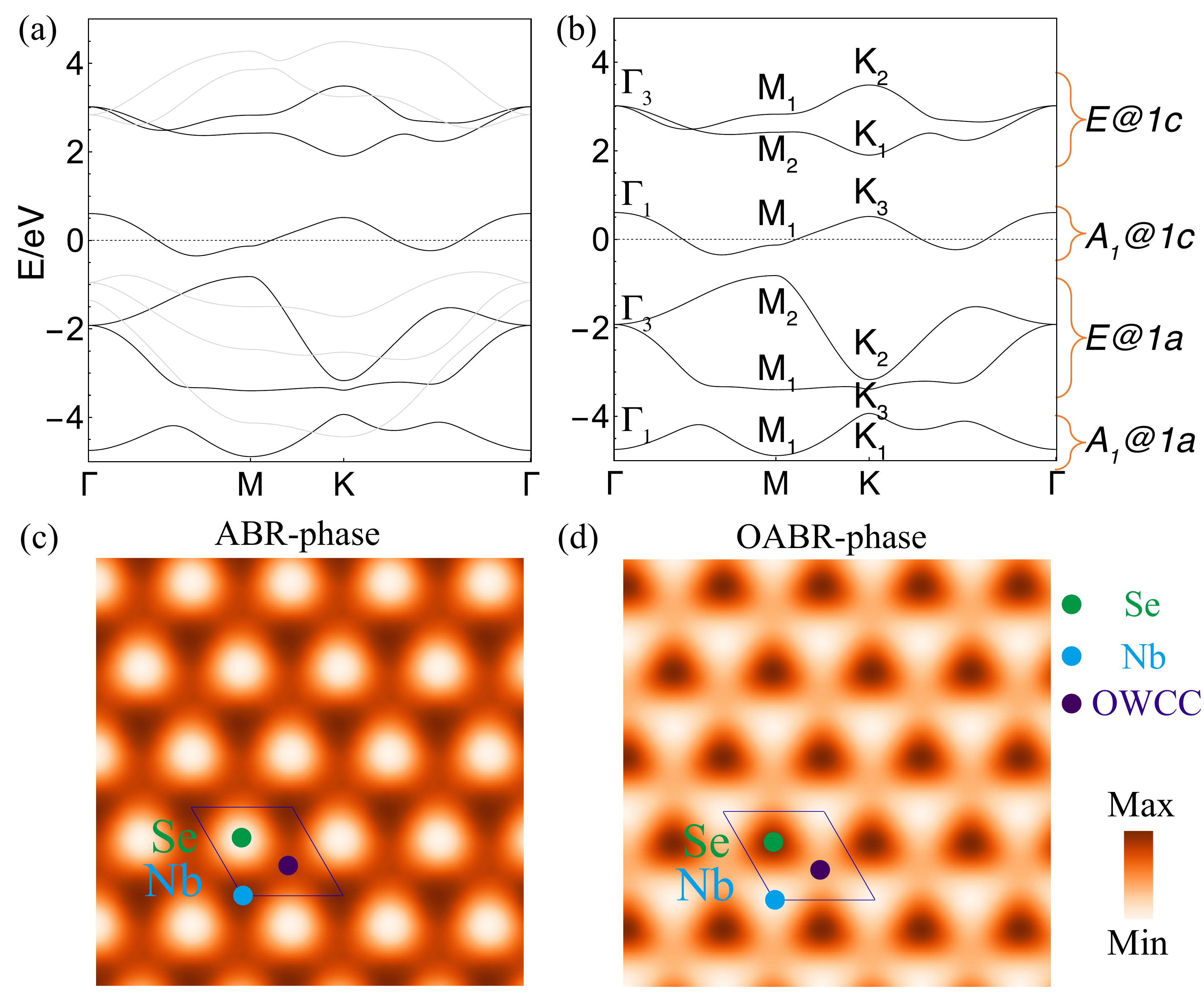}
    \caption{(a)NbSe$_2$ electron band structure from the DFT calculation. We plot the closest 11 bands to the Fermi level, as they are isolated. 
    The black lines are $m_z$ even, while the gray lines are $m_z$ odd.
    (b) The symmetry reps of electron states in the $m_z$-even sector.
    The symmetry reps at the high-symmetry momenta and the corresponding elementary band representation (EBR) of each isolated band set are labeled according to $p3m1$, since we consider the $m_z$-even sector where $m_z$ trivially holds. 
    ((c-d) Simulated scanning tunneling microscopy (STM) images for \ch{NbSe2}. (c) illustrates the atomic band representation (ABR) phase originating from the top two $m_z$-odd bands (grey) in (a), corresponding to the EBR $E@1a$. The scanning tunneling microscopy (STM) peak aligns with the Nb atomic site, consistent with the band representation. (d) depicts the obstructed atomic band representation (OABR) phase arising from the quasi-flat band at the Fermi level, associated with the EBR $A_1@1c$. This reflects an obstructed Wannier charge center (OWCC) located at the empty $1c$ site. The peak in STM is located at the Se site, while the second brightest spot is the OWCC site. 
    }
    \label{fig:DFT_el}
\end{figure}

\section{Introduction}
Band topologies or quantum geometries of electronic structure have played an essential role in understanding rich exotic phenomena in quantum materials~\cite{Qi2010TITSC,Hasan2010TI,Provost1980FSMetric,Fubini1904,Study1905,Martin.Souza.1999,Yu2025QGReview}, including fractional Chern insulators (FCIs)~\cite{neupert, sheng, regnault,Sun2011,Tang11,Jie2021IdealBands,Parker2023IdealBands,Valentin2023IdealBands,liu2024theorygeneralizedlandaulevels,cai2023signatures,zeng2023integer,park2023observation,Xu2023FCItMoTe2,Ji2024LocalProbetMoTe2,Young2024MagtMoTe2,Kang2024_tMoTe2_2.13,xu2024interplaytopologycorrelationssecond,park_Ferromagnetism_2024,Park_2025_tMoTe2_gap,Xu_2025_SC_MoTe2,Lu2024fractional,xie_Even_2024, choi_Electric_2024, lu_Extended_2024,xiao_coupled_2012,wu_topological_2019,pan_band_2020,PhysRevResearch.3.L032070,zhang_electronic_2021,devakul_magic_2021,wang_topological_2023,reddy_fractional_2023,dong_composite_2023,qiu_interaction-driven_2023,wang_topology_2023,reddy_toward_2023,wang_fractional_2024,yu_fractional_2024,xu_maximally_2024,abouelkomsan_band_2024,jia_moire_2024,zhang_polarization-driven_2024,PhysRevResearch.6.L032063,PhysRevB.109.245131,park_Topological_2023,herzog-arbeitman_Moire_2024,kwan_Moire_2023,yu_Moire_2024,guo_Fractional_2024,zhou_Fractional_2024,dong_Anomalous_2024,soejima_Anomalous_2024,huang_Impurityinduced_2024,tan_Wavefunction_2024,dong_Theory_2024,huang_Selfconsistent_2024,dassarma_Thermal_2024,xie_Integer_2024,dong_Stability_2024,kudo_Quantum_2024,zhou_New_2024}, optical conductivity~\cite{SWM2000,Resta2006Polarization,Martin2004ElectronicStructure,Noack.Aebischer.2001,Queiroz.Verma.2024.instantaneous,Fu.Onishi.2024.Dielectric,Stengel.Souza.2024}, the superfluid weight of superconductors~\cite{Torma2015SWBoundChern,Torma2016SuperfluidWeightLieb,Liang2017SWBandGeo,Hu2019MATBGSW,Xie2020TopologyBoundSCTBG,Torma2020SFWTBG,Rossi2021CurrentOpinion,Yu2022EOCPTBG,Torma2023WhereCanQuantumGeometryLeadUs,Tian2023QuantumGeoSC}, electron-phonon coupling~\cite{Yu05032023GeometryEPC,Alexandradinata2024ShiftCurrent}, and correlated charge fluctuations~\cite{Yu2024_QG_Charge_Fluctuation,Wu2024QGCornerChargeFluctuationsManyBody}.
Based on the topological quantum chemistry \cite{Bradlyn2017TQC,MTQC} or symmetry indicator theory \cite{Po2017SymIndi,Watanabe2018SIMSG,Kruthoff2017TCI,Song2017HOTI}, the symmetry-protected band topology of a gapped band structure can be diagnosed as strong topology, fragile topology, or trivial topology, according to the expansion coefficients of its symmetry eigenvalues on the basis of elementary band representations. Recently, the topologically trivial and fragile bands not diagnosed by symmetry eigenvalues were further classified by real space invariants (RSI) \cite{song2020,xu2021three}. The trivial bands are sub-classified into two types: atomic band representation (ABR, whose Wannier charge center is at the atom’s site) and obstructed atomic band representation (OABR, whose Wannier charge center is pinned away from atoms). The wave function of a topological (strong or fragile) band can not be Wannierized and is, therefore, an extended state. 

Although the OABR band is Wannierizable, its Wannier charge center resides away from the atomic sites and typically exhibits a large spatial spread across the lattice. As a result, electronic states in OABR topologies inherently possess non-trivial quantum geometry. However, the potential connection between the quantum geometry of OABR bands and many-body quantum states, such as density waves or superconductivity, remains under-explored in real materials or beyond tuned theoretical models~\cite{Torma2015SWBoundChern,Herzog-Arbeitman2021QGOAI}.

The hexagonal phase of layered transition metal dichalcogenides (TMDs) exhibits both charge density wave (CDW) order~\cite{Wilson2001,Rossnagel_2011,zhu2015classification,WIL75,CHA15,UGE16,LIA18,LIN20,DRE21} and superconductivity at lower temperatures~\cite{NbSe2SC,NbSe2SC2,CAO15,UGE16,XI16,LIA18,ZHA19a,DRE21,WAN22a}. These orders display dimensionality-dependent behavior upon exfoliation to the two-dimensional limit~\cite{lin2020patterns,monoNbSe2,CAL09,LER15,FLI15,FLI16}. Previous experiments and first-principles calculations have suggested that the CDW phase and the associated phonon softening in NbSe$_2$ arise from strong electron-phonon coupling~\cite{zhu2015classification,lian2018unveiling,lin2020patterns}. However, despite the fact that \textit{ab initio} calculations can reproduce the characteristic CDW wavevector of $2\pi/3$ in NbSe$_2$, its precise physical origin and numerical value remain not understood. Recently, obstructed atomic phases in \ch{WSe2} have been directly observed in experiments~\cite{holbrook2024real}, and were also proposed in other systems~\cite{liu2024massive1ddiracline}.

In this work, we study the obstructed atomic phases in a series of TMD monolayers, including \ch{NbSe2}, accompanying our experimental paper~\cite{cualuguaru2025probing} that provides direct evidence of obstructed atomic phases in \ch{NbSe2}.
This study is the first in a series aimed at understanding the CDW phase and phonon softening.
Here, we revisit the electronic structure and band topology of the monolayer 1H-NbSe$_2$ and its related compounds 1H-MoS$_2$, NbS$_2$, TaS$_2$, TaSe$_2$ and WS$_2$.  
Combined with first-principle calculations and TQC theory \cite{Bradlyn2017TQC}, we construct a 6-band, 3-band, and the simplest single-band models for all of them by numerically and analytically Wannierizing the DFT bands.
We find that the band at the Fermi energy is obstructed atomic, which is consistent with the previous theoretical analysis in TMD materials \cite{xu2021three,li2022obstructed,yang2023superconductivity, barnett2006coexistence, zeng2021multiorbital, jung2022hidden, qian2022c, sodequist2022abundance}.
In the 3-band model, the obstructed atomic Wannier function can be well approximated by the most compact Wannier function with more than 90\% accuracy for all compounds and in particular with a remarkable 94\% accuracy for NbSe$_2$.
(Here being compact means that the Wannier function has only nonzero probability on a finite number of lattice sites in the model of interest~\cite{Schindler2021CompactWannier,PhysRevB.108.195102,PhysRevLett.128.087002}.)
We find that the simplest single-band model has NNN hopping larger than the NN hopping (by nearly an order of magnitude for MoS$_2$, NbSe$_2$, TaSe$_2$ and WS$_2$), which comes from the cancellation between the atomic onsite terms and the atomic NN hopping after projecting to the obstructed atomic Wannier functions.
We further provide a new perturbation theory to  construct an effective 3-band model that well-captures the three DFT bands originating from the Nb atoms as well as the analytical Wannier states. 
Our results set the foundation for future analytic study and understanding of the effect of quantum geometry on the correlated phases in this family of materials.

\section{DFT Results}

In this section, we analyze the electronic band structure and topological properties of 1H-NbSe$_2$. The crystal symmetry of 1H-NbSe$_2$ is characterized by the 3D plane group $P\bar6m2$, where Nb and Se occupy the Wyckoff positions $1a(0,0,0)$ and $2c(\frac{1}{3},\frac{2}{3},z)$, respectively. As detailed in Appendix A, we first obtained the band structure of 1H-NbSe$_2$ from ab initio calculations \cite{vasp1,vasp2}. Combined with the Wannier90 package \cite{wannier90}, the eleven bands near the Fermi level are wannierized by five $d$ orbitals on Nb and 6 $p$ orbitals on the two Se atoms.
As shown in Fig.~\ref{fig:DFT_el}(a), there are 6 bands of a M$_z$-even eigenvalue and 5 bands of a M$_z$-odd eigenvalue, which are plotted with black and gray lines, respectively~\cite{barnett2006coexistence, liu2013three, fang2015ab, cappelluti2013tight}.
Since the isolated band on the Fermi level is M$_z$-even and there is no coupling between the two sectors of different M$_z$ eigenvalues in the one-layer system, here we only consider the 6 M$_z$-even bands. By ignoring M$_z$ symmetry, the 3D space group $P\bar6m2$ can be projected onto the layer group $p3m1$, where Nb and Se occupy the Wyckoff positions $1a(0,0)$ and $1b(\frac{1}{3},\frac{2}{3})$ of $p3m1$, respectively.
In Fig.~\ref{fig:DFT_el}(b), we calculate the elementary band representations (EBRs) of the four isolated band sets with M$_z$-even eigenvalue. As the band representation of the single band on the Fermi level is induced from an $A_1$ orbital at $1c(\frac{2}{3},\frac{1}{3})$, which is not occupied by any atom, its topology is OABR.
We note that the obstructed atomic band occurs not only in NbSe$_2$, but also in other 2D TMD materials.
As shown by \figref{fig:el-band_8TMD} of \appref{app:other_materials}, 1H-MoS$_2$, NbS$_2$, TaS$_2$, TaSe$_2$, and WS$_2$ also have one isolated $m_z$-even band near the Fermi energy which is obstructed atomic as $A_1$@1c. We remark that the 1T phase TMD materials have different band structures. For example, 1T-\ch{TaS2} has three isolated bands near the Fermi level with a different CDW phase compared with 1H-TaS$_2$.

Although both ABR and OABR are topologically trivial, their wavefunctions differ significantly, as revealed by their distinct real-space charge density distribution (CDD) in simulated scanning tunneling microscopy (STM) images. In Fig.\ref{fig:DFT_el}(c-d), we present the CDD for ABR and OABR bands separately, with a tip distance of $d = 4$\AA, highlighting their unique features (see Appendix. \ref{app:sec:DFT_STM} for more details). 
We note that the relative intensity of each site in CDD depends on the orbital weight and the shape of the Wannier orbital, when the tip distance is fixed.
Figure~\ref{fig:DFT_el}(c) shows the CDD from the top two $m_z$-odd bands in Fig.~\ref{fig:DFT_el}(a), associated with the EBR $E@1a$. These bands are primarily derived from the $(d_{xz}, d_{yz})$ orbitals of Nb. Due to the out-of-plane nature of the $d_{xz/yz}$ orbitals, the STM signal is strongest at the Nb atomic site. 
In contrast, Fig.~\ref{fig:DFT_el}(d) presents the CDD for the quasi-flat band at the Fermi level, corresponding to the EBR $A_1@1c$. This band features an obstructed Wannier charge center (OWCC) at the empty site $1c$. The quasi-flat band predominantly arises from the $d_{z^2}$ orbital of Nb, with smaller contributions from Nb $d_{xy}, d_{x^2-y^2}$ orbitals and $m_z$-even combinations of $p$ orbitals from the Se atoms. In this case, the simulated CDD peak appears at the Se site, which is closest to the STM tip. 
The second brightest site corresponds to the empty $1c$ position, while the Nb site is the darkest. This observation aligns with the EBR $A_1@1c$, providing clear evidence of the obstructed nature of the quasi-flat band in \ch{NbSe2}. In Ref.~\cite{cualuguaru2025probing}, the authors directly probed the this OWCC from the quasi-flat band in \ch{NbSe2} using STM, providing an unambiguous quantitative experimental identification of the obstructed atomic phase.

\section{Single-Particle Model: 6-band, 3-band and 1-band}

In this section, we present three effective tight-binding (TB) models for monolayer 1H-\ch{NbSe2}, focusing on deriving a simplified, analytically tractable model for the band at the Fermi level. This model provides explicit closed-form expressions for both the energy and eigenstates.

\subsection{6-band Model}

The 6 $m_z$-even combinations of the 11 orbitals are the focus of our work, which reads 
\eqa{
\label{main_eq:6band_basis}
& c^\dagger_{\bsl{R},d_{z^2}} = \hat{c}^\dagger_{\bsl{R},d_{z^2}} \\
& c^\dagger_{\bsl{R}, d_{xy}} = \hat{c}^\dagger_{\bsl{R}, d_{xy}} \\
& c^\dagger_{\bsl{R}, d_{x^2-y^2}} = \hat{c}^\dagger_{\bsl{R}, d_{x^2-y^2}}\\
&c^\dagger_{\bsl{R}+\bsl{\tau}_{\Se},z} = \frac{\hat{c}^\dagger_{\bsl{R}+\bsl{\tau}_{\Se,1},p_z}-\hat{c}^\dagger_{\bsl{R}+\bsl{\tau}_{\Se,2},p_z}}{\sqrt{2}} \\
&c^\dagger_{\bsl{R}+\bsl{\tau}_{\Se},x} = \frac{\hat{c}^\dagger_{\bsl{R}+\bsl{\tau}_{\Se,1},p_x}+\hat{c}^\dagger_{\bsl{R}+\bsl{\tau}_{\Se,2},p_x}}{\sqrt{2}}\\
&c^\dagger_{\bsl{R}+\bsl{\tau}_{\Se},y} = \frac{\hat{c}^\dagger_{\bsl{R}+\bsl{\tau}_{\Se,1},p_y}+\hat{c}^\dagger_{\bsl{R}+\bsl{\tau}_{\Se,2},p_y}}{\sqrt{2}}\ ,
}
where 
\eq{
\bsl{\tau}_{\Se} = \frac{\bsl{\tau}_{\Se,1}+\bsl{\tau}_{\Se,2}}{2} =  (0,\frac{a}{\sqrt{3}}, 0)
}
is the projection of Se sublattices onto the $x-y$ plane.
For convenience of constructing the model, we group the basis of the 6-band model as
\eq{
\label{main_eq:c_dagger_basis}
c^\dagger_{\bsl{R}+\bsl{\tau}} = \left\{ 
\begin{array}{cc}
  (c^\dagger_{\bsl{R},d_{z^2}}, c^\dagger_{\bsl{R}, d_{xy}}, c^\dagger_{\bsl{R}, d_{x^2-y^2}})  & ,\bsl{\tau} = \bsl{\tau}_{\Nb}\\
 (c^\dagger_{\bsl{R}+\bsl{\tau}_{\Se},x},c^\dagger_{\bsl{R}+\bsl{\tau}_{\Se},y},c^\dagger_{\bsl{R}+\bsl{\tau}_{\Se},z})  & , \bsl{\tau} = \bsl{\tau}_{\Se} 
\end{array}
\right.\ ,
}
and then the general 6-band TB model reads
\eq{
\label{main_eq:HTB6band}
H_6 = \sum_{\bsl{R}\bsl{R}',\bsl{\tau}\bsl{\tau}'}^{\Lambda}c^\dagger_{\bsl{R}+\bsl{\tau}}  t_{\bsl{\tau}\bsl{\tau}'}(\bsl{R}+\bsl{\tau}-\bsl{R}'-\bsl{\tau}') c_{\bsl{R}'+\bsl{\tau}'}\ ,
}
where $t_{\bsl{\tau}\bsl{\tau}'}(\bsl{R}+\bsl{\tau}-\bsl{R}'-\bsl{\tau}')$ is a $3\times 3$ matrix (there are $3$ orbitals at both Nb and projected Se sites), and $\Lambda$ is a range cutoff which means we only include $t_{\bsl{\tau}\bsl{\tau}'}(\bsl{R}+\bsl{\tau}-\bsl{R}'-\bsl{\tau}')$ with $|\bsl{R}+\bsl{\tau}-\bsl{R}'-\bsl{\tau}'|\leq \Lambda$.
In the momentum space, we have 
\eq{
\label{main_eq:HTB6band_k}
H_6 = \sum_{\bsl{k}} \mat{c^\dagger_{\Nb,\bsl{k}} &  c^\dagger_{\Se,\bsl{k}} } h_6(\bsl{k})  \mat{c_{\Nb,\bsl{k}} \\  c_{\Se,\bsl{k}} } \ ,
}
where
\eqa{
\label{eq:c_k_6band}
& c^\dagger_{\Nb,\bsl{k}} = \frac{1}{\sqrt{N}} \sum_{\bsl{R}} e^{\ii \bsl{k}\cdot\bsl{R} } c^\dagger_{\bsl{R}} \\
& c^\dagger_{\Se,\bsl{k}} = \frac{1}{\sqrt{N}} \sum_{\bsl{R}} e^{\ii \bsl{k}\cdot\left(\bsl{R} + \bsl{\tau}_{\Se} \right)} c^\dagger_{\bsl{R} + \bsl{\tau}_{\Se}} \ ,
}
and 
\eq{
\left[ h_6(\bsl{k}) \right]_{\bsl{\tau}\bsl{\tau}'} = \sum_{\bsl{R}} e^{ - \ii \bsl{k}\cdot \left( \bsl{R} + \bsl{\tau} - \bsl{\tau}' \right) } t_{\bsl{\tau}\bsl{\tau}'}(\bsl{R}+\bsl{\tau}-\bsl{\tau}')\ .
}
The symmetries constrain the form of $t_{\bsl{\tau}\bsl{\tau}'}(\bsl{R}+\bsl{\tau}-\bsl{R}'-\bsl{\tau}') $, and the parameters are directly determined by Wannier90 without any fitting. (See details in \appref{app:TB_Models}.)
Moreover, the sign of the NN hopping parameters can be directly understood from the sign of the overlap of the Wannier functions.
As shown in \figref{fig:DFT_el_H_6} in \appref{app:TB_Models}, we need to at least include terms up to 4NN to obtain a mean absolute error of  $0.0503$eV for the band dispersion of the single band at the Fermi level.
Therefore, the 6-band model is rather complicated. 
To reduce the complexity, we now build a 3-band from the full DFT 6-band model.

\subsection{3-Band Model}

\label{sec:3-band_model}

To build a 3-band model from the full 6-band model, we first numerically build the Wannier states for the top three bands from Wannier90. Our resulting Wannier states will be "renormalized" Nb even $d$-orbitals since their orbitals have much higher energies ($\sim 2 \eV$) than the Se $p$ orbitals. (See \appref{app:TB_Models}.) 
We label creation operators for the resultant Wannier states as 
\eq{
\label{main_eq:3band_basis}
\widetilde{c}^\dagger_{\bsl{R}} = ( \widetilde{c}^\dagger_{\bsl{R},d_{z^2}}, \widetilde{c}^\dagger_{\bsl{R},d_{xy}}, \widetilde{c}^\dagger_{\bsl{R},d_{x^2-y^2}})\ .
}
The Wannier states are just ``renormalized" $d_{z^2}$, $d_{xy}$ and $d_{x^2-y^2}$ orbitals with major (about 64\%) overlap on $d_{z^2}$, $d_{xy}$ and $d_{x^2-y^2}$, as shown by the approximate form of $\widetilde{c}^\dagger_{\bsl{R}}$ in \eqref{eq:3-band_model_basis} of \appref{app:TB_Models}.

With the basis in \eqref{main_eq:3band_basis}, the 3-band model reads
\eq{
\label{main_eq:3-band_model_el}
H_3 = \sum_{\bsl{R}\bsl{R}'} \widetilde{c}^\dagger_{\bsl{R}} \widetilde{t}(\bsl{R}-\bsl{R}') \widetilde{c}_{\bsl{R}'}\ ,
}
where $\widetilde{t}(\bsl{R}-\bsl{R}')$ is a $3\times 3$ matrix.
The explicit form of the hopping terms can be derived from symmetries and the values of the parameters can be determined from Wannier90 without fitting.
We find that the band structure of the one band crossing the Fermi level, given by the 3-band NNN model is close to the DFT one, as shown in \figref{fig:DFT_el_H_3_H_1}a.
The representations of this one band at high symmetry points show that its Wannier center is located \emph{away} from the atomic Wycoff position of Nb. As such, this band is an obstructed atomic band representation (OABR)~\cite{Bradlyn2017TQC,xu2021three}.

Within the 3-band model, we can have a simple understanding of the OABR at the Fermi level.
To achieve this, we rotate $\widetilde{c}^\dagger_{\bsl{R}}$ to a new form:
\eq{
\label{main_eq:3-band_model_rotated_basis}
(\widetilde{c}^\dagger_{\bsl{R},1}, \widetilde{c}^\dagger_{\bsl{R},2}, \widetilde{c}^\dagger_{\bsl{R},3} ) =  (\widetilde{c}^\dagger_{\bsl{R},z^2}, \widetilde{c}^\dagger_{\bsl{R},xy}, \widetilde{c}^\dagger_{\bsl{R},x^2-y^2}) R\ , 
}
where 
\eq{
\label{main_eq:R_rotation}
R = \left(
\begin{array}{ccc}
 \frac{1}{\sqrt{3}} & \frac{1}{\sqrt{3}} & \frac{1}{\sqrt{3}} \\
 0 & \frac{1}{\sqrt{2}} & -\frac{1}{\sqrt{2}} \\
 -\sqrt{\frac{2}{3}} & \frac{1}{\sqrt{6}} & \frac{1}{\sqrt{6}} \\
\end{array}
\right)\ .
}
This is in analogy to the $sp_2$ hybridization, if we replace $s$, $p_x$ and $p_y$ orbitals by $d_{z^2}$, $d_{xy}$ and $d_{x^2 - y^2}$, as they have the same symmetry representations under $p3m1$.
In the rotated basis, the numerical values of the parameters in the 3-band model suggest that
\eq{
\label{main_eq:H_3_rewritten}
H_3 = \sum_{\bsl{R}} H_3(\bsl{R}) + ... \ ,
}
where
\eqa{
\label{main_eq:3-band_local}
H_3(\bsl{R})  = \mat{\widetilde{c}^\dagger_{\bsl{R}+\bsl{a}_1+\bsl{a}_2,1} &  \widetilde{c}^\dagger_{\bsl{R},2} & \widetilde{c}_{\bsl{R}+\bsl{a}_1,3}^\dagger }  M \mat{\widetilde{c}_{\bsl{R}+\bsl{a}_1+\bsl{a}_2,1} \\  \widetilde{c}_{\bsl{R},2} \\ \widetilde{c}_{\bsl{R}+\bsl{a}_1,3} }\ ,
}
\eq{
M = \mat{ 
E_0 & t & t \\
 t & E_0 & t\\
 t & t & E_0
}\ ,
}
$E_0 = 1.733$eV, $t = -0.7840$eV, and ``..." includes terms with coefficients with amplitudes no larger than $0.3$eV. 
The terms in \eqref{main_eq:3-band_local}, besides ``...", are strictly local, \ie,
\eq{
\left[ H_3(\bsl{R}), H_3(\bsl{R}') \right] = 0 
}
owing to the definition of $\widetilde{c}_{\bsl{R},1}$, $\widetilde{c}_{\bsl{R},2}$ and  $\widetilde{c}_{\bsl{R},3}$ in \eqref{main_eq:3-band_model_rotated_basis}. (See also \figref{fig:DFT_el_H_3_H_1}c.)
Diagonalizing $M$ gives one eigenvalue $E_0 + 2 t$ with eigenvector $\frac{1}{\sqrt{3}} (1,1,1)$ and the doubly-degenerate eigenvalue  $E_0 - t$ with two eigenvectors $\frac{1}{\sqrt{6}} (-2,1,1)$ and $\frac{1}{\sqrt{2}} (0,1,-1)$.
The eigenvector $\frac{1}{\sqrt{3}} (1,1,1)$ corresponds to the OABR $A_1@1c$, whose creation operator reads
\eq{
\label{main_eq:w_compact_1band_el}
w^\dagger_{compact,\bsl{R}} = \frac{1}{\sqrt{3}} (\widetilde{c}^\dagger_{\bsl{R}+\bsl{a}_1+\bsl{a}_2,1} +  \widetilde{c}^\dagger_{\bsl{R},2} + \widetilde{c}_{\bsl{R}+\bsl{a}_1,3}^\dagger )\ ,
}
where being compact means that the Wannier function has nonzero probability on a finite number of lattice sites in the basis of the 3-band model. 
Here $w^\dagger_{compact,\bsl{R}} $ is most compact since it involves the smallest number of lattice sites for any Wannier functions with $1c$ Wannier center.
In the momentum space, 
\eqa{
w^\dagger_{compact,\bsl{k}} & = \frac{1}{\sqrt{N}}\sum_{\bsl{R}} e^{\ii \bsl{k}\cdot\bsl{R}} w^\dagger_{compact,\bsl{R}} \\
& = (\widetilde{c}^\dagger_{\bsl{k},1}, \widetilde{c}^\dagger_{\bsl{k},2}, \widetilde{c}^\dagger_{\bsl{k},3}) v_{compact,\bsl{k}} \ ,
}
where 
\eq{
v_{compact,\bsl{k}} = \frac{1}{\sqrt{3}} \mat{ e^{-\ii \left( \bsl{a}_1+\bsl{a}_2 \right) \cdot \bsl{k}} \\ 1  \\ e^{-\ii \bsl{a}_1 \cdot \bsl{k}} }
}
We expect $w^\dagger_{compact,\bsl{R}}$ to be a good approximation of the Wannier state $w^\dagger_{\bsl{R}}$ of the OABR; indeed, the probability overlap between their corresponding momentum-space eigenvectors of the 3-band model is a remarkable 0.94 (in average of the momentum), \ie, 
\eq{
\sqrt{\frac{1}{N}\sum_{\bsl{k}}^{\BZ} \left| v_{compact,\bsl{k}}^\dagger v_{w,\bsl{k}} \right|^2} = 0.970\ ,
}
where $v_{w,\bsl{k}}$ is the momentum-space eigenvector of the 3-band model
\eqa{
w^\dagger_{\bsl{k}} & = \frac{1}{\sqrt{N}}\sum_{\bsl{R}} e^{\ii \bsl{k}\cdot\bsl{R}} w^\dagger_{\bsl{R}}  = (\widetilde{c}^\dagger_{\bsl{k},1}, \widetilde{c}^\dagger_{\bsl{k},2}, \widetilde{c}^\dagger_{\bsl{k},3}) v_{w,\bsl{k}} \ .
}
Therefore, we have the following approximate relation
\eq{
\label{main_eq:w_approx_1band_el}
w^\dagger_{\bsl{R}} \approx w^\dagger_{compact,\bsl{R}} \ ,
}
where $w^\dagger_{compact,\bsl{R}}$ is defined in \eqref{main_eq:w_compact_1band_el}.

We note that for 1H-MoS$_2$, NbS$_2$, TaS$_2$, TaSe$_2$, and WS$_2$ which also exhibit one isolated $m_z$-even A$_1$@1c band near the Fermi energy,  their Wannier functions can be approximated by the most compact Wannier functions in the three band model with more than 90\% probability, as shown in \tabref{tab:1-band_hoppings} in \appref{app:other_materials}.

\subsection{1-Band Model}

From the full 3-band \emph{DFT model}, we can use Wannier90 to construct a one-band model for the band at the Fermi level.
The resultant Wannier function (recall that its creation operator is labeled by $w^\dagger_{\bsl{R}}$) is an $A_1$ irrep at the $1c$ Wycoff position ($A_1@1c$), which is obstructed atomic, as no atom is located at that position.
The one-band model reads 
\eq{
H_1 = \sum_{\bsl{R},\bsl{R}'} w^\dagger_{\bsl{R}}  w_{\bsl{R}'} t_{w}(\bsl{R}-\bsl{R}')\ .
}
To NNN, Wannier90 suggests 
\eqa{
\label{eq:one-band_hopping}
& t_{w}(\bsl{0}) = 0.003303\\ 
& t_{w}(\bsl{a}_1) = 0.01779\\ 
& t_{w}(2\bsl{a}_1+\bsl{a}_2) = 0.09553
}
in units of eV.
The simple NNN 1-band model produces a band that is very close to the DFT band structure as shown in \figref{fig:DFT_el_H_3_H_1}b. The interesting feature in \eqref{eq:one-band_hopping} is that the NNN hopping $t_{w}(2\bsl{a}_1+\bsl{a}_2)$ is much larger than the NN hopping $t_{w}(\bsl{a}_1)$.

We understand this feature within the 3-band model in \eqref{main_eq:3-band_model_el} within the NNN approximation.
From the approximate form of the OABR Wannier function in \eqref{main_eq:w_approx_1band_el}, we can derive the approximate form of the NN $t_{w}(\bsl{a}_1) $ and NNN hopping $t_{w}(\bsl{a}_1+2\bsl{a}_2)$ in the 1-band model:
\eqa{
 & t_{w}(\bsl{a}_1)  \approx f_{w,onsite}(\bsl{a}_1) + f_{w,NN}(\bsl{a}_1) + f_{w,NNN}(\bsl{a}_1) \\
 & t_{w}(\bsl{a}_1+2\bsl{a}_2) \approx f_{w,NN}(\bsl{a}_1+2\bsl{a}_2) + f_{w,NNN}(\bsl{a}_1+2\bsl{a}_2) \ ,
}
where the definition of each term can be found in \eqnref{eq:f_expressions} in \appref{app:TB_Models}.
Numerically, we find the approximate values $t_{w}(\bsl{a}_1)\approx -0.006167\eV$ and $t_{w}(2\bsl{a}_1+\bsl{a}_2)\approx 0.08258 \eV$, which capture well the qualitative difference between the NN and NNN hopping and are in good agreement with our Wannier calculation..
In the approximate expression, $t_{w}(\bsl{a}_1)$ mainly comes from the onsite and NN terms in the 3-band model, and the 3-band onsite contribution has opposite signs to and similar amplitude as the 3-band NN contribution. This canceling effect makes $t_{w}(\bsl{a}_1)$ small.
On the other hand, $t_{w}(\bsl{a}_1+2\bsl{a}_2)$ mainly comes from the NN and NNN terms in the 3-band model, where the 3-band NN and NNN contributions have the same signs.
Owing to the cancelling effect in $t_{w}(\bsl{a}_1)$, we eventually have $|t_{w}(\bsl{a}_1)| \ll |t_{w}(\bsl{a}_1+2\bsl{a}_2)|$.

We note that for MoS$_2$, NbS$_2$, TaS$_2$, TaSe$_2$, and WS$_2$ that have one isolated $m_z$-even A$_1$@1c band near the Fermi energy, they also have NN hoppings smaller than the NNN hoppings among the obstructed Wannier functions, especially for MoS$_2$, TaSe$_2$, NbSe$_2$ and WS$_2$ which have NN hoppings nearly one-order-of-magnitude smaller than the NNN hoppings. (See \tabref{tab:1-band_hoppings} in \appref{app:other_materials}.)
The fact that NN hoppings are smaller than the NNN hoppings in these materials for the flat band can also be explained approximately as the cancellation between atomic onsite terms and atomic NN terms, as shown in \tabref{tab:1-band_hoppings_from_6band} in \appref{app:other_materials}.

\subsection{Lower 3-Band Model}

From the full DFT 6-band model, we can also numerically build the Wannier states for the lowest three bands (below the Fermi level) from Wannier90.
The discussion of this part is analogous to that of \secref{sec:3-band_model}.
The trial states are chosen to be three Se p orbitals in \eqref{eq:6band_basis}, since we know the Se $m_z$-even combinations of p-orbitals have much lower energies ($\sim 2 \eV$) than the Nb d orbitals from DFT. Hence our resulting Wannier states will be ``renormalized" Se $m_z$-even combinations of p-orbitals, as shown by the approximated forms of the Waniner states in \eqref{eq:lower-3-band_model_rotated_basis} of \appref{app:TB_Models}.
We label creation operators for the resultant Wannier states as 
\eq{
\label{main_eq:lower_3band_basis}
\widetilde{c}^\dagger_{\bsl{R}+\bsl{\tau}_{\Se}} = ( \widetilde{c}^\dagger_{\bsl{R}+\bsl{\tau}_{\Se},z}, \widetilde{c}^\dagger_{\bsl{R}+\bsl{\tau}_{\Se},x}, \widetilde{c}^\dagger_{\bsl{R}+\bsl{\tau}_{\Se},y})\ .
}
With the basis in \eqref{eq:lower_3band_basis}, the lower-3-band model has the general form of
\eq{
\label{main_eq:lower_3-band_model_el}
H_{lower-3} = \sum_{\bsl{R}\bsl{R}'} \widetilde{c}^\dagger_{\bsl{R}+\bsl{\tau}_{\Se}} \widetilde{t}(\bsl{R}-\bsl{R}') \widetilde{c}_{\bsl{R}'+\bsl{\tau}_{\Se}}\ ,
}
where $\widetilde{t}(\bsl{R}-\bsl{R}')$ is a $3\times 3$ matrix.
Based on symmetries and Wannier90, we can obtain a NNN model for the lower 3 bands, where the expressions of the $\widetilde{t}(\bsl{R}-\bsl{R}')$ to NNN order are shown in \eqnref{eq:lower_3band_hopping_values} in \appref{app:lower-3-band_model}.
The band structure of the lower three bands, given by the lower-3-band NNN model is close to the DFT one, as shown in \figref{fig:lower_3band_DFT_el_6bandsim}(a).

\section{New Perturbation Method}

\subsection{New Perturbation Method}
\label{sec:new_perturb}

In this section, we introduce a new perturbation method for deriving an analytical effective model for \ch{NbSe2}. Our primary goal is to obtain a 3-band and a 1-band model from the more complex 6-band model. Achieving the implied few band model is not feasible using conventional perturbation approaches due to the intricate coupling between the bands and the need to accurately capture the physics near the Fermi level. In the following, we first outline the formalism of the new perturbation method, and then apply it to \ch{NbSe2} to obtain minimal analytical models. 

Suppose we have a Hamiltonian with the following matrix form
\eq{
h= \mat{ H_0 & S \\ S^\dagger & H_1 }\ ,
}
where $H_0$ is a $n\times n$ matrix, $H_1$ is $m\times m$, and $S$ is $m\times n$.
We choose $H_0$ and $H_1$ such that the eigenvalues of $H_0$ are larger than those of $H_1$. 
The corresponding eigenequation reads
\eq{
\mat{ H_0 & S \\ S^\dagger & H_1 } \mat{ \psi_0 \\ \psi_1 } = E \mat{ \psi_0 \\ \psi_1 } \ .
} 

Suppose the top $n$ bands of $h$ are separated from the bottom $m$ bands by a gap $G$ that is (i) much larger than the elements of $S$, and (ii) the spread of the $n$ eigenvalues of $H_0$ (\ie, the difference between the highest and lowest eigenvalues of $H_0$) is much smaller than $G$.
Then, the conventional way to derive the approximated effective model for the highest $n$ eigenvalues and eigenvectors of $h$ is to first re-write the eigenequation into
\eqa{
\label{main_eq:gen_perturb}
& \left( H_0 + S \frac{1}{E-H_1} S^\dagger \right) \psi_0 = E \psi_0 \\
& \psi_1 = (E-H_1)^{-1} S^\dagger \psi_0\ .
}
Here $E-H_1$ is invertible, since we are considering the highest $n$ eigenvalues of $h$, which makes $E$ larger than all eigenvalues of $H_1$. In practice, we can directly replace $E$ in $(E-H_1)^{-1}$ by the average of the eigenvalues (trace divided by the number of bands)  of $H_0$ (labeled as $E_0$), leading to an approximate effective model for the highest $n$ eigenvalues of $h$ as $H_0 + S \frac{1}{E_0-H_1} S^\dagger$.
The approximate effective Hamiltonian would become exact if the highest $n$ eigenvalues of $h$ are exactly the same and we use its value as $E_0$.

The conventional perturbation theory would fail when the two conditions are violated, which is the case in \ch{NbSe2} due to the small gap between the first $n=3$ bands and the next $m=3$ bands as well as he relatively large bandwidth of the first $n=3$ bands.
Nevertheless, to address this, we propose a new perturbation method for the case where (i) the top $n$ eigenvalues of $h$ are $E_0'$ with degeneracy $D<n$ and $E_0''$ with degeneracy $n-D$, and (ii) $E-H_1$ is invertible for $E = E_0', E_0''$.
This case is relevant for {\nbse} as discussed in \secref{sec:perturbation_three_band}.
In this case, we perform the following replacement in \eqref{main_eq:gen_perturb}:
\eq{
\label{main_eq:replacement_new_perturb}
\frac{1}{E-H_1} \rightarrow a + b E \ ,
}
where 
\eqa{
a & = - \frac{1}{E_0' - H_1 } \frac{E_0''}{ E_0' - E_0''} + \frac{1}{E_0'' - H_1} \frac{E_0'}{E_0' - E_0''} \\ 
b & = \left( \frac{1}{E_0' - H_1} - \frac{1}{E_0'' - H_1} \right)  \frac{1}{E_0 ' - E_0''}
}
are chosen to make sure the replacement is exact for $E= E_0', E_0''$.
With the replacement in \eqref{main_eq:replacement_new_perturb}, \eqref{main_eq:gen_perturb} becomes 
\eqa{
\label{main_eq:new_perturb}
& \left( H_0 + S a S^\dagger \right) \psi_0 = E (1- S b S^\dagger) \psi_0 \\
& \psi_1 = (a + b E) S^\dagger \psi_0 \ .
}
Suppose $1- S b S^\dagger$ is invertible, and by defining 
\eq{
1- S b S^\dagger = U \Lambda U^\dagger
}
with unitary $U$ and diagonal $\Lambda$, we end up with
\eq{
 (\sqrt{\Lambda})^{-1} U^\dagger \left( H_0 + S a S^\dagger \right) U (\sqrt{\Lambda})^{-1}\widetilde{\psi}_0 = E \widetilde{\psi}_0
}
with $\widetilde{\psi}_0 =\sqrt{\Lambda}  U^\dagger \psi_0$
meaning that the effective model for the highest n eigenvalues is 
\eq{
\label{main_eq:new_perturb_heff}
h_{eff} = (\sqrt{\Lambda})^{-1} U^\dagger \left( H_0 + S a S^\dagger \right) U (\sqrt{\Lambda})^{-1} \ .
}
A similar method can be applied to determine the lowest $m$ eigenvalues in this scenario. It is important to note that this approach does not rely on the smallness of the matrix elements of S, but it does require prior knowledge of the energies $E_0'$ and $E_0''$. For almost flat bands around some energies, these values can be found numerically. In the case of \ch{NbSe2}, this requirement is straightforward to satisfy, as $E'_0$ is zero to first order. Using this information, the wavefunction and the corresponding Hamiltonian can be self-consistently obtained.

\subsection{Perturbation theory For the 6-Band Model}
\label{sec:perturbation_three_band}

\begin{figure*}[t]
    \centering
    \includegraphics[width=1.5\columnwidth]{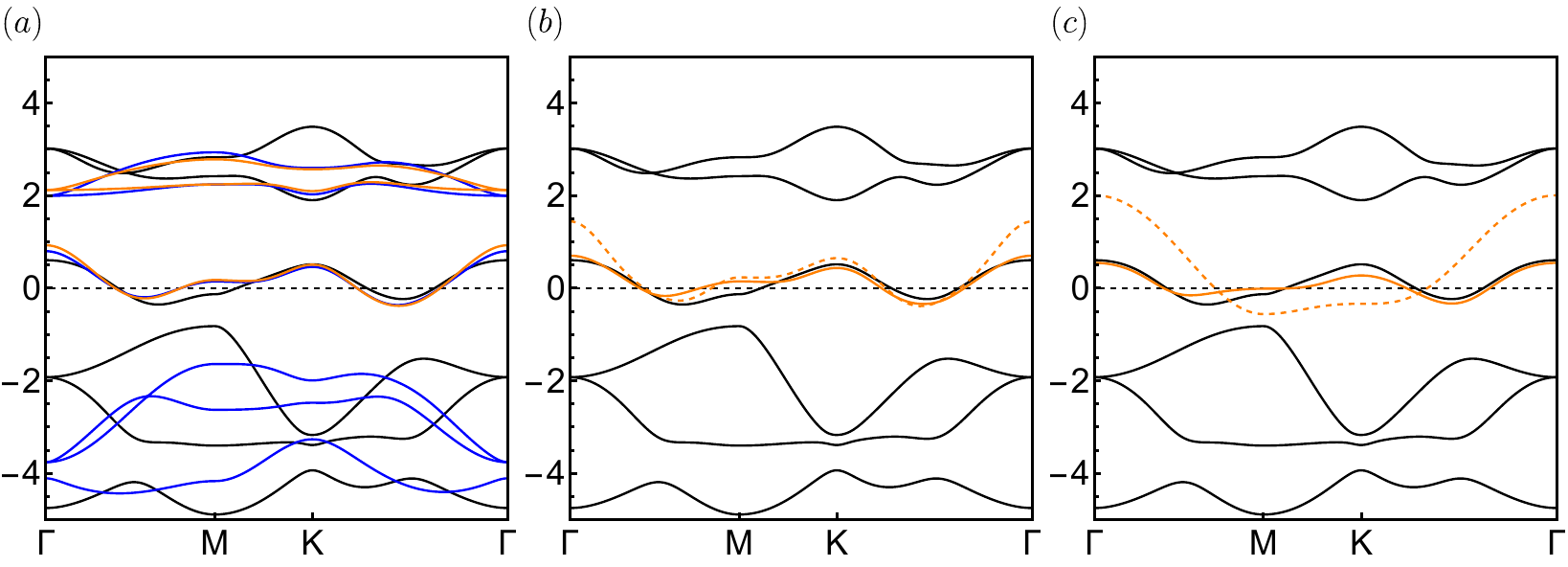}
    \caption{
    The band structures of DFT are plotted in black.
    (a) The band structures of the 6-band Se-onsite NNN Hamiltonian are plotted in blue, and the band structures of the effective 3-band model in \eqref{main_eq:Nb_eff_final} are plotted in orange.
    (b) The band structure given by \eqref{main_eq:approx_E_w_k} and \eqnref{main_eq:exp_value} for $H_{6,\text{Se-onste,NNN}}$ is plotted as orange dashed and solid lines.
    (c) The band structure given by \eqref{main_eq:approx_E_w_k} and \eqnref{main_eq:exp_value} for the simplification in \eqref{eq:sim2_approx1} and \eqref{eq:sim2_approx2} is plotted as orange dashed and solid lines, respectively, in (c).
    }
    \label{main_fig:approx_El}
\end{figure*}

We now build an effective model for the upper three bands.
In the momentum space, the matrix Hamiltonian of the 6-band model can be written as  
\eq{
h_6(\bsl{k}) = \left( \begin{array}{cc}
 H_{\Nb}(\bsl{k})  & S(\bsl{k}) \\
 S^\dagger(\bsl{k}) & H_{\Se}(\bsl{k}) \\
\end{array}
\right) \ ,
} 
where $H_{\Nb}(\bsl{k})$, $H_{\Se}(\bsl{k})$ and $S(\bsl{k})$ are $3\times 3$ matrices.
The eigenvalue equations read
\eqa{
    \left( \begin{array}{cc}
 H_{\Nb}(\bsl{k})  & S(\bsl{k}) \\
 S^\dagger(\bsl{k}) & H_{\Se}(\bsl{k}) \\
\end{array}
\right)\left(\begin{array}{cc}
\psi_{\Nb, \bsl{k} }  \\
 \psi_{\Se,\bsl{k}}  \\
\end{array}
\right)= E_{\bsl{k}}\left(\begin{array}{cc}
\psi_{\Nb, \bsl{k} }  \\
 \psi_{\Se,\bsl{k}}  \\
\end{array}
\right)\ ,
}
where $\psi_{\Nb, \bsl{k} }$ and $\psi_{\Se, \bsl{k} }$ are three-component vectors.
According to \eqref{main_eq:gen_perturb}, the eigen-equation can be written as
\eqa{
\label{main_eq:Nb_eff_gen_expression}
& \left[ H_{\Nb}(\bsl{k}) + S(\bsl{k})  \frac{1}{ E_{\bsl{k}}- H_{\Se}(\bsl{k})} S^\dagger(\bsl{k}) \right] \psi_{\Nb,\bsl{k}}= E_{\bsl{k}} \psi_{Nb,\bsl{k}} \\
& \psi_{Se,\bsl{k}} = (E_{\bsl{k}} - H_{\Se}(\bsl{k}))^{-1} S^\dagger(\bsl{k}) \psi_{Nb,\bsl{k}}
}
If we use the DFT-precise $h_6(\bsl{k})$, \eqref{main_eq:Nb_eff_gen_expression} is invalid for the single band near the Fermi level as $E_{\bsl{k}}- H_{\Se}(\bsl{k})$ is not invertible for that band.
Therefore, to derive the effective model, we use a simplified $h_6(\bsl{k})$ which contains only the onsite terms for Nb and Se, the NN hopping between Nb and Se, and the NNN hopping between Nb and Nb---we refer to this simplified $h_{6}(\bsl{k})$ as the Se-onsite NNN 6-band model. These terms are by far the largest in the Hamiltonian. 
As shown in \figref{main_fig:approx_El}(a), the Se-onsite NNN 6-band model maintains the shape of the relevant 1 band near the Fermi energy and has the highest two bands roughly at the correct energy, although it does not match the lower 3 bands well.
The $H_{\Se}(\bsl{k})$ part of the Se-onsite NNN 6-band model $h_{6}(\bsl{k})$ reads $H_{\Se}(\bsl{k}) = \diag(E_{z},E_{x}, E_{x}) = \diag ( -2.4102 ,  -1.6090 , -1.6090 )\eV$; thus, $E_{\bsl{k}}- H_{\Se}(\bsl{k})$ is invertible for all three upper bands.

Yet, for the Se-onsite NNN 6-band model, \eqref{main_eq:Nb_eff_gen_expression} is not an eigenvalue problem due to $E_{\bsl{k}}$ in the denominator.
To make it an eigenvalue problem, we observe that (i) the single band of the Se-onsite NNN 6-band model around the Fermi energy is quasi-flat, ranging from $-0.3555$eV to $0.8016$eV with mean energy at $0$eV because the band is half-filled, and (ii) the highest two bands range from $1.997$eV to $2.937$eV with mean energy at $E_1=2.417\eV\approx 2.4$eV.
Therefore, as an approximation, we can first approximate the upper three bands as a single exactly flat band at $0$eV and a doubly-degenerate flat band at $E_1$, and then we can apply new perturbation method in \eqref{main_eq:new_perturb} on  the Se-onsite NNN 6-band model to derive the effective model for the upper three bands, which, according to \eqref{main_eq:new_perturb_heff}, reads
\eqa{
\label{main_eq:Nb_eff_final}
& H_{\Nb,eff}(\bsl{k}) = \\
& \quad (\sqrt{\Lambda_{\bsl{k}}})^{-1} U_{\bsl{k}}^\dagger \left( H_{\Nb}(\bsl{k}) + S(\bsl{k}) a_{\bsl{k}} S^\dagger(\bsl{k}) \right) U_{\bsl{k}} (\sqrt{\Lambda_{\bsl{k}}})^{-1}\ ,
}
where $a_{\bsl{k}}=-H_{Se}^{-1}(\bsl{k}) $,  $\left[ b_{\bsl{k}} \right]_{ij} = \frac{E^{-1}_{i} + (E_1 - E_{i})^{-1}}{E_1} \delta_{ij}$ for $i=p_x,p_y,p_z$ of Se in $H_{Se}(\mathbf{k})$, and $U_{\bsl{k}} \Lambda_{\bsl{k}} U_{\bsl{k}}^\dagger = 1- S(\bsl{k})^\dagger b_{\bsl{k}} S(\bsl{k})$ with unitary $U_{\bsl{k}}$ and diagonal $\Lambda_{\bsl{k}}$.
Here we have used the fact that $\Lambda_{\bsl{k}}$ has non-negative diagonal elements for the Se-onsite NNN 6-band model.
As shown in \figref{main_fig:approx_El}(a), the dispersion obtained from the approximated effective model in \eqref{main_eq:Nb_eff_final} matches the upper three bands of the Se-onsite NNN 6-band model extremely  well, validating the perturbation method.

By diagonalizing the effective Hamiltonian in \eqref{main_eq:Nb_eff_final}, we can obtain $\psi_{Nb,\bsl{\bsl{k}}}$ for the single band at the Fermi level, as $\sqrt{\Lambda_{\bsl{k}}}  U_{\bsl{k}}^\dagger \psi_{Nb,\bsl{\bsl{k}}}$ is the eigenvector of \eqref{main_eq:Nb_eff_final}.
Combined with the second equation in \eqref{main_eq:Nb_eff_gen_expression}, we can obtain the Bloch state of the single band at the Fermi level, which reads
\eq{
\gamma_{approx,w,\bsl{k}}^\dagger = N_{\bsl{k}} \left[ c^\dagger_{\Nb,\bsl{k}} +   c^\dagger_{\Se,\bsl{k}}  (- H_{\Se}(\bsl{k}))^{-1} S^\dagger(\bsl{k}) \right] \psi_{Nb,\bsl{k}} \ ,
}
where $H_{\Se}(\bsl{k})$ and $S(\bsl{k})$ are taken from the Se-onsite NNN 6-band model, $\psi_{Nb,\bsl{k}}$ is solved from the effective model in \eqref{main_eq:Nb_eff_final}, and $N_{\bsl{k}}$ is the normalization factor, and $c^\dagger_{\Nb,\bsl{k}}$ and $c^\dagger_{\Se,\bsl{k}}$ are defined in \eqref{eq:c_k_6band}.
The $\gamma_{approx,w,\bsl{k}}^\dagger$ has probability overlap with the DFT-precise Bloch state $w_{\bsl{k}}^\dagger$ for the single band at the Fermi level, \ie,
\eq{
\sqrt{\frac{1}{N}\sum_{\bsl{k}} \left| \bra{0} \gamma_{approx,w,\bsl{k}} w_{\bsl{k}}^\dagger \ket{0} \right|^2} =  0.980\ ,
}
where $\ket{0}$ is the vacuum state.
However, this approach remains overly complicated due to the fact that we end up with a $3$-band model; therefore, we simplify it further in the following discussion.

\subsection{Perturbative Analysis of the 1-band Wavefunction}

We now focus on the analytical derivation of the 1-band wavefunction from perturbation theory.
For this purpose, the conventional perturbation theory is good enough.
The wavefunction of one band reads
\eq{
\label{main_eq:psi_w}
\psi_{w,\bsl{k},approx} = N_{\bsl{k}}\mat{ \psi_{\Nb, \bsl{k}} \\
- H^{-1}_{\Se}(\bsl{k}) S^\dagger(\bsl{k}) \psi_{\Nb, \bsl{k}}}\ ,
} 
and we can solve for $\psi_{\Nb, \bsl{k}}$ with the effective 3-band model
\eqa{
\label{main_eq:H_eff_for_psi_Nb}
   H_{\Nb,eff}(\bsl{k}) = H_{\Nb}(\bsl{k}) - S(\bsl{k}) H^{-1}_{\Se}(\bsl{k}) S^\dagger(\bsl{k})\ ,
}
where we have already approximated the single band around the Fermi energy by an exactly flat band with zero energy.

To further perturbatively solve the 3-band Hamiltonian $H_{\Nb,eff}(\bsl{k})$, we now perform perturbation theory around the compact obstructed atomic band at the Fermi level.
Specifically, the zeroth-order Hamiltonian we choose describes a compact obstructed atomic orbital at 1c position, which reads 
\eq{
\overline{H}_{eff,0}(\bsl{k}) 
= 
\mat{ 
 E_0(\bsl{k}) & & \\
& E_+(\bsl{k}) & \\
& & E_-(\bsl{k})
}\ .
}
The $\mathbf{k}$-dependence in $E_i(\bsl{k})$ comes from a unitary transformation that combines orbital from different unit cells:
\eq{
U_w(\bsl{k}) = \mat{ e^{-\ii \bsl{k}\cdot(\bsl{a}_1+\bsl{a}_2)} & & \\ & 1 & \\ & & e^{-\ii \bsl{k}\cdot\bsl{a}_1}}U\ ,
}
and 
\eq{
U=  \frac{1}{\sqrt{3}}\left( \begin{array}{ccc}
 1 & e^{i \frac{4\pi}{3}} & e^{-i \frac{4\pi}{3}} \\
 1 & 1 & 1 \\
 1 & e^{i \frac{2\pi}{3}} & e^{-i \frac{2\pi}{3}} \\
\end{array}\right) \ .
}
(See \ref{app:1-band_approx} for details)
The basis of the compact obstructed atomic orbital is 
\eq{
\begin{aligned}
&(c^\dagger_{\bsl{k},0}, c^\dagger_{\bsl{k},+}, c^\dagger_{\bsl{k},-})\\
&= (c^\dagger_{Nb, \bsl{k},d_{z^2}}, c^\dagger_{Nb, \bsl{k},d_{xy}},c^\dagger_{Nb, \bsl{k} ,d_{x^2-y^2}}) R U_w(\bsl{k})\ .    
\end{aligned}
\label{main_eq:OAIkbasis}
}
We can also transform $ H_{\Nb,eff}(\bsl{k})$ to the basis \eqref{main_eq:OAIkbasis}, which is labeled as 
\eq{
\begin{aligned}
\overline{H}_{\Nb,eff}(\bsl{k}) &= (RU_w(\bsl{k}))^{-1} H_{\Nb,eff} RU_w(\bsl{k}) \\
&= \overline{H}_{eff,0}(\bsl{k}) + \overline{H}_{eff,1}(\bsl{k}) \ ,
\end{aligned}
}
where 
\eq{
\overline{H}_{eff,1}(\bsl{k}) = \overline{\Nb,eff}(\bsl{k}) - \overline{H}_{eff,0}(\bsl{k}) \ .
}
We treat $\overline{H}_{eff,1}(\bsl{k})$ as a perturbation since its matrix elements have maximum absolute values of about 1.2721eV over the BZ, while the minimum absolute value of the matrix elements of $\overline{H}_{eff,0}(\bsl{k})$ is about 3.6970eV.
Then, the approximated dispersion for the obstructed atomic band reads
\eq{
\label{main_eq:approx_E_w_k}
E_{w,\bsl{k}} = E_0(\bsl{k}) + \left[ \overline{H}_{eff,1}(\bsl{k}) \right]_{11} + \sum_{n=2,3} \frac{ \left| [\overline{H}_{eff,1 }( \bsl{k})]_{n1} \right|^2}{E_0(\bsl{k})-E_1(\bsl{k})}\ ,
}
and the corresponding eigenstate is created by
\eqa{
c^\dagger_{eff,0,\bsl{k}} & = (c^\dagger_{\bsl{k},0}, c^\dagger_{\bsl{k},+}, c^\dagger_{\bsl{k},-}) \left( \begin{array}{c}
1  \\
 \frac{1}{E_0(\bsl{k})-E_+(\bsl{k})} [\overline{H}_{eff,1 }( \bsl{k})]_{21}\\
\frac{1}{E_0(\bsl{k})-E_-(\bsl{k})} [\overline{H}_{eff,1 }( \bsl{k})]_{31}  \\
\end{array}
\right) \ ,
}
meaning that the approximate expression of $\psi_{\Nb}(\bsl{k})$ derived from the perturbation theory reads
\eq{
\label{main_eq:approx_psi_NB}
\psi_{\Nb}(\bsl{k}) = R U_w(\bsl{k}) \left( \begin{array}{c}
1  \\
 \frac{1}{E_0(\bsl{k})-E_+(\bsl{k})} [\overline{H}_{eff,1 }( \bsl{k})]_{21}\\
\frac{1}{E_0(\bsl{k})-E_-(\bsl{k})} [\overline{H}_{eff,1 }( \bsl{k})]_{31}  \\
\end{array}
\right) \ .
}
By substituting \eqnref{main_eq:approx_psi_NB} into \eqref{main_eq:psi_w}, we can obtain the approximated expression of the creation operator of the obstructed atomic band.
With the parameter values for the 6-band Se-onsite NNN model, we find that the perturbation gives an approximated state of the obstructed atomic band that has a remarkable probability overlap with the DFT-precise one:
\eq{
\sqrt{\frac{1}{N}\sum_{\bsl{k}} \left| \bra{0} c_{eff,0,\bsl{k}} w_{\bsl{k}}^\dagger \ket{0} \right|^2} =  0.977\ .
}
As shown by the orange dashed line in \figref{main_fig:approx_El}(b), the approximated dispersion given by \eqnref{main_eq:approx_E_w_k} qualitatively captures the obstructed atomic band near Fermi energy of the DFT precise Hamiltonian.
If we compare the expectation value
\eq{
\label{main_eq:exp_value}
\bra{0} c_{eff,0,\bsl{k}} H_{6,\text{Se-onste,NNN}} c_{eff,0,\bsl{k}}^\dagger \ket{0} 
}
to the dispersion of the obstructed atomic band of the DFT precise Hamiltonian, we can see a reasonable match in \figref{main_fig:approx_El}(b) as the orange line.

In \appref{app:1-band_approx}, we consider further simplifications of the 6-band Hamiltonian beyond the Se-onsite NNN approximation and provide simpler analytical expressions of $\psi_{\Nb}(\bsl{k})$, which gives approximated wavefunctions that have about 0.9 probability overlap with the DFT wavefunction.
Although the resultant $E_{w,\bsl{k}}$ has a large deviation from the DFT bands, the band given by the \eqnref{main_eq:exp_value} can still have a reasonable match with the DFT band as exemplified in \figref{main_fig:approx_El}(c), providing a reasonable starting point for any further correlated study.

\section{Discussion}

We have studied the obstructed atomic Wannier functions in monolayer NbSe$_2$ and related compounds. Our results show that the obstructed atomic band at the Fermi level in NbSe$_2$ exhibits large quantum geometry due to its obstructed nature. 
Qantum geometry has been recently found to contribute crucially to the electron-phonon coupling~\cite{Yu05032023GeometryEPC}.
As the electron-phonon coupling can give rise to CDW, one future direction is to find the connection between CDW and the obstructed atomic Wannier functions in this material.
The relation between superconductivity and the obstructed atomic Wannier functions is also worth studying.

\begin{acknowledgments}

B.A.B. was supported by the Gordon and Betty Moore Foundation through Grant No. GBMF8685 towards the Princeton theory program, the Gordon and Betty Moore Foundation’s EPiQS Initiative (Grant No. GBMF11070), the Office of Naval Research (ONR Grant No. N00014-20-1-2303), the Global Collaborative Network Grant at Princeton University, the Simons Investigator Grant No. 404513, the BSF Israel US foundation No. 2018226, the NSF-MERSEC (Grant No. MERSEC DMR 2011750), the Simons Collaboration on New Frontiers in Superconductivity (Grant no. SFI-MPS-NFS- 00006741-01), and the Schmidt Foundation at the Princeton University. J. Y.'s work at Princeton University is supported by the Gordon and Betty Moore Foundation through Grant No. GBMF8685 towards the Princeton theory program. J. Y.'s  work at University of Florida is supported by startup funds at University of Florida. M.M.U. acknowledges support from the European Union ERC Starting grant LINKSPM (Grant \#758558) and by the grant PID2023-153277NB-I00 funded by the Spanish Ministry of Science, Innovation and Universities. H.G. acknowledges funding from the EU NextGenerationEU/PRTR-C17.I1, as well as by the IKUR Strategy under the collaboration agreement between Ikerbasque Foundation and DIPC on behalf of the Department of Education of the Basque Government. F. J. acknowledges support from the grant PID2021-128760NB-I00 funded by the Spanish Ministry of Science, Innovation and Universities.
Y.J. was supported by the European Research Council (ERC) under the European Union’s Horizon 2020 research and innovation program (Grant Agreement No. 101020833), as well as by the IKUR Strategy under the collaboration agreement between Ikerbasque Foundation and DIPC on behalf of the Department of Education of the Basque Government.
D.C. acknowledges support from the DOE Grant No. DE-SC0016239 and the hospitality of the Donostia International Physics Center, at which this work was carried out. D.C. also gratefully acknowledges the support provided by the Leverhulme Trust. 
Y.X. was supported by the Fundamental Research Funds for the Central Universities (grant no. 226-2024-00200) and the National Natural Science Foundation of China (General Program no. 12374163). 
H. H. was supported by the European Research Council (ERC) under the European Union’s Horizon 2020 research and innovation
program (Grant Agreement No. 101020833). H. H. was also supported by the Gordon and Betty Moore Foundation through
Grant No.GBMF8685 towards the Princeton theory program, the Gordon and Betty Moore Foundation’s EPiQS Initiative (Grant
No. GBMF11070).

\end{acknowledgments}

\onecolumngrid
\newpage

\appendix

\renewcommand{\thetable}{S\arabic{table}}
\renewcommand{\thefigure}{S\arabic{figure}}
\renewcommand{\theequation}{S\arabic{section}.\arabic{equation}}

\tableofcontents

\section{DFT calculation and STM simulation methods}\label{app:sec:DFT_STM}

In this section, we describe the methods of DFT calculation and STM simulation for the monolayer 1H-NbSe$_2$.

\subsection{DFT calculation}
The first-principles calculations were performed on the Vienna ab initio simulation package\cite{vasp1,vasp2}. The generalized gradient approximation with the Perdew-Burke-Ernzerhof type exchange-correlation potential was adopted\cite{PBE}. The convergence accuracy of self-consistent calculations is $10^{-6}$ eV per unit cell by using $k$ grids with a $11\times11\times1$ mesh. We constructed an 11-orbital tight-binding Hamiltonian of the monolayer NbSe$_2$ using the Wannier90 package\cite{wannier90} and using the maximally localized Wannier functions of five $d$ orbitals of Nb and three $p$ orbitals of two Se.

\subsection{STM Simulation}
Using the tight-binding Hamiltonian and Wannier functions, we calculated the real-space charge density distribution (CDD) contributed by states within a specified energy window. The CDD provides a direct basis for comparison with scanning tunneling microscopy (STM) experiments. The calculation method is detailed below.

We start from the \textit{ab initio} Bloch wavefunctions $\psi_{n\vk}^{0}(\vrr)$ expressed in the plane wave basis:
\begin{equation}
\begin{aligned}
\psi_{n\vk}^{0}(\vrr) &= \frac{1}{\sqrt{\Omega_0}} \sum_{\vG} e^{i(\vk+\vG)\cdot \vrr} C_{n\vk}^{0,\vG},
\end{aligned}
\end{equation}
where $\Omega_0$ is the system column and $C_{n\vk}^{0,\vG}$ is the plane wave coefficients, which are assumed to be normalized, i.e., $\sum_{\vG} |C_{n\vk}^{0,\vG}|^2= \Omega_0$. 
However, the \textit{ab initio} Bloch functions $\psi_{n\vk}^{0}(\vrr)$ have random gauges over the BZ. In order to obtain a smooth-gauged $\psi_{n\vk}(\vrr)$, we use \textit{Wannier90} to obtain a unitary transformation $U_{\vk}$, i.e., 
\begin{equation}
\begin{aligned}
\psi_{i\vk}(\vrr)&=\sum_m U_{\vk,im} \psi_{m\vk}^{0}(\vrr) 
\end{aligned}    
\end{equation}
$U_{\vk}$ is composed of two transformations $U_{\vk} = U_{\vk}^{ML} U_{\vk}^{Dis}$, 
where $U_{\vk}^{Dis}$ is for disentanglement and $U_{\vk}^{ML}$ for obtaining maximally localized Wannier functions (MLWFs). As $U_{\vk}$ is unitary, the transformed Bloch wavefunction $\psi_{i\vk}(\vrr)$ is also normalized for each $(i,\vk)$. 
The MLWFs are then constructed using the Fourier transformation 
\begin{equation}
\begin{aligned}
W_{i,\vR}(\vrr)&= \frac{1}{\sqrt{N}} \sum_{\vk}  \psi_{i\vk}(\vrr) e^{i\vk\cdot(\vR-\mathbf{t}_i)},
\end{aligned}    
\end{equation}
where $N$ is the number of unit cells and $\mathbf{t}_i$ is the center of the $i$-th Wannier function. 

From the MLWFs $W_{i,\vR}(\vrr)$, we obtain the tight-binding Hamiltonian $h(\vk)$ from Wannier90 with eigenstates $u_{n\vk}^i$ for the $n$-th band $\epsilon_{n\vk}$ ($i$ denotes the $i$-th Wannier component).
Then the CDD measured within an energy window $[E_1,E_2]$ and tip distance $d_0$ is given by
\begin{equation}
\begin{aligned}
A(\vrr_{\parallel})= \sum_{\substack{n\vk\\E_1 \le \epsilon_{n\vk}\le E_2}} |\sum_i u_{n\vk}^i \psi_{i\vk}(\vrr_{\parallel}, d_0)|^2,
\end{aligned}
\end{equation}
where $\vrr_{\parallel}$ denotes inplane coordinates.

\section{Single-Particle Models: 6-band, 3-band and 1-band}
\label{app:TB_Models}

In this section, we describe the three models for the monolayer 1H-MoSe$_2$.

\begin{figure}[t]
    \centering
    \includegraphics[width=7.0in]{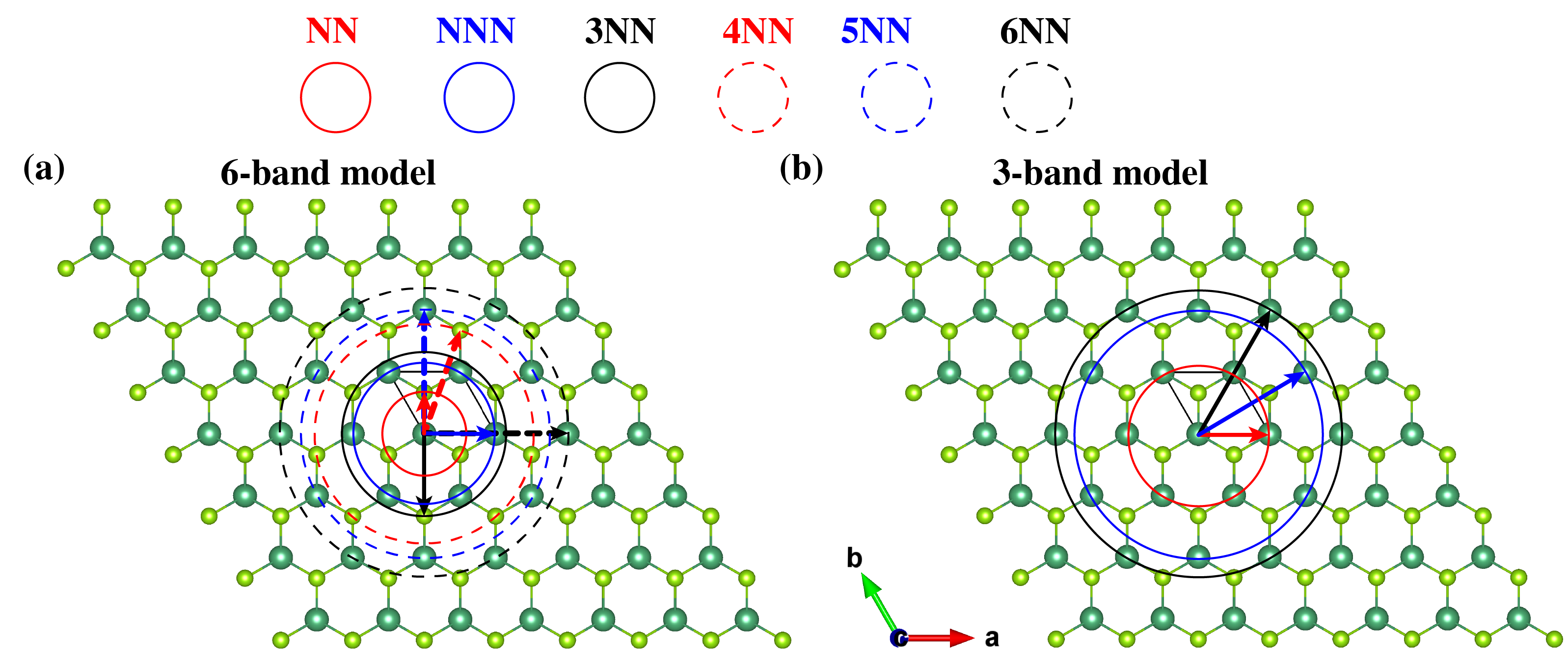}
    \caption{Crystal Structure of 1H-\nbse viewed from the top. The large and small spheres are Nb and Se atoms, respectively, where the two Se atoms related by $m_z$ symmetry are projected onto the same position. The lattice vectors are given in \eqref{eq:primitive_lattice_vectors}. The hopping distances (with respect to an Nb atom at the origin) that are used in constructing the tight-binding models are indicated by the circles of different radii in (a) for the 6-band model in \eqref{eq:HTB6band} and (b) for the 3-band model.
    }
    \label{fig:CrystalStr}
\end{figure}

In each unit cell of the monolayer 1H-MoSe$_2$, there is one Nb atom at 
\eq{
\label{eq:tau_Nb}
\bsl{\tau}_{\Nb} = ( 0,0,0)
}
and two Se atoms at 
\eq{
\label{eq:tau_Se}
\bsl{\tau}_{\Se,1} = (0,\frac{a}{\sqrt{3}},c)\text{ and }\bsl{\tau}_{\Se,2} = (0,\frac{a}{\sqrt{3}},-c)\ ,
}
where $a=3.474\AA$ is the 2D lattice constant, and $c=1.680\AA$.
The two primitive lattice vectors are 
\eqa{
\label{eq:primitive_lattice_vectors}
& \bsl{a}_1=a(1,0,0) \\
& \bsl{a}_2=a(-\frac{1}{2},\frac{\sqrt{3}}{2},0)\ .
}
The crystal structure of 1H-{\nbse} viewed from the top is shown in \figref{fig:CrystalStr}.
1H-{\nbse} has plane group $p3m1$ generated by the three-fold rotational symmetry $C_3$ along $z$ and the mirror symmetry $m_x$ that flips $x$; it also has the mirror symmetry $m_z$ that flips $z$ direction, and the time-reversal (TR) symmetry.

The 11 DFT electron bands that are closest to the Fermi level are shown in \figref{fig:DFT_el}(a) in the main text.
These 11 bands form an isolated set with nonzero direct gaps from the bands above and below.
The 11 bands are split into $m_z$-even 6 bands and $m_z$-odd 5 bands (see \eqref{fig:DFT_el}(a)); since the Fermi energy only cuts the bands in the $m_z$-even sector --- and since the closest mirror odd band is almost 1eV away from the Fermi level--- from now on we will only consider the electron bands in the $m_z$-even sector. 
Using Wannier90~\cite{wannier90}, we numerically build the Wannier functions directly from the DFT Bloch states for the 11 bands without any fitting.
We find that the 11 Wannier functions are very close to 11 atomic orbitals in one unit cell, namely $(d_{z^2}, d_{xz}, d_{yz}, d_{x^2-y^2}, d_{xy})$ of one Nb atom and $(p_z, p_x, p_y)$ of both two Se atoms. This is expected due to the well-separated set of 11 bands from the low-energy core states. Thus, we may directly treat the 11 Wannier functions as the 11 atomic orbitals.
Specifically, the creation operators for the 11 atomic orbitals read
\eq{
\label{eq:11band_basis}
\hat{c}^\dagger_{\bsl{R},d_{z^2}}, \hat{c}^\dagger_{\bsl{R},d_{xz}}, \hat{c}^\dagger_{\bsl{R}, d_{yz}}, \hat{c}^\dagger_{\bsl{R}, d_{x^2-y^2}}, \hat{c}^\dagger_{\bsl{R}, d_{xy}}, \hat{c}^\dagger_{\bsl{R}+\bsl{\tau}_{\Se,1},p_z}, \hat{c}^\dagger_{\bsl{R}+\bsl{\tau}_{\Se,1},p_x}, \hat{c}^\dagger_{\bsl{R}+\bsl{\tau}_{\Se,1},p_y}, \hat{c}^\dagger_{\bsl{R}+\bsl{\tau}_{\Se,2},p_z}, \hat{c}^\dagger_{\bsl{R}+\bsl{\tau}_{\Se,2},p_x}, \hat{c}^\dagger_{\bsl{R}+\bsl{\tau}_{\Se,2},p_y}\ ,
}
where $\bsl{R}$ is the lattice vector.

\subsection{6-band Model}

The 6 $m_z$-even combinations of these orbitals are the focus of our work:
\eq{
\label{eq:6_mz_even_basis_electron}
\hat{c}^\dagger_{\bsl{R},d_{z^2}}, \hat{c}^\dagger_{\bsl{R}, d_{xy}}, \hat{c}^\dagger_{\bsl{R}, d_{x^2-y^2}}, \frac{\hat{c}^\dagger_{\bsl{R}+\bsl{\tau}_{\Se,1},p_z}-\hat{c}^\dagger_{\bsl{R}+\bsl{\tau}_{\Se,2},p_z}}{\sqrt{2}}, \frac{\hat{c}^\dagger_{\bsl{R}+\bsl{\tau}_{\Se,1},p_x}+\hat{c}^\dagger_{\bsl{R}+\bsl{\tau}_{\Se,2},p_x}}{\sqrt{2}}, \frac{\hat{c}^\dagger_{\bsl{R}+\bsl{\tau}_{\Se,1},p_y}+\hat{c}^\dagger_{\bsl{R}+\bsl{\tau}_{\Se,2},p_y}}{\sqrt{2}}\ .
}

Wannier90 directly gives the 11-band hopping model with the Wannier functions as the basis.
By projecting the 11-band model to the 6-band $m_z$-even basis, we can obtain a 6-band tight-binding (TB) model, whose parameters are directly determined by Wannier90 without any fitting.
\begin{figure}
    \centering
    \includegraphics[width=0.8\columnwidth]{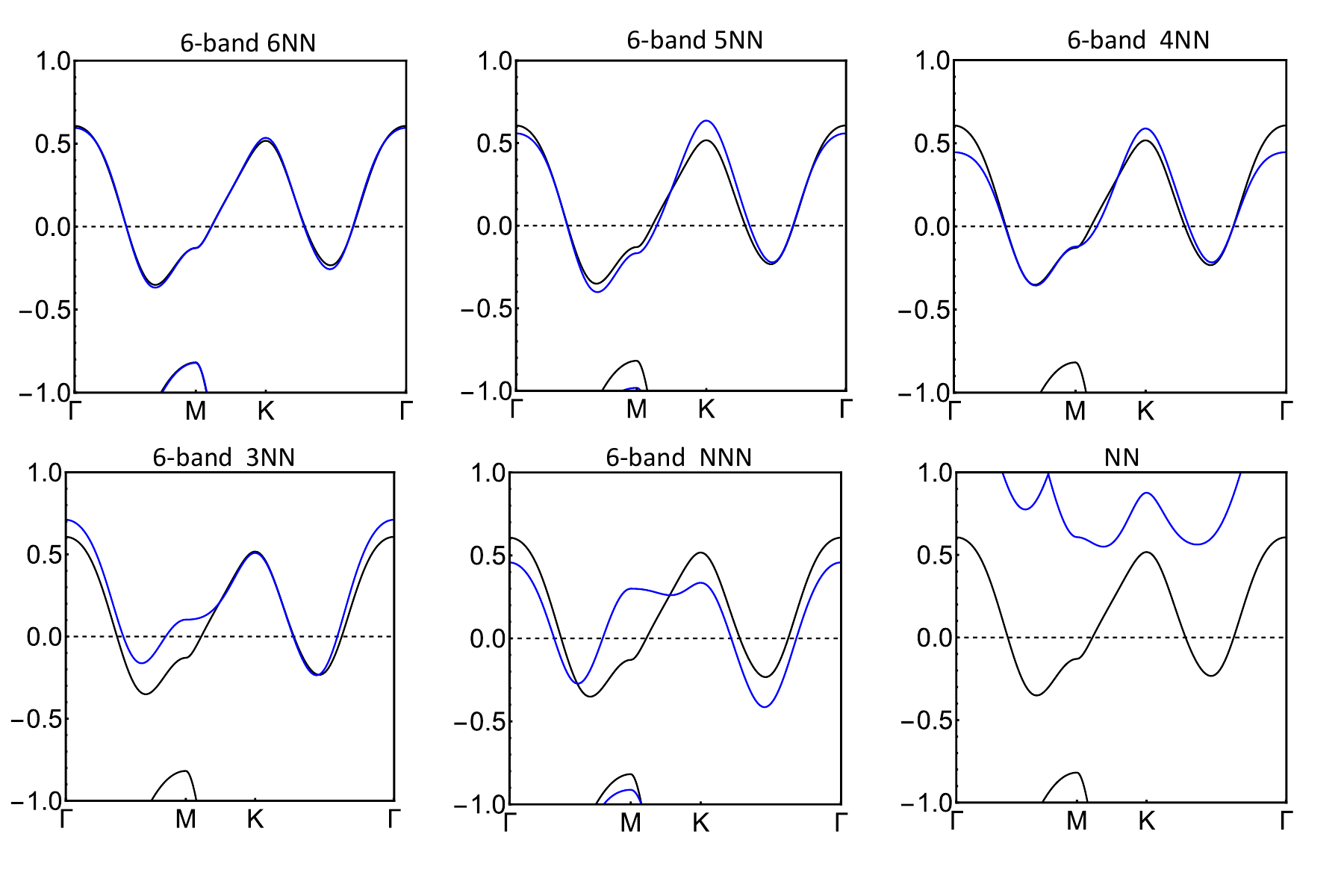}
    \caption{The black lines are the electron bands from the DFT calculation, and the blue lines are given by the truncated 6-band model in \eqref{eq:HTB6band}.
    $NN$, $NNN$, $3NN$, $4NN$, $5NN$ and $6NN$ specifies the cutoff $\Lambda$, as shown in Fig. \ref{fig:CrystalStr}(a).
    For example, $4NN$ means that we choose $\Lambda$ such that we only need hopping terms up to $4NN$.
    The mean absolute error of the single band at the Fermi level along the high-symmetry line is $0.0087$eV, $0.0469$eV, $0.0503$eV, $0.0898$eV, $0.1816$eV and $0.7391$eV for 6NN, 5NN, 4NN, 3NN, NNN and NN models respectively.
    Here the mean absolute error is defined as the mean value $|E_{\bsl{k}}^{model} - E_{\bsl{k}}^{DFT}|$ on the high-symmetry line.
    }
    \label{fig:DFT_el_H_6}
\end{figure}
However, the DFT 6-band TB model is not short-range enough for either analytical discussions or the deep analytical understanding of the physics.
To see this, let us first write out the general form of the 6-band TB model.
For convenience, we define 
\eqa{
\label{eq:6band_basis}
& (c^\dagger_{\bsl{R},d_{z^2}}, c^\dagger_{\bsl{R}, d_{xy}}, c^\dagger_{\bsl{R}, d_{x^2-y^2}}) = (\hat{c}^\dagger_{\bsl{R},d_{z^2}}, \hat{c}^\dagger_{\bsl{R}, d_{xy}}, \hat{c}^\dagger_{\bsl{R}, d_{x^2-y^2}})\\
& (c^\dagger_{\bsl{R}+\bsl{\tau}_{\Se},z},c^\dagger_{\bsl{R}+\bsl{\tau}_{\Se},x},c^\dagger_{\bsl{R}+\bsl{\tau}_{\Se},y}) = (\frac{\hat{c}^\dagger_{\bsl{R}+\bsl{\tau}_{\Se,1},p_z}-\hat{c}^\dagger_{\bsl{R}+\bsl{\tau}_{\Se,2},p_z}}{\sqrt{2}}, \frac{\hat{c}^\dagger_{\bsl{R}+\bsl{\tau}_{\Se,1},p_x}+\hat{c}^\dagger_{\bsl{R}+\bsl{\tau}_{\Se,2},p_x}}{\sqrt{2}}, \frac{\hat{c}^\dagger_{\bsl{R}+\bsl{\tau}_{\Se,1},p_y}+\hat{c}^\dagger_{\bsl{R}+\bsl{\tau}_{\Se,2},p_y}}{\sqrt{2}})
}
with 
\eq{
\bsl{\tau}_{\Se} = \frac{\bsl{\tau}_{\Se,1}+\bsl{\tau}_{\Se,2}}{2} =  (0,\frac{a}{\sqrt{3}}, 0)
}
the projection of Se sublattices onto the $x-y$ plane, and
\eq{
\label{eq:c_dagger_basis}
c^\dagger_{\bsl{R}+\bsl{\tau}} = \left\{ 
\begin{array}{cc}
  (c^\dagger_{\bsl{R},d_{z^2}}, c^\dagger_{\bsl{R}, d_{xy}}, c^\dagger_{\bsl{R}, d_{x^2-y^2}})  & ,\bsl{\tau} = \bsl{\tau}_{\Nb}\\
 (c^\dagger_{\bsl{R}+\bsl{\tau}_{\Se},x},c^\dagger_{\bsl{R}+\bsl{\tau}_{\Se},y},c^\dagger_{\bsl{R}+\bsl{\tau}_{\Se},z})  & , \bsl{\tau} = \bsl{\tau}_{\Se} \ .
\end{array}
\right.\ .
}
Then, a general  6-band model reads
\eq{
\label{eq:HTB6band}
H_6 = \sum_{\bsl{R}\bsl{R}',\bsl{\tau}\bsl{\tau}'}^{\Lambda}c^\dagger_{\bsl{R}+\bsl{\tau}}  t_{\bsl{\tau}\bsl{\tau}'}(\bsl{R}+\bsl{\tau}-\bsl{R}'-\bsl{\tau}') c_{\bsl{R}'+\bsl{\tau}'}\ ,
}
where $t_{\bsl{\tau}\bsl{\tau}'}(\bsl{R}+\bsl{\tau}-\bsl{R}'-\bsl{\tau}')$ is a $3\times 3$ matrix (there are $3$ $m_z$-even orbitals at both Nb and projected Se sites), and $\Lambda$ is a range cutoff which means we only include $t_{\bsl{\tau}\bsl{\tau}'}(\bsl{R}+\bsl{\tau}-\bsl{R}'-\bsl{\tau}')$ with $|\bsl{R}+\bsl{\tau}-\bsl{R}'-\bsl{\tau}'|\leq \Lambda$.
The symmetry representations (reps) furnished by the basis of the 6-band model (\eqref{eq:6band_basis}) are
\eqa{
& g c^\dagger_{\bsl{R}+\bsl{\tau}} g^\dagger = c^\dagger_{g(\bsl{R}+\bsl{\tau})} U_{g}^{\bsl{\tau}\bsl{\tau}} \\
& \TR c^\dagger_{\bsl{R}+\bsl{\tau}} \TR^\dagger = c^\dagger_{\bsl{R}+\bsl{\tau}}\ ,
}
where $g=C_3, m_x$ (recall that $C_3$ is the three-fold rotation symmetry along $z$ and $m_x$ is the mirror symmetry that flips $x$), 
\eqa{
\label{eq:6band_sym_rep}
& U_{C_3}^{\bsl{\tau}_{Nb}\bsl{\tau}_{Nb}} = e^{-\ii L_z^{Nb} \frac{2\pi}{3} } = \left(
\begin{array}{ccc}
 1 & 0 & 0 \\
 0 & -\frac{1}{2} & -\frac{\sqrt{3}}{2} \\
 0 & \frac{\sqrt{3}}{2} & -\frac{1}{2} \\
\end{array}
\right)\\ 
& U_{C_3}^{\bsl{\tau}_{Se}\bsl{\tau}_{Se}} = e^{-\ii L_z^{Se} \frac{2\pi}{3} } =  \left(
\begin{array}{ccc}
 1 & 0 & 0 \\
 0 & -\frac{1}{2} & -\frac{\sqrt{3}}{2} \\
 0 & \frac{\sqrt{3}}{2} & -\frac{1}{2} \\
\end{array}
\right) \ ,\\
& U_{m_x}^{\bsl{\tau}_{Nb}\bsl{\tau}_{Nb}} = U_{m_x}^{\bsl{\tau}_{Se}\bsl{\tau}_{Se}} = \left(
\begin{array}{ccc}
 1 & 0 & 0 \\
 0 & -1 & 0 \\
 0 & 0 & 1 \\
\end{array}
\right)\ ,
}
$L_z^{Nb}$ is the z-component angular momentum matrix projected to the three $d$ orbitals of Nb atoms which reads
\eq{
\label{eq:L_z_Nb}
L_z^{Nb} 
= 
\mat{ 
0 & 0 & 0 \\
0 & 0 &  2\ii \\
0 & -2\ii & 0
}\ ,
}
and $L_z^{Se}$ is the z-component angular momentum matrix projected to the three $p$ orbitals of Se atoms which reads
\eq{
\label{eq:L_z_Se}
L_z^{Se} 
= 
\mat{ 
0 & 0 & 0 \\
0 & 0 & - \ii \\
0 & \ii & 0
}\ .
}
As a result, the hopping $t_{\bsl{\tau}\bsl{\tau}'}(\bsl{R}+\bsl{\tau}-\bsl{R}'-\bsl{\tau}')$ satisfies
\eqa{
\label{eq:6band_sym_t}
& U_{g}^{\bsl{\tau}\bsl{\tau}}  t_{\bsl{\tau}\bsl{\tau}'}(\bsl{R}+\bsl{\tau}-\bsl{R}'-\bsl{\tau}') \left[U_{g}^{\bsl{\tau}'\bsl{\tau}'}\right]^\dagger = t_{\bsl{\tau}\bsl{\tau}'}(g(\bsl{R}+\bsl{\tau}-\bsl{R}'-\bsl{\tau}')) \\
& t_{\bsl{\tau}\bsl{\tau}'}^*(\bsl{R}+\bsl{\tau}-\bsl{R}'-\bsl{\tau}') = t_{\bsl{\tau}\bsl{\tau}'}(\bsl{R}+\bsl{\tau}-\bsl{R}'-\bsl{\tau}') \\
& t_{\bsl{\tau}\bsl{\tau}'}^\dagger(\bsl{R}+\bsl{\tau}-\bsl{R}'-\bsl{\tau}') = t_{\bsl{\tau}'\bsl{\tau}}(\bsl{R}'+\bsl{\tau}'-\bsl{R}-\bsl{\tau})\ ,
}
where the second equality comes from time-reversal and the last comes from the Hermiticity.

The parameter values of the 6-band model obtained from DFT are listed in the following.
For on-site energies, we find
\eqa{\label{onsiteSeNb6Band}
& t_{\bsl{\tau}_{\Nb}\bsl{\tau}_{\Nb}}(\bsl{0}) =\left(
\begin{array}{ccc}
 E_{d_{z^2}} & 0 &  \\
 0 & E_{d_{xy}} & 0 \\
 0 & 0 & E_{d_{x^2-y^2}} \\
\end{array}
\right)= \left(
\begin{array}{ccc}
 0.4787 & 0 & 0 \\
 0 & 0.5575 & 0 \\
0 & 0 & 0.5575 \\
\end{array}
\right) \\
& t_{\bsl{\tau}_{\Se}\bsl{\tau}_{\Se}}(\bsl{0}) = \left(
\begin{array}{ccc}
 E_{z} & 0 &  \\
 0 & E_{x} & 0 \\
 0 & 0 & E_{y} \\
\end{array}
\right)
=
\left(
\begin{array}{ccc}
 -2.4102 & 0 & 0 \\
 0 & -1.6090 & 0 \\
 0 & 0 & -1.6090 \\
\end{array}
\right)\ .
}
Here the Se onsite energies (from $4p$ orbitals) are smaller than those of Nb onsite energies (from $4d$ orbitals) because $4p$ electrons penetrate deeper towards core than $4d$ orbitals.
For NN hoppings, we find
\eqa{ \label{tNNSeNb6Band}
& t_{\bsl{\tau}_{\Se}\bsl{\tau}_{\Nb}}(\bsl{\tau}_{\Se}) =  \left(
\begin{array}{ccc}
 t_{NN,z,d_{z^2}} & 0 & t_{NN,z,d_{x^2-y^2}} \\
 0 & t_{NN,x,d_{xy}} & 0 \\
 t_{NN,y,d_{z^2}} & 0 & t_{NN,y,d_{x^2-y^2}} \\
\end{array}
\right) = \left(
\begin{array}{ccc}
 0.7775 & 0 & -0.9739 \\
 0 & -1.2246 & 0 \\
 -0.8363 & 0 & -0.6312 \\
\end{array}
\right) \ .
}
For NNN hoppings, we find
\eqa{
& t_{\bsl{\tau}_{\Nb}\bsl{\tau}_{\Nb}}(\bsl{a}_1) = \left(
\begin{array}{ccc}
 t_{NNN,d_{z^2},d_{z^2}} &  t_{NNN,d_{z^2},d_{xy}}  & t_{NNN,d_{z^2},d_{x^2-y^2}} \\
- t_{NNN,d_{z^2},d_{xy}} & t_{NNN,d_{xy},d_{xy}} & t_{NNN,d_{xy}, d_{x^2-y^2}} \\
t_{NNN,d_{z^2},d_{x^2-y^2}} & -t_{NNN,d_{xy}, d_{x^2-y^2}} & t_{NNN,d_{x^2-y^2},d_{x^2-y^2}} \\
\end{array}
\right)= \left(
\begin{array}{ccc}
 -0.2285 & -0.0842 & -0.3625 \\
 0.0842 & 0.2364 & 0.1603 \\
 -0.3625 & -0.1603 & -0.4730 \\
\end{array}
\right)\\
& t_{\bsl{\tau}_{\Se}\bsl{\tau}_{\Se}}(\bsl{a}_1)= \left(
\begin{array}{ccc}
 t_{NNN,z,z} &  t_{NNN,z,x}  & t_{NNN,z,y} \\
- t_{NNN,z,x} & t_{NNN,x,x} & t_{NNN,x,y} \\
t_{NNN,z,y} & -t_{NNN,x, y} & t_{NNN,y,y} \\
\end{array}
\right) = \left(
\begin{array}{ccc}
 -0.1967 & 0.1448 & -0.0143 \\
 -0.1448 & 0.8330 & 0.0265 \\
 -0.0143 & -0.0265 & 0.0355 \\
\end{array}
\right)\ ;
\label{SeSeHoppinga16Band}
}
for 3NN hoppings, we find
\eqa{
& t_{\bsl{\tau}_{\Se}\bsl{\tau}_{\Nb} }(-\bsl{a}_1-2\bsl{a}_2+\bsl{\tau}_{\Se}) =  \left(
\begin{array}{ccc}
 t_{3NN,z,d_{z^2}} & 0 & t_{3NN,z,d_{x^2-y^2}} \\
 0 & t_{3NN,x,d_{xy}} & 0 \\
 t_{3NN,y,d_{z^2}} & 0 & t_{3NN,y,d_{x^2-y^2}} \\
\end{array}
\right)  = \left(
\begin{array}{ccc}
 0.0962 & 0 & -0.2000 \\
 0 & 0.0071 & 0 \\
 -0.1381 & 0 & 0.2048 \\
\end{array}
\right) \ ;
}
for 4NN hoppings, we find
\eqa{
& t_{\bsl{\tau}_{\Se}\bsl{\tau}_{\Nb} }(\bsl{a}_1+\bsl{a}_2+\bsl{\tau}_{\Se}) =  \left(
\begin{array}{ccc}
 t_{4NN,z,d_{z^2}} &  t_{4NN,z,d_{xy}}  & t_{4NN,z,d_{x^2-y^2}} \\
 t_{4NN,x,d_{z^2}} & t_{4NN,x,d_{xy}} & t_{4NN,x, d_{x^2-y^2}} \\
t_{4NN,y,d_{z^2}} & t_{4NN,y,d_{xy}} & t_{4NN,y,d_{x^2-y^2}} \\
\end{array}
\right) = \left(
\begin{array}{ccc}
 -0.0502 & -0.0793 & 0.0229 \\
 -0.0005 & 0.0116 & 0.0019 \\
 -0.0367 & -0.0530 & 0.0191 \\
\end{array}
\right)\ ;
}
for 5NN hoppings, we find
\eqa{
& t_{\bsl{\tau}_{\Nb}\bsl{\tau}_{\Nb}}(\bsl{a}_1+2\bsl{a}_2) = \left(
\begin{array}{ccc}
 t_{5NN,d_{z^2},d_{z^2}} & 0 & t_{5NN,d_{z^2},d_{x^2-y^2}} \\
 0 & t_{5NN,d_{xy},d_{xy}} & 0 \\
 t_{5NN,d_{x^2-y^2},d_{z^2}} & 0 & t_{5NN,d_{x^2-y^2},d_{x^2-y^2}} \\
\end{array}
\right) = \left(
\begin{array}{ccc}
 0.0094 & 0 & -0.0059 \\
 0 & -0.0051 & 0 \\
 -0.0104 & 0 & 0.0282 \\
\end{array}
\right)\\
& t_{\bsl{\tau}_{\Se}\bsl{\tau}_{\Se}}(\bsl{a}_1+2\bsl{a}_2) = \left(
\begin{array}{ccc}
 t_{5NN,z,z} & 0 & t_{5NN,z,y} \\
 0 & t_{NN,x,x} & 0 \\
 t_{5NN,y,z} & 0 & t_{5NN,y,y} \\
\end{array}
\right) =\left(
\begin{array}{ccc}
 0.0445 & 0 & -0.0509 \\
 0 & -0.0079 & 0 \\
 0.0169 & 0 & -0.0222 \\
\end{array}
\right)\ ;
}
for 6NN hoppings, we find
\eqa{
& t_{\bsl{\tau}_{\Nb}\bsl{\tau}_{\Nb}}(2\bsl{a}_1) = \left(
\begin{array}{ccc}
 t_{6NN,d_{z^2},d_{z^2}} &  t_{6NN,d_{z^2},d_{xy}}  & t_{6NN,d_{z^2},d_{x^2-y^2}} \\
- t_{6NN,d_{z^2},d_{xy}} & t_{6NN,d_{xy},d_{xy}} & t_{6NN,d_{xy}, d_{x^2-y^2}} \\
t_{6NN,d_{z^2},d_{x^2-y^2}} & -t_{6NN,d_{xy}, d_{x^2-y^2}} & t_{6NN,d_{x^2-y^2},d_{x^2-y^2}} \\
\end{array}
\right) = \left(
\begin{array}{ccc}
 0.0110 & 0.0056 & 0.0213 \\
 -0.0056 & 0.0057 & -0.0091 \\
 0.0213 & 0.0091 & 0.0322 \\
\end{array}
\right) \\
& t_{\bsl{\tau}_{\Se}\bsl{\tau}_{\Se}}(2\bsl{a}_1) = \left(
\begin{array}{ccc}
 t_{6NN,z,z} &  t_{6NN,z,x}  & t_{6NN,z,y} \\
- t_{6NN,z,x} & t_{6NN,x,x} & t_{6NN,x,y} \\
t_{6NN,z,y} & -t_{6NN,x, y} & t_{6NN,y,y} \\
\end{array}
\right) = \left(
\begin{array}{ccc}
 -0.0089 & 0.0125 & 0.0073 \\
 -0.0125 & 0.0611 & -0.0010 \\
 0.0073 & 0.0010 & -0.0060 \\
\end{array}
\right)\ ,
}
where the unit is eV.

\begin{figure}
    \centering
    \includegraphics[width=7.0in]{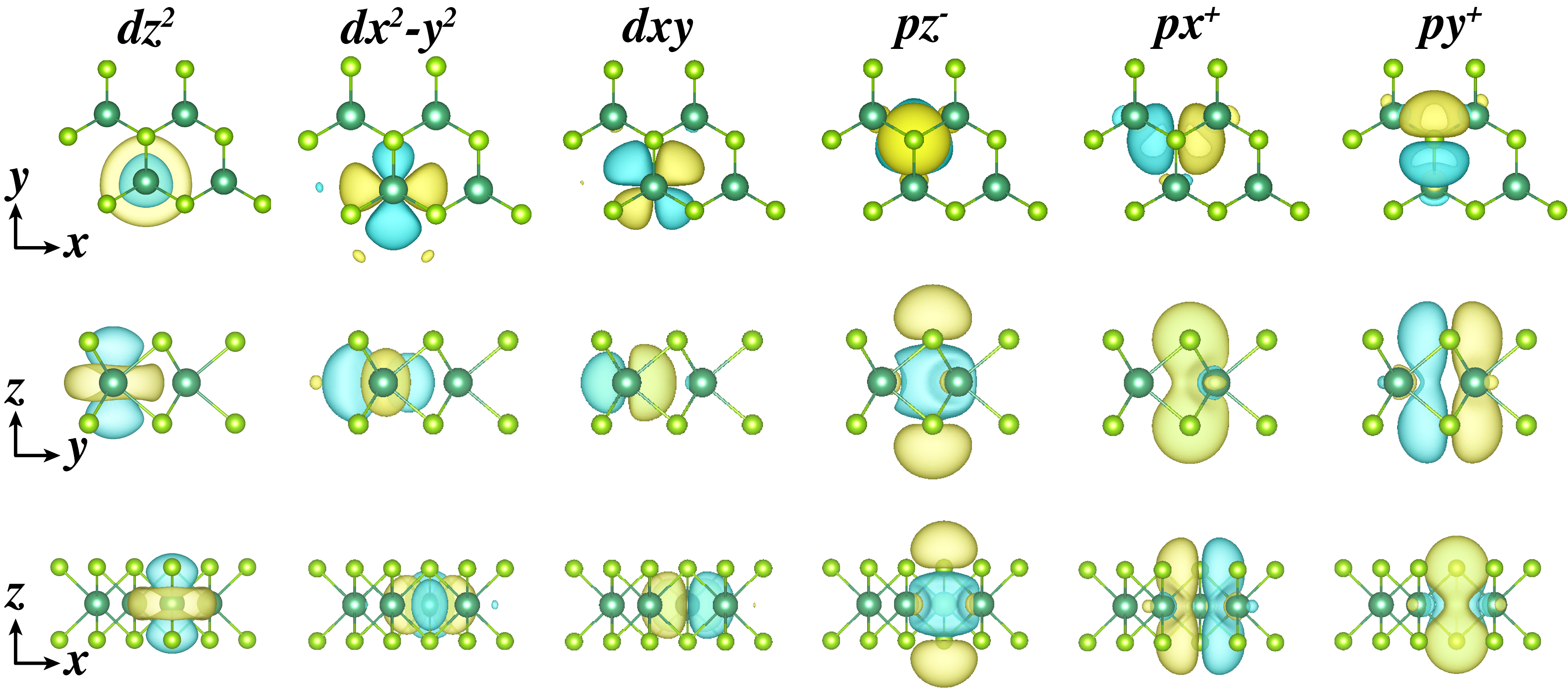}
    \caption{Real-space distribution of the six mirror-even Wannier functions $\{d_{z^2},d_{x^2-y^2},d_{xy}\}$ on Nb and $\{p_{z^{-}},p_{x^+},p_{y^+}\}$, as defined in Eq. \ref{main_eq:6band_basis}. For each Wannier function in one column, we present it from the top view along the $z$ direction and side view from both $x$ and $y$ directions. The transparent yellow (positive) and blue (negative) colors indicate opposite signs of wavefunction values. Note that the $d_{z^2}$ orbital in the plot has the opposite sign from the atomic $d_{z^2}$ orbital. }
    \label{fig:NbSe2_orbitals1}
\end{figure}

\begin{figure}
    \centering
    \includegraphics[width=7.0in]{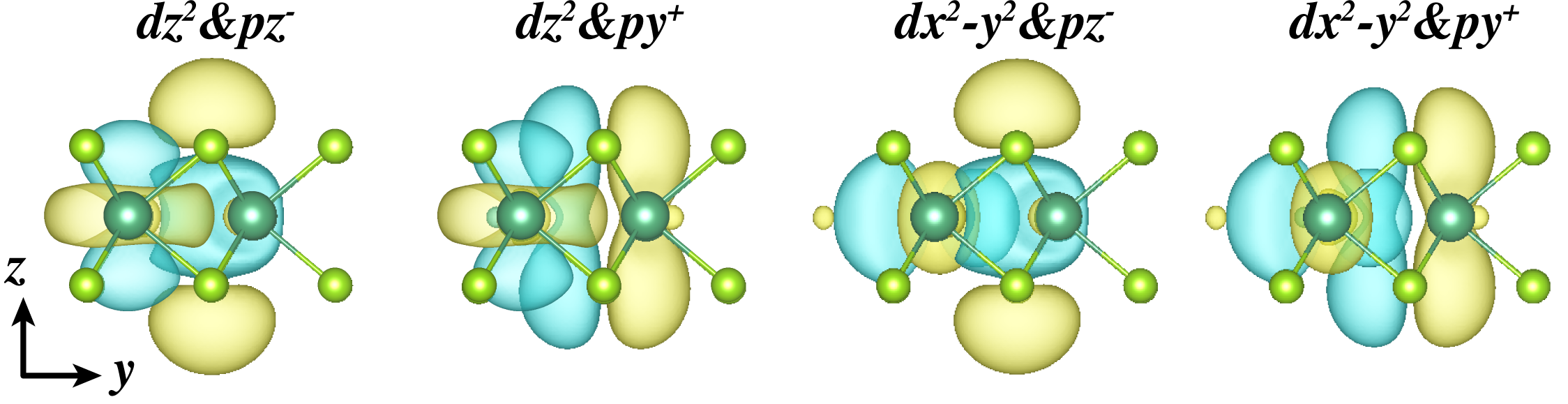}
    \caption{The overlap between two different Wannier functions (as labeled on the top of each plot) from the view along $x$ direction. In the plot, $d$ orbitals are from Nb, while $p_i^{\pm}=\frac{1}{\sqrt{2}}(p_i^{\text{Se}_1}\pm p_i^{\text{Se}_2)})$ are effective $m_z$-even and odd $p$ orbitals from two Se atoms, as defined in Eq. \ref{main_eq:6band_basis}. }
    \label{fig:NbSe2_orbitals2}
\end{figure}

For the NN hopping parameters,  the signs of $t_{NN,z,d_{z^2}}$, $t_{NN,y,d_{z^2}}$, $t_{NN,z,d_{x^2-y^2}}$ and $t_{NN,y,d_{x^2-y^2}}$ can be directly understood from the sign of the overlap of the Wannier functions.
As shown \figref{fig:NbSe2_orbitals2}, the Se $z$ and Nb $d_{z^2}$ have mainly the opposite-sign overlap (\ie, the overlap is between the part of Wannier functions of the opposite sign), while Se $p_z$ to Nb $d_{x^2-y^2}$, Se $p_y$ to Nb $d_{z^2}$, and Se $p_y$ to Nb $d_{{x^2-y^2}}$ all have mainly same-sign overlap (\ie, the overlap is between the part of Wannier functions of the same signs); thus $t_{NN,z,d_{z^2}}$ is positive while $t_{NN,y,d_{z^2}}$, $t_{NN,z,d_{x^2-y^2}}$ and $t_{NN,y,d_{x^2-y^2}}$ are negative.
We note that, although we use the notation of atomic orbitals, some of the \textit{ab initio} Wannier functions of the 6-band model exhibit sign differences from the atomic orbitals. These sign differences arise from the numerical Wannierization procedure in \textit{Wannier90}, which generally results in random signs during the calculation. 

As shown in \figref{fig:DFT_el_H_6}, if we do not change the values of any parameters obtained from DFT, including terms up to 4NN is required to have a reasonably good description of the band dispersion of the single band at the Fermi level.
Therefore, the 6-band model is complicated, unless we change the parameter values (as we will do in \appref{app:analytic} to simplify the hoppings).
Another way to reduce the complexity is to build an analytic 3-band from the full DFT 6-band model, as we will do in the next section (\appref{app:3-band_model}).
Specifically, we will use Wannier90 to construct three Wannier functions from the full DFT 6-band model and obtain the 3-band TB models from Wannier90. We then derive an analytic approximation of the Wannier functions from the 6-band model. From the analytic model, we unravel the surprising effects of a Wannier center moving off the atom site.

\subsection{3-Band Model}
\label{app:3-band_model}

From the full 6-band model, we first numerically build the Wannier states for the top three bands from Wannier90.
The trial states are chosen to be three Nb d orbitals in \eqref{eq:6band_basis}, since we know the Nb d orbitals have much higher energies ($\sim 2 \eV$ more) than the Se p orbitals according to the parameters' values in \eqref{onsiteSeNb6Band}. Hence our resulting Wannier states will be "renormalized" Nb even d-orbitals.
We label creation operators for the resultant Wannier states as 
\eq{
\label{eq:3band_basis}
\widetilde{c}^\dagger_{\bsl{R}} = ( \widetilde{c}^\dagger_{\bsl{R},d_{z^2}}, \widetilde{c}^\dagger_{\bsl{R},d_{xy}}, \widetilde{c}^\dagger_{\bsl{R},d_{x^2-y^2}})\ .
}
The symmetry reps furnished by \eqref{eq:3band_basis} read
\eqa{
\label{eq:3band_sym_rep}
& g \widetilde{c}^\dagger_{\bsl{R}} g^{-1} = \widetilde{c}^\dagger_{g \bsl{R}} U_g \ ,\  \TR \widetilde{c}^\dagger_{\bsl{R}} \TR^{-1} = \widetilde{c}^\dagger_{\bsl{R}} \ ,
}
where $g=C_3,m_x$,
\eqa{
\label{eq:3band_sym_rep_Ug}
& U_{C_3} =\left(
\begin{array}{ccc}
 1 & 0 & 0 \\
 0 & -\frac{1}{2} & -\frac{\sqrt{3}}{2} \\
 0 & \frac{\sqrt{3}}{2} & -\frac{1}{2} \\
\end{array}
\right) \ ,\  U_{m_x} = \left(
\begin{array}{ccc}
 1 & 0 & 0 \\
 0 & -1 & 0 \\
 0 & 0 & 1 \\
\end{array}
\right) \ .
}
With the basis in \eqref{eq:3band_basis}, the 3-band model reads
\eq{
\label{eq:3-band_model_el}
H_3 = \sum_{\bsl{R}\bsl{R}'} \widetilde{c}^\dagger_{\bsl{R}} \widetilde{t}(\bsl{R}-\bsl{R}') \widetilde{c}_{\bsl{R}'}\ ,
}
where $\widetilde{t}(\bsl{R}-\bsl{R}')$ is a $3\times 3$ matrix.
As a result, $\widetilde{t}(\bsl{R}-\bsl{R}')$ satisfies
\eqa{
\label{eq:3band_sym_t}
& U_g \widetilde{t}(\bsl{R}-\bsl{R}') U_g^\dagger = \widetilde{t}(g(\bsl{R}-\bsl{R}')) \\
& t^*(\bsl{R}-\bsl{R}') = t(\bsl{R}-\bsl{R}') \\
& t^\dagger(\bsl{R}-\bsl{R}') = t(\bsl{R}'-\bsl{R})\ ,
}
where the last equality comes from the Hermiticity.
Up to NNN, the independent hopping terms are
\eqa{
\label{eq:3band_hopping_form}
& \widetilde{t}(0) = \left(
\begin{array}{ccc}
 E_{z^2} & 0 &  \\
 0 & E_{xy} & 0 \\
 0 & 0 & E_{xy} \\
\end{array}
\right)\ ,\ \widetilde{t}(\bsl{a}_1) = \left(
\begin{array}{ccc}
 t_{NN,z^2,z^2} & t_{NN,z^2,xy} & t_{NN,z^2,x^2-y^2} \\
 - t_{NN,z^2,xy} & t_{NN,xy,xy} & t_{NN,xy,x^2-y^2} \\
 t_{NN,z^2,x^2-y^2} & - t_{NN,xy,x^2-y^2} & t_{NN,x^2-y^2,x^2-y^2} \\
\end{array}
\right)\\
& \widetilde{t}(\bsl{a}_1+2\bsl{a}_2) = \left(
\begin{array}{ccc}
 t_{NNN,z^2,z^2} & 0 & t_{NNN,z^2,x^2-y^2} \\
 0 & t_{NNN,xy,xy} & 0 \\
 t_{NNN,x^2-y^2,z^2} & 0 & t_{NNN,x^2-y^2,x^2-y^2} \\
\end{array}
\right)\ .
}
Numerically, we find that the values of independent hoppings in $\widetilde{t}(\bsl{R})$ are: 
\eqa{
\label{eq:3band_hopping_values}
& \left(
\begin{array}{ccc}
 E_{z^2} & 0 &  \\
 0 & E_{xy} & 0 \\
 0 & 0 & E_{xy} \\
\end{array}
\right) = \left(
\begin{array}{ccc}
 1.3078 & 0 & 0 \\
 0 & 1.9459 & 0 \\
 0 & 0 & 1.9459 \\
\end{array}
\right)\\
& \left(
\begin{array}{ccc}
 t_{NN,z^2,z^2} & t_{NN,z^2,xy} & t_{NN,z^2,x^2-y^2} \\
 - t_{NN,z^2,xy} & t_{NN,xy,xy} & t_{NN,xy,x^2-y^2} \\
 t_{NN,z^2,x^2-y^2} & - t_{NN,xy,x^2-y^2} & t_{NN,x^2-y^2,x^2-y^2} \\
\end{array}
\right) = \left(
\begin{array}{ccc}
 -0.1436 & -0.3317 & -0.3593 \\
 0.3317 & 0.2446 & 0.2754 \\
 -0.3593 & -0.2754 & -0.0794 \\
\end{array}
\right)\\
&  \left(
\begin{array}{ccc}
 t_{NNN,z^2,z^2} & 0 & t_{NNN,z^2,x^2-y^2} \\
 0 & t_{NNN,xy,xy} & 0 \\
 t_{NNN,x^2-y^2,z^2} & 0 & t_{NNN,x^2-y^2,x^2-y^2} \\
\end{array}
\right) =\left(
\begin{array}{ccc}
 0.0370 & 0 & 0.0617 \\
 0 & 0.0380 & 0 \\
 -0.0102 & 0 & 0.0916 \\
\end{array}
\right)\ .
}
The band structure of the one band crossing the Fermi level, given by the 3-band NNN model is close to the DFT one, as shown in \figref{fig:DFT_el_H_3_H_1}(a). The representations of this one band at high symmetry points tell us that its Wannier center is located \emph{away} from the atomic Wycoff position of Nb. As such, this band is an obstructed atomic band representation (OABR)~\cite{Bradlyn2017TQC,xu2021three}. This will be analytically clearer once we obtain the analytic form of the Wannier states, through a basis rotation, as below.

\subsubsection{Approximated Analytical Forms of the Wannier Basis of the 3-band Model}

The creation operators for Wannier basis ($\widetilde{c}^\dagger_{\bsl{R}_0}$ in \eqref{eq:3band_basis}) of the 3-band model are
\eq{
\widetilde{c}^\dagger_{\bsl{R}_0} = \sum_{\bsl{R},\bsl{\tau}} c^\dagger_{\bsl{R}-\bsl{R}_0+\bsl{\tau}} \widetilde{\xi}_{\bsl{R}+\bsl{\tau}}\ ,
}
where $c^\dagger_{\bsl{R}+\bsl{\tau}}$ is defined in \eqref{eq:6band_basis} (which contains both Nb and Se orbitals depending on $\bsl{\tau}$), and $\widetilde{\xi}_{\bsl{R}+\bsl{\tau}}$ is a $3 \times 3$ matrix.
From the symmetry reps in \eqref{eq:3band_sym_rep}, we obtain the constraints on $\xi_{\bsl{R}+\bsl{\tau}}$, which read
\eqa{
\label{eq:3band_basis_xi_tilde_sym}
& U_{g}^{\bsl{\tau}\bsl{\tau}} \widetilde{\xi}_{\bsl{R}+\bsl{\tau}} = \widetilde{\xi}_{g(\bsl{R}+\bsl{\tau})} U_{g}\\
& \widetilde{\xi}_{\bsl{R}+\bsl{\tau}}^* = \widetilde{\xi}_{\bsl{R}+\bsl{\tau}}\\
& \sum_{\bsl{R},\bsl{\tau}} \widetilde{\xi}_{\bsl{R}+\bsl{\tau}}^\dagger \widetilde{\xi}_{\bsl{R}+\bsl{\tau}} = 1\ ,
}
where $U_{g=C_3,m_x}^{\bsl{\tau}\bsl{\tau}} $ are in \eqref{eq:6band_sym_rep}, and $U_{g=C_3,m_x}$ are in \eqref{eq:3band_sym_rep_Ug}. The last equality is the normalization condition of the Wannier orbitals.
Up to NNN, numerical calculation suggestion the following forms for the independent $\widetilde{\xi}_{\bsl{R}+\bsl{\tau}}$:
\eq{
\widetilde{\xi}_{\bsl{ 0}} =\left(
\begin{array}{ccc}
 0.8971 & 0 & 0 \\
 0 & 0.8446 & 0 \\
 0 & 0 & 0.8446 \\
\end{array}
\right)\ ,
}
\eq{
\widetilde{\xi}_{\bsl{\tau}_{\Se}} = \left(
\begin{array}{ccc}
 0.1872 & 0 & -0.1957 \\
 0 & -0.3209 & 0 \\
 -0.1296 & 0 & -0.1670 \\
\end{array}
\right) \ ,
}
\eq{
\widetilde{\xi}_{\bsl{a}_1} =
\left(
\begin{array}{ccc}
 -0.0051 & 0.0068 & -0.0251 \\
 -0.0184 & -0.0106 & -0.0112 \\
 0.0032 & 0.0087 & -0.0486 \\
\end{array}
\right)\ .
}

To derive the approximated form of $\widetilde{\xi}_{\bsl{R}+\bsl{\tau}}$, we rotate $\widetilde{c}^\dagger_{\bsl{R},z^2/xy/x^2-y^2}$ to a new form:
\eq{
\label{eq:3-band_model_rotated_basis}
(\widetilde{c}^\dagger_{\bsl{R},1}, \widetilde{c}^\dagger_{\bsl{R},2}, \widetilde{c}^\dagger_{\bsl{R},3} ) =  (\widetilde{c}^\dagger_{\bsl{R},z^2}, \widetilde{c}^\dagger_{\bsl{R},xy}, \widetilde{c}^\dagger_{\bsl{R},x^2-y^2}) R\ , 
}
where 
\eq{
\label{eq:R_rotation}
R = \left(
\begin{array}{ccc}
 \frac{1}{\sqrt{3}} & \frac{1}{\sqrt{3}} & \frac{1}{\sqrt{3}} \\
 0 & \frac{1}{\sqrt{2}} & -\frac{1}{\sqrt{2}} \\
 -\sqrt{\frac{2}{3}} & \frac{1}{\sqrt{6}} & \frac{1}{\sqrt{6}} \\
\end{array}
\right)\ .
}
In the rotated basis, the symmetry rep reads
\eqa{
& C_3 (\widetilde{c}^\dagger_{\bsl{R},1}, \widetilde{c}^\dagger_{\bsl{R},2}, \widetilde{c}^\dagger_{\bsl{R},3} ) C_3^{-1} = (\widetilde{c}^\dagger_{C_3\bsl{R},1}, \widetilde{c}^\dagger_{C_3\bsl{R},2}, \widetilde{c}^\dagger_{C_3\bsl{R},3} ) \mat{ 0 & 0 & 1\\ 1 & 0 & 0 \\ 0 & 1 & 0 } \\
& m_x (\widetilde{c}^\dagger_{\bsl{R},1}, \widetilde{c}^\dagger_{\bsl{R},2}, \widetilde{c}^\dagger_{\bsl{R},3} ) m_x^{-1} = (\widetilde{c}^\dagger_{m_x\bsl{R},1}, \widetilde{c}^\dagger_{m_x\bsl{R},2}, \widetilde{c}^\dagger_{m_x\bsl{R},3} ) \mat{ 1 & 0 & 0\\ 0 & 0 & 1 \\ 0 & 1 & 0 } \\
& \TR (\widetilde{c}^\dagger_{\bsl{R},1}, \widetilde{c}^\dagger_{\bsl{R},2}, \widetilde{c}^\dagger_{\bsl{R},3} )\TR^{-1} = (\widetilde{c}^\dagger_{\bsl{R},1}, \widetilde{c}^\dagger_{\bsl{R},2}, \widetilde{c}^\dagger_{\bsl{R},3} ) \ .
}
Note the $C_3$ and $m_x$ are permutation matrices. 
After the rotation, the numerical data suggest that
\eqa{
\label{eq:3-band_model_basis}
& \widetilde{c}^\dagger_{\bsl{R},1} = x_1 ( \frac{1}{\sqrt{3}}  c^\dagger_{\bsl{R},d_{z^2}} -\frac{\sqrt{2}}{\sqrt{3}}  c^\dagger_{\bsl{R},d_{x^2-y^2}}  ) + x_2 c^\dagger_{\bsl{R}+\bsl{\tau}_{\Se}, z} + x_2 c^\dagger_{\bsl{R}-\bsl{a}_2+\bsl{\tau}_{\Se}, y} + x_2 c^\dagger_{\bsl{R}-\bsl{a}_1-\bsl{a}_2+\bsl{\tau}_{\Se}, y} + ... \\
& \widetilde{c}^\dagger_{\bsl{R},2} = x_1 ( \frac{1}{\sqrt{3}}  c^\dagger_{\bsl{R},d_{z^2}}  +\frac{1}{\sqrt{2}} c^\dagger_{\bsl{R},d_{xy}}  +\sqrt{\frac{1}{6}} c^\dagger_{\bsl{R},d_{x^2-y^2}} ) + x_2 c^\dagger_{\bsl{R}-\bsl{a}_1-\bsl{a}_2+\bsl{\tau}_{\Se}, z}  + x_2 (-\frac{\sqrt{3}}{2} c^\dagger_{\bsl{R}+\bsl{\tau}_{\Se}, x} - \frac{1}{2} c^\dagger_{\bsl{R}+\bsl{\tau}_{\Se}, y})\\
& \quad \quad + x_2 (-\frac{\sqrt{3}}{2} c^\dagger_{\bsl{R}-\bsl{a}_2+\bsl{\tau}_{\Se}, x} - \frac{1}{2} c^\dagger_{\bsl{R}-\bsl{a}_2+\bsl{\tau}_{\Se}, y}) + ... \\
& \widetilde{c}^\dagger_{\bsl{R},3} = x_1 ( \frac{1}{\sqrt{3}}  c^\dagger_{\bsl{R},d_{z^2}}  - \frac{1}{\sqrt{2}} c^\dagger_{\bsl{R},d_{xy}}  + \sqrt{\frac{1}{6}} c^\dagger_{\bsl{R},d_{x^2-y^2}} ) + x_2 c^\dagger_{\bsl{R}-\bsl{a}_2+\bsl{\tau}_{\Se}, z} + x_2 (\frac{\sqrt{3}}{2} c^\dagger_{\bsl{R}+\bsl{\tau}_{\Se}, x} - \frac{1}{2} c^\dagger_{\bsl{R}+\bsl{\tau}_{\Se}, y}) \\
& \quad \quad +  x_2 (\frac{\sqrt{3}}{2} c^\dagger_{\bsl{R}-\bsl{a}_1-\bsl{a}_2+\bsl{\tau}_{\Se}, x} - \frac{1}{2} c^\dagger_{\bsl{R}-\bsl{a}_1-\bsl{a}_2+\bsl{\tau}_{\Se}, y}) + ... \ ,
}
where
\eq{
\label{eq:values_x_1_x_2_3-band_basis}
x_1 = 0.8615\ ,\ x_2 = 0.2702\ ,
}
``$...$ "labels the terms that are smaller than $0.1$, and the coefficients of $c^\dagger_{\bsl{R}+\bsl{\tau}_{\Se}, z}$ and $c^\dagger_{\bsl{R}-\bsl{a}_2+\bsl{\tau}_{\Se}}$ in $\widetilde{c}^\dagger_{\bsl{R},1}$ are approximately equal (the difference of them is less than 0.01 and is included in ``$...$ ").
We note that if we neglect the small terms in ``$...$ " in \eqref{eq:3-band_model_basis}, the Wannier functions are not exactly orthonormal anymore (but they are almost so).
From the expression, we can see that the Wannier function is dominated by the orbitals on Nb atoms.

In the rotated basis, the onsite and hoppings terms become
\eqa{
\label{eq:3band_hopping_values_Rotated}
 & R^\dagger \widetilde{t}(0) R = R^\dagger \left(
\begin{array}{ccc}
 E_{z^2} & 0 &  \\
 0 & E_{xy} & 0 \\
 0 & 0 & E_{xy} \\
\end{array}
\right) R =\left(
\begin{array}{ccc}
 1.7332 & -0.2127 & -0.2127 \\
 -0.2127 & 1.7332 & -0.2127 \\
 -0.2127 & -0.2127 & 1.7332 \\
\end{array}
\right) \\
& R^\dagger \widetilde{t}(\bsl{a}_1) R =  R^\dagger \left(
\begin{array}{ccc}
 t_{NN,z^2,z^2} & t_{NN,z^2,xy} & t_{NN,z^2,x^2-y^2} \\
 - t_{NN,z^2,xy} & t_{NN,xy,xy} & t_{NN,xy,x^2-y^2} \\
 t_{NN,z^2,x^2-y^2} & - t_{NN,xy,x^2-y^2} & t_{NN,x^2-y^2,x^2-y^2} \\
\end{array}
\right) R = \left(
\begin{array}{ccc}
 0.2379 & 0.0869 & 0.0397 \\
 0.0397 & -0.1082 & 0.0771 \\
 0.0869 & -0.7826 & -0.1082 \\
\end{array}
\right) \\
 & R^\dagger \widetilde{t}(\bsl{a}_1+2\bsl{a}_2) R = 
 R^\dagger \left(
\begin{array}{ccc}
 t_{NNN,z^2,z^2} & 0 & t_{NNN,z^2,x^2-y^2} \\
 0 & t_{NNN,xy,xy} & 0 \\
 t_{NNN,x^2-y^2,z^2} & 0 & t_{NNN,x^2-y^2,x^2-y^2} \\
\end{array}
\right) R =\left(
\begin{array}{ccc}
 0.0491 & 0.0012 & 0.0012 \\
 -0.0497 & 0.0588 & 0.0207 \\
 -0.0497 & 0.0207 & 0.0588 \\
\end{array}
\right)\ ,
}
where the unit is $\eV$.

\begin{figure}
    \centering
    \includegraphics[width=0.8\columnwidth]{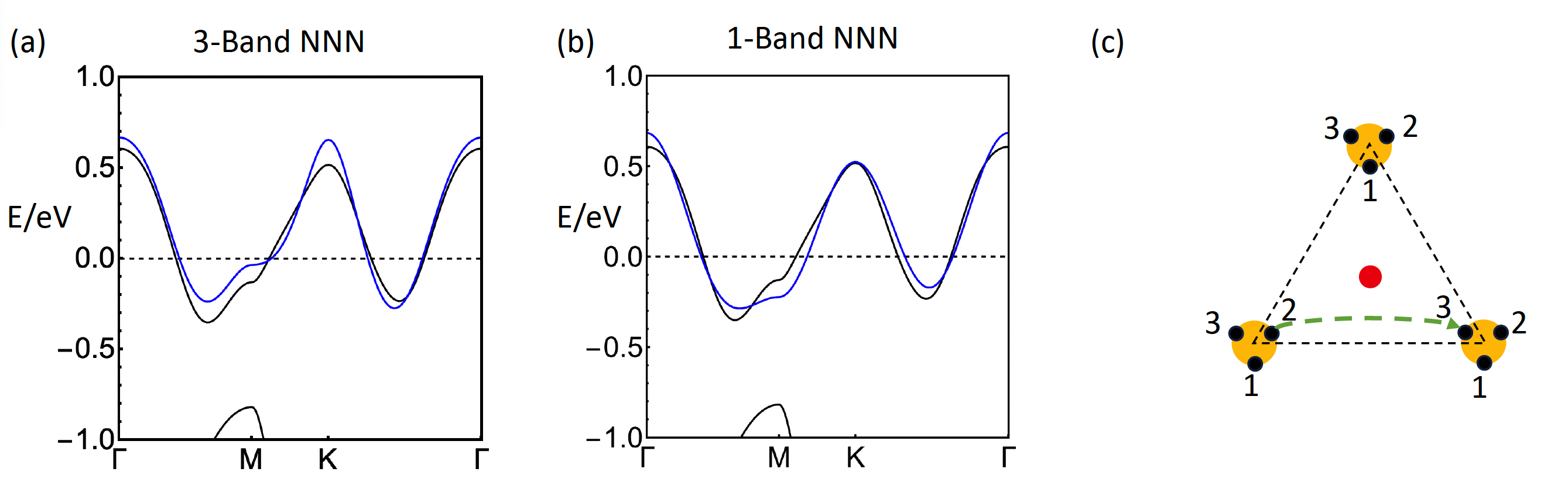}
    \caption{Analytic tight-binding models obtained from Wannier functions.
    The black lines are the electron bands from the DFT calculation.
    The blue line in (a) is given by the 3-band NNN model \eqref{eq:3-band_model_el} with hopping form in \eqref{eq:3band_hopping_form} and parameter values in \eqref{eq:3band_hopping_values}.
    The blue line in (b) is given by the 1-band NNN model \eqref{eq:one-band_model} with parameter values in \eqref{eq:app:one-band_hopping}.
    Note that the band structure plots in (a,b) involve no parameter tuning. Matching will improve if we allow tuning.
    (c) shows the basis of the local 3-band model in \eqref{eq:3-band_local} in the dashed triangle, whose eigenstates are good approximations of the Wannier basis of the 1-band model in \eqref{eq:one-band_model}.
    Specifically, the orange dot is the lattice site, and the black dots labeled by $1$, $2$ and $3$ are the rotated basis $\widetilde{c}^\dagger_{\bsl{R},1}, \widetilde{c}^\dagger_{\bsl{R},2}, \widetilde{c}^\dagger_{\bsl{R},3}$ defined in \eqref{eq:3-band_model_rotated_basis}, respectively.  
    The red dot marks the Wannier center of \eqref{eq:w_compact_1band_el}
    The green dashed arrow and its symmetry-related partners would correspond to the dominant hoppings in the rotated basis, according to \eqref{eq:3band_hopping_values_Rotated}.
    }
    \label{fig:DFT_el_H_3_H_1}
\end{figure}

According to \eqref{eq:3band_hopping_values}, we can see the dominant hoppings in the 3-band model are those for $\widetilde{c}^\dagger_{\bsl{R}+\bsl{a}_1,3}\widetilde{c}_{\bsl{R},2}$ (and its symmetry-related partners with value -0.7826 eV).
The dominant hopping can be intuitively understood as sp$^2$ hybridization: 
$d_{z^2}$, $d_{xy}$ and $d_{x^2-y^2}$ orbitals furnish the same symmetry reps as $s$, $p_x$ and $p_y$ under $C_3$ and $m_x$; if we intuitively treat $d_{z^2}$, $d_{xy}$ and $d_{x^2-y^2}$ as $s$, $p_x$ and $p_y$, the matrix $R$ in \eqref{eq:R_rotation} would just correspond to the sp$^2$ hybridization, and $\widetilde{c}^\dagger_{\bsl{R}+\bsl{a}_1,3}\widetilde{c}_{\bsl{R},2}$ and its symmetry-related partners would correspond to the hopping along the same $sp^2$ bond (e.g., the hopping denoted by the green arrow in Fig.\ref{fig:DFT_el_H_3_H_1}(c)), which has the largest wavefunction overlaps, naturally giving the largest hoppings.
We note that this is just an intuitive (but symmetry-based)  picture---the shapes of the wavefunctions of $\widetilde{c}^\dagger_{\bsl{R},1...3}$ are in reality more 
complicated than the simple sp$^2$ hybridization.
Nevertheless, the dominant hoppings are consistent with the intuitive picture.

\subsection{1-Band Model}
\label{app:1band_model}

From the full 3-band DFT model (not just the NNN model in Eq.\ref{eq:3band_hopping_values_Rotated}), we can use Wannier90 to construct a one-band model for the band at the Fermi level.
The resultant Wannier function is an $A_1$ irrep at the $1c$ Wycoff position ($A_1@1c$), which is obstructed atomic, as no atom is located at that position.
We use $w^\dagger_{\bsl{R}}$ to label the creation operator of the Wannier function, and the symmetry rep furnished by $w^\dagger_{\bsl{R}}$ reads
\eqa{
& C_3 w^\dagger_{\bsl{R}} C_3^{-1}  = w^\dagger_{C_3 \bsl{R}-\bsl{a}_1} \\
& m_x w^\dagger_{\bsl{R}} m_x^{-1}  = w^\dagger_{m_x \bsl{R}-\bsl{a}_1} \\
& \TR w^\dagger_{\bsl{R}} \TR^{-1}  = w^\dagger_{ \bsl{R}} \ .
}
The one-band model reads 
\eq{
\label{eq:one-band_model}
H_1 = \sum_{\bsl{R},\bsl{R}'} w^\dagger_{\bsl{R}}  w_{\bsl{R}'} t_{w}(\bsl{R}-\bsl{R}')\ ,
}
where
\eqa{
& C_3: t_{w}(\bsl{R}-\bsl{R}') = t_{w}(C_3(\bsl{R}-\bsl{R}')) \\
& m_x:t_{w}(\bsl{R}-\bsl{R}') = t_{w}(m_x(\bsl{R}-\bsl{R}')) \\
& \TR:t_{w}^*(\bsl{R}-\bsl{R}') = t_{w}(\bsl{R}-\bsl{R}') \\
& \text{Hermiticity}: t_{w}^*(\bsl{R}-\bsl{R}') = t_{w}(-\bsl{R}+\bsl{R}') \ .
}
To NNN, the numerical results suggest
\eq{
\label{eq:app:one-band_hopping}
t_{w}(\bsl{0}) = 0.0033\ ,\ t_{w}(\bsl{a}_1) = 0.0178\ ,\ t_{w}(\bsl{a}_1+2\bsl{a}_2) = 0.0955\ .
}
The simple NNN 1-band model produces a band that is very close to the DFT band structure as shown in \figref{fig:DFT_el_H_3_H_1}(b). The interesting feature in \eqref{eq:app:one-band_hopping} is that the NNN hopping $t_{w}(2\bsl{a}_1+\bsl{a}_2)$ is much larger than the NN hopping $t_{w}(\bsl{a}_1)$.
To understand this feature, we need to first derive an approximate analytical form of the Wannier state $w^\dagger_{\bsl{R}}$ based on the basis of the 3-band model.

To do so, let us go back to the 3-band model in \eqref{eq:3-band_model_el}, but re-write it in the rotated ("hybridized") basis $\widetilde{c}^\dagger_{\bsl{R},1}, \widetilde{c}^\dagger_{\bsl{R},2}, \widetilde{c}^\dagger_{\bsl{R},3}$ defined in \eqref{eq:3-band_model_rotated_basis}.
Based on the numerical values of the hopping in the rotated basis (\eqref{eq:3band_hopping_values_Rotated}), we find that in the rotated basis,
\eq{
\label{eq:H_3_rewritten}
H_3 = E_0 \sum_{\bsl{R}}\sum_{\alpha=1,2,3}  \widetilde{c}^\dagger_{\bsl{R},\alpha} \widetilde{c}_{\bsl{R},\alpha} + t \sum_{\bsl{R}} \left[  \widetilde{c}^\dagger_{\bsl{R},2} \widetilde{c}_{\bsl{R}+\bsl{a}_1,3} + \widetilde{c}^\dagger_{\bsl{R}+\bsl{a}_1,3} \widetilde{c}_{\bsl{R}+\bsl{a}_1+\bsl{a}_2,1}  + \widetilde{c}^\dagger_{\bsl{R}+\bsl{a}_1+\bsl{a}_2,1} \widetilde{c}_{\bsl{R},2}  + h.c. \right] + ... \ ,
}
where $E_0 = 1.7332$eV, $t = -0.7826$eV, and ``..." includes terms with coefficients with amplitudes no larger than $0.3$eV. 
The terms besides the ones denoted by ``..." in \eqref{eq:H_3_rewritten} are strictly local and compact, \ie,
\eqa{
H_3 = \sum_{\bsl{R}} H_3(\bsl{R}) + ... \ ,
}
where
\eqa{
\label{eq:3-band_local}
H_3(\bsl{R}) & = E_0 \left[ \widetilde{c}^\dagger_{\bsl{R}+\bsl{a}_1+\bsl{a}_2,1} \widetilde{c}_{\bsl{R}+\bsl{a}_1+\bsl{a}_2,1} + \widetilde{c}^\dagger_{\bsl{R},2} \widetilde{c}_{\bsl{R},2} + \widetilde{c}^\dagger_{\bsl{R}+\bsl{a}_1,3} \widetilde{c}_{\bsl{R}+\bsl{a}_1,3} \right] \\
& \quad + t \left[  \widetilde{c}^\dagger_{\bsl{R},2} \widetilde{c}_{\bsl{R}+\bsl{a}_1,3} + \widetilde{c}^\dagger_{\bsl{R}+\bsl{a}_1,3} \widetilde{c}_{\bsl{R}+\bsl{a}_1+\bsl{a}_2,1}  + \widetilde{c}^\dagger_{\bsl{R}+\bsl{a}_1+\bsl{a}_2,1} \widetilde{c}_{\bsl{R},2}  + h.c. \right] \\
& = \mat{\widetilde{c}^\dagger_{\bsl{R}+\bsl{a}_1+\bsl{a}_2,1} &  \widetilde{c}^\dagger_{\bsl{R},2} & \widetilde{c}_{\bsl{R}+\bsl{a}_1,3}^\dagger }  M \mat{\widetilde{c}_{\bsl{R}+\bsl{a}_1+\bsl{a}_2,1} \\  \widetilde{c}_{\bsl{R},2} \\ \widetilde{c}_{\bsl{R}+\bsl{a}_1,3} }\ ,
}
\eq{
M = \mat{ 
E_0 & t & t \\
 t & E_0 & t\\
 t & t & E_0
}
}
and 
\eq{
\left[ H_3(\bsl{R}), H_3(\bsl{R}') \right] = 0 
}
owing to the definition of $\widetilde{c}_{\bsl{R},1}$, $\widetilde{c}_{\bsl{R},2}$ and  $\widetilde{c}_{\bsl{R},3}$ in \eqref{eq:3-band_model_rotated_basis}. (See also \figref{fig:DFT_el_H_3_H_1}c.)
Diagonalizing $M$ gives one eigenvalue $E_0 + 2 t$ with eigenvector $\frac{1}{\sqrt{3}} (1,1,1)$ and the doubly-degenerate eigenvalue  $E_0 - t$ with two eigenvectors $\frac{1}{\sqrt{6}} (-2,1,1)$ and $\frac{1}{\sqrt{2}} (0,1,-1)$.
Since the band of interest is a rank-1 elementary band representation (EBR), we expect  $\frac{1}{\sqrt{3}} (1,1,1)$, which corresponds to 
\eq{
\label{eq:w_compact_1band_el}
w^\dagger_{compact,\bsl{R}} = \frac{1}{\sqrt{3}} (\widetilde{c}^\dagger_{\bsl{R}+\bsl{a}_1+\bsl{a}_2,1} +  \widetilde{c}^\dagger_{\bsl{R},2} + \widetilde{c}_{\bsl{R}+\bsl{a}_1,3}^\dagger )\ ,
}
to be a good approximation of the 3-band DFT Wannier state $w^\dagger_{\bsl{R}}$. (See also \figref{fig:DFT_el_H_3_H_1}c.)
Indeed, the probability overlap between $w^\dagger_{compact,\bsl{R}}$ and $w^\dagger_{\bsl{R}}$ is  a remarkable 0.9379, \ie,
\eq{
\label{eq:wannier_overlap_prob}
\frac{1}{N}\sum_{\bsl{k}}\left| \bra{0} w_{compact,\bsl{k}} w^\dagger_{\bsl{k}} \ket{0}
\right|^2 = 0.9379\ ,
}
where
\eqa{
& w^\dagger_{\bsl{k}} = \frac{1}{\sqrt{N}} \sum_{\bsl{R}} e^{\ii \bsl{R}\cdot \bsl{k} } w^\dagger_{\bsl{R}} \ ,\  w^\dagger_{compact, \bsl{k}} = \frac{1}{\sqrt{N}} \sum_{\bsl{R}} e^{\ii \bsl{R}\cdot \bsl{k} } w^\dagger_{compact,\bsl{R}} \ ,
}
resulting in 
\eq{
\label{eq:w_approx_1band_el}
w^\dagger_{\bsl{R}} \approx w^\dagger_{compact,\bsl{R}} \ .
}

From the approximated form of the Wannier function in \eqref{eq:w_approx_1band_el}, we can perform $\bra{0} w^\dagger_{compact,\bsl{R}'} H_3 w_{compact,\bsl{R}}$ to obtain the approximated relation between the hoppings in the 1-band model and those in the 3-band NNN model in \eqref{eq:3band_hopping_form}, which reads
\eqa{
\label{eq:t_1band_from_3band}
 & t_{w}(\bsl{a}_1)  \approx f_{w,onsite}(\bsl{a}_1) + f_{w,NN}(\bsl{a}_1) + f_{w,NNN}(\bsl{a}_1) \\
 & t_{w}(\bsl{a}_1+2\bsl{a}_2) \approx f_{w,NN}(\bsl{a}_1+2\bsl{a}_2) + f_{w,NNN}(\bsl{a}_1+2\bsl{a}_2) \ ,
}
where
\eqa{
\label{eq:f_expressions}
& f_{w,onsite}(\bsl{a}_1) = \frac{1}{9}\left[ E_{z^2}-E_{xy} \right] \\
& f_{w,NN}(\bsl{a}_1)=\frac{1}{9}\left[ t_{NN,x^2-y^2,x^2-y^2}+2 \sqrt{3} t_{NN,xy,x^2-y^2}+3 t_{NN,xy,xy}-\sqrt{2} t_{NN,z^2,x^2-y^2}+\sqrt{6} t_{NN,z^2,xy}+5 t_{NN,z^2,z^2} \right] \\
& f_{w,NNN}(\bsl{a}_1)=\frac{1}{9}\left[  -2 t_{NNN,x^2-y^2,x^2-y^2}-2 \sqrt{2} t_{NNN,x^2-y^2,z^2}+\sqrt{2} t_{NNN,z^2,x^2-y^2}+2 t_{NNN,z^2,z^2} \right] \\
& f_{w,NN}(\bsl{a}_1 + 2\bsl{a}_2 )= \frac{1}{9} \left[  -2 t_{NN,x^2-y^2,x^2-y^2} -2 \sqrt{3} t_{NN,xy,x^2-y^2} -\sqrt{2} t_{NN,z^2,x^2-y^2}-\sqrt{6} t_{NN,z^2,xy}+2 t_{NN,z^2,z^2} \right] \\
& f_{w,NNN}(\bsl{a}_1 + 2\bsl{a}_2 )  = \frac{1}{9}\left[ 3 t_{NNN,x^2-y^2,x^2-y^2}+3 t_{NNN,xy,xy}+3 t_{NNN,z^2,z^2} \right]\ .
}
With the parameter values in \eqref{eq:3band_hopping_values}, we have 
\eqa{ 
& f_{w,onsite}(\bsl{a}_1)= -0.0709 \eV \\
& f_{w,NN}(\bsl{a}_1) = 0.0651 \eV\\
& f_{w,NNN}(\bsl{a}_1) = 0.0008 \eV\ ,
}
leading to 
\eq{
t_{w}(\bsl{a}_1) \approx -0.0050\ .
}
Therefore, the small NN $t_{w}(\bsl{a}_1)$ comes from the cancellation between the on-site term and the NN hoppings in the 3-band model. 
On the other hand, 
\eqa{ 
& f_{w,NN}(\bsl{a}_1 + 2\bsl{a}_2 )   = 0.0265 \eV \\
& f_{w,NNN}(\bsl{a}_1 + 2\bsl{a}_2 ) = 0.0556 \eV\ ,
}
leading to 
\eq{
t_{w}(\bsl{a}_1+2\bsl{a}_2) \approx 0.08203 \eV 
}
without any cancellation between NN and NNN terms in the 3-band model. 
As will be shown in \appref{app:other_materials}, this cancellation of the on-site term and the NN hoppings also happens in other materials.

\subsection{Lower 3-Band Model}
\label{app:lower-3-band_model}

\begin{figure}
    \centering
    \includegraphics[width=0.8\columnwidth]{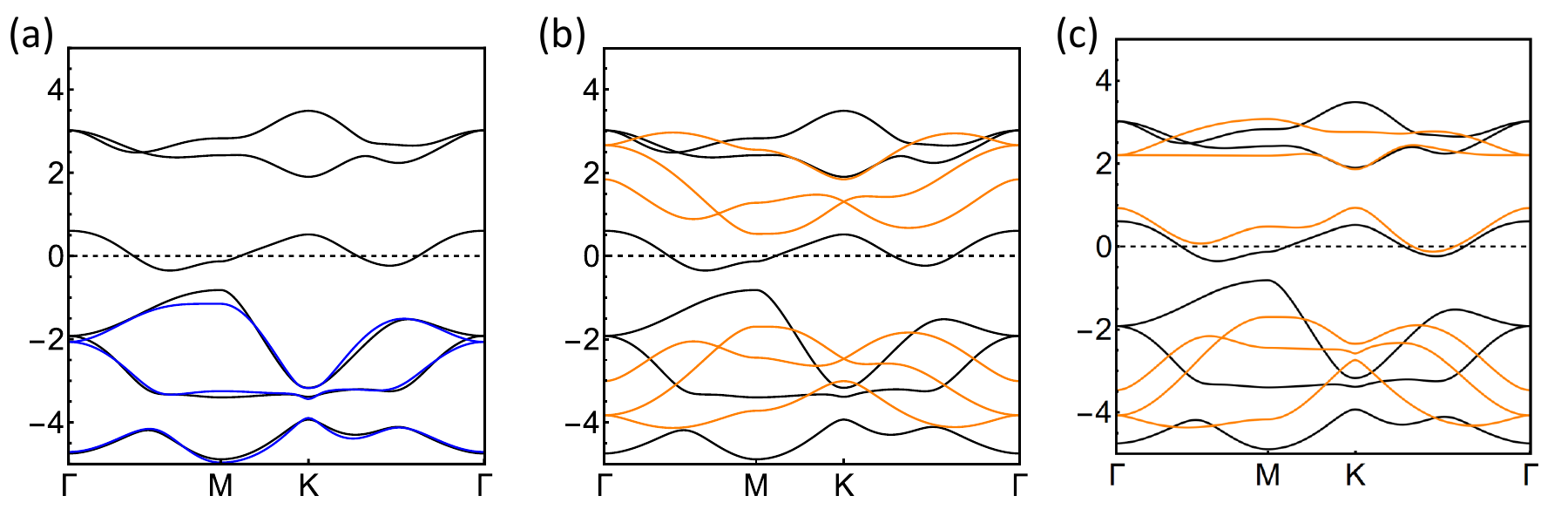}
    \caption{
    The black lines are the electron bands from the DFT calculation.
    The blue line in (a) is given by the lower-3-band NNN model \eqref{eq:lower_3-band_model_el} with hopping form in \eqref{eq:lower_3band_hopping_form} and parameter values in \eqref{eq:lower_3band_hopping_values}.
    The orange line in (b) is given by the simplified 6-band model in \eqref{eq:6-band_simplified}.
    The orange line in (c) is given by the simplified 6-band model in \eqref{eq:6-band_simplified} together with $t_{\bsl{\tau}_{\Nb}\bsl{\tau}_{\Nb}}(\bsl{a}_1)$ and its symmetry-related hoppings in \eqref{SeSeHoppinga16Band}.
    }
    \label{fig:lower_3band_DFT_el_6bandsim}
\end{figure}

From the full DFT 6-band model, we can also numerically build the Wannier states for the lowest three bands from Wannier90, \ie, the lower 3 bands in \figref{fig:DFT_el}(b).
We add this model just for completeness, and the discussion of this part is analogous to that of \appref{app:3-band_model}.
The trial states are chosen to be three Se p orbitals in \eqref{eq:6band_basis}, since we know the Se $m_z$-even combinations of p-orbitals have much lower energies ($\sim 2 \eV$) than the Nb d orbitals~\eqref{onsiteSeNb6Band}. Hence our resulting Wannier states will be ``renormalized" Se $m_z$-even combinations of p-orbitals.
We label creation operators for the resultant Wannier states as 
\eq{
\label{eq:lower_3band_basis}
\widetilde{c}^\dagger_{\bsl{R}+\bsl{\tau}_{\Se}} = ( \widetilde{c}^\dagger_{\bsl{R}+\bsl{\tau}_{\Se},z}, \widetilde{c}^\dagger_{\bsl{R}+\bsl{\tau}_{\Se},x}, \widetilde{c}^\dagger_{\bsl{R}+\bsl{\tau}_{\Se},y})\ .
}
The symmetry reps furnished by \eqref{eq:lower_3band_basis} read
\eqa{
\label{eq:lower_3band_sym_rep}
& g \widetilde{c}^\dagger_{\bsl{R}+\bsl{\tau}_{\Se}} g^{-1} = \widetilde{c}^\dagger_{g \bsl{R}+ g \bsl{\tau}_{\Se}} U_g \ ,\  \TR \widetilde{c}^\dagger_{\bsl{R}+\bsl{\tau}_{\Se}} \TR^{-1} = \widetilde{c}^\dagger_{\bsl{R}+\bsl{\tau}_{\Se}} \ ,
}
where $g=C_3,m_x$, and $U_{C_3}$ and $U_{m_x}$ are in \eqref{eq:3band_sym_rep_Ug}.
With the basis in \eqref{eq:lower_3band_basis}, the lower-3-band model reads
\eq{
\label{eq:lower_3-band_model_el}
H_{lower-3} = \sum_{\bsl{R}\bsl{R}'} \widetilde{c}^\dagger_{\bsl{R}+\bsl{\tau}_{\Se}} \widetilde{t}(\bsl{R}-\bsl{R}') \widetilde{c}_{\bsl{R}'+\bsl{\tau}_{\Se}}\ ,
}
where $\widetilde{t}(\bsl{R}-\bsl{R}')$ is a $3\times 3$ matrix.
As a result, $\widetilde{t}(\bsl{R}-\bsl{R}')$ satisfies
\eqa{
\label{eq:lower_3band_sym_t}
& U_g \widetilde{t}(\bsl{R}-\bsl{R}') U_g^\dagger = \widetilde{t}(g(\bsl{R}-\bsl{R}')) \\
& \widetilde{t}^*(\bsl{R}-\bsl{R}') = \widetilde{t}(\bsl{R}-\bsl{R}') \\
& \widetilde{t}^\dagger(\bsl{R}-\bsl{R}') = \widetilde{t}(\bsl{R}'-\bsl{R})\ ,
}
where the last equality comes from Hermiticity.
Up to NNN, the independent hopping terms are
\eqa{
\label{eq:lower_3band_hopping_form}
& \widetilde{t}(0) = \left(
\begin{array}{ccc}
 E_{z} & 0 &  \\
 0 & E_{x} & 0 \\
 0 & 0 & E_{x} \\
\end{array}
\right)\ ,\ \widetilde{t}(\bsl{a}_1) = \left(
\begin{array}{ccc}
 t_{NN,z,z} & t_{NN,z,x} & t_{NN,z,y} \\
 - t_{NN,z,x} & t_{NN,x,x} & t_{NN,x,y} \\
 t_{NN,z,y} & - t_{NN,x,y} & t_{NN,y,y} \\
\end{array}
\right)\\
& \widetilde{t}(2\bsl{a}_1+\bsl{a}_2) = \left(
\begin{array}{ccc}
 t_{NNN,z,z} & 0 & t_{NNN,z,y} \\
 0 & t_{NNN,x,x} & 0 \\
 t_{NNN,y,z} & 0 & t_{NNN,y,y} \\
\end{array}
\right)\ .
}
Numerically, we find that the values of independent hoppings in $\widetilde{t}(\bsl{R})$ read: 
\eqa{
\label{eq:lower_3band_hopping_values}
& \left(
\begin{array}{ccc}
 E_{z} & 0 &  \\
 0 & E_{x} & 0 \\
 0 & 0 & E_{x} \\
\end{array}
\right) = \left(
\begin{array}{ccc}
 -3.6641 & 0 & 0 \\
 0 & -2.7848 & 0 \\
 0 & 0 & -2.7848 \\
\end{array}
\right)\\
& \left(
\begin{array}{ccc}
 t_{NN,z,z} & t_{NN,z,x} & t_{NN,z,y} \\
 - t_{NN,z,x} & t_{NN,x,x} & t_{NN,x,y} \\
 t_{NN,z,y} & - t_{NN,x,y} & t_{NN,y,y} \\
\end{array}
\right) = \left(
\begin{array}{ccc}
 -0.1703 & -0.1297 & 0.0880 \\
 0.1297 & 0.6473 & 0.0440 \\
 0.0880 & -0.0440 & -0.2919 \\
\end{array}
\right)\\
&  \left(
\begin{array}{ccc}
 t_{NNN,z,z} & 0 & t_{NNN,z,y} \\
 0 & t_{NNN,x,x} & 0 \\
 t_{NNN,y,z} & 0 & t_{NNN,y,y} \\
\end{array}
\right)=\left(
\begin{array}{ccc}
 -0.0038 & 0 & 0.0486 \\
 0 & -0.0591 & 0 \\
 -0.0580 & 0 & -0.0569 \\
\end{array}
\right)\ .
}
The band structure of the lower three bands, given by the lower-3-band NNN model is close to the DFT one, as shown in \figref{fig:lower_3band_DFT_el_6bandsim}(a). 

\subsubsection{Approximated Analytical Forms of the Wannier Basis of the lower-3-band Model}

To show the approximated form of $\widetilde{c}^\dagger_{\bsl{R}+\bsl{\tau}_{\Se}}$, we rotate $\widetilde{c}^\dagger_{\bsl{R}+\bsl{\tau}_{\Se}}$ to a new form:
\eq{
\label{eq:lower-3-band_model_rotated_basis}
(\widetilde{c}^\dagger_{\bsl{R}+\bsl{\tau}_{\Se},1}, \widetilde{c}^\dagger_{\bsl{R}+\bsl{\tau}_{\Se},2}, \widetilde{c}^\dagger_{\bsl{R}+\bsl{\tau}_{\Se},3} ) =  (\widetilde{c}^\dagger_{\bsl{R}+\bsl{\tau}_{\Se},z}, \widetilde{c}^\dagger_{\bsl{R}+\bsl{\tau}_{\Se},x}, \widetilde{c}^\dagger_{\bsl{R}+\bsl{\tau}_{\Se},y}) R\ , 
}
where $R$ is in \eqref{eq:R_rotation}.
In the rotated basis, the symmetry rep reads
\eqa{
& C_3 (\widetilde{c}^\dagger_{\bsl{R}+\bsl{\tau}_{\Se},1}, \widetilde{c}^\dagger_{\bsl{R}+\bsl{\tau}_{\Se},2}, \widetilde{c}^\dagger_{\bsl{R}+\bsl{\tau}_{\Se},3} ) C_3^{-1} = (\widetilde{c}^\dagger_{C_3\bsl{R}+C_3\bsl{\tau}_{\Se},1}, \widetilde{c}^\dagger_{C_3\bsl{R}+C_3\bsl{\tau}_{\Se},2}, \widetilde{c}^\dagger_{C_3\bsl{R}+C_3\bsl{\tau}_{\Se},3} ) \mat{ 0 & 0 & 1\\ 1 & 0 & 0 \\ 0 & 1 & 0 } \\
& m_x (\widetilde{c}^\dagger_{\bsl{R}+\bsl{\tau}_{\Se},1}, \widetilde{c}^\dagger_{\bsl{R}+\bsl{\tau}_{\Se},2}, \widetilde{c}^\dagger_{\bsl{R}+\bsl{\tau}_{\Se},3} ) m_x^{-1} = (\widetilde{c}^\dagger_{m_x\bsl{R}+m_x\bsl{\tau}_{\Se},1}, \widetilde{c}^\dagger_{m_x\bsl{R}+m_x\bsl{\tau}_{\Se},2}, \widetilde{c}^\dagger_{m_x\bsl{R}+m_x\bsl{\tau}_{\Se},3} ) \mat{ 1 & 0 & 0\\ 0 & 0 & 1 \\ 0 & 1 & 0 } \\
& \TR (\widetilde{c}^\dagger_{\bsl{R}+\bsl{\tau}_{\Se},1}, \widetilde{c}^\dagger_{\bsl{R}+\bsl{\tau}_{\Se},2}, \widetilde{c}^\dagger_{\bsl{R}+\bsl{\tau}_{\Se},3} )\TR^{-1} = (\widetilde{c}^\dagger_{\bsl{R}+\bsl{\tau}_{\Se},1}, \widetilde{c}^\dagger_{\bsl{R}+\bsl{\tau}_{\Se},2}, \widetilde{c}^\dagger_{\bsl{R}+\bsl{\tau}_{\Se},3} ) \ .
}
After the rotation, the numerical data suggest that
\eqa{
\label{eq:lower_3-band_model_basis}
& \widetilde{c}^\dagger_{\bsl{R}+\bsl{\tau}_{\Se},1} = y_1 ( \frac{1}{\sqrt{3}}  c^\dagger_{\bsl{R}+\bsl{\tau}_{\Se},z} -\frac{\sqrt{2}}{\sqrt{3}}  c^\dagger_{\bsl{R}+\bsl{\tau}_{\Se},y}  ) + y_2 c^\dagger_{\bsl{R}, d_{z^2}} + y_3 c^\dagger_{\bsl{R}+\bsl{a}_1+\bsl{a}_2, d_{x^2-y^2}} + y_3 c^\dagger_{\bsl{R}+\bsl{a}_2, d_{x^2-y^2}} + ... \\
& \widetilde{c}^\dagger_{\bsl{R}+\bsl{\tau}_{\Se},2} = y_1 ( \frac{1}{\sqrt{3}}  c^\dagger_{\bsl{R}+\bsl{\tau}_{\Se},z}  +\frac{1}{\sqrt{2}} c^\dagger_{\bsl{R}+\bsl{\tau}_{\Se},x}  +\sqrt{\frac{1}{6}} c^\dagger_{\bsl{R}+\bsl{\tau}_{\Se},y} ) + y_2 c^\dagger_{\bsl{R}+\bsl{a}_1+\bsl{a}_2, d_{z^2}}  + y_3 (-\frac{\sqrt{3}}{2} c^\dagger_{\bsl{R}+\bsl{a}_2, d_{xy}} - \frac{1}{2} c^\dagger_{\bsl{R}+\bsl{a}_2, d_{x^2-y^2}})\\
& \quad \quad + y_3 (-\frac{\sqrt{3}}{2} c^\dagger_{\bsl{R}, d_{xy}} - \frac{1}{2} c^\dagger_{\bsl{R}, d_{x^2-y^2}}) + ... \\
& \widetilde{c}^\dagger_{\bsl{R}+\bsl{\tau}_{\Se},3} = y_1 ( \frac{1}{\sqrt{3}}  c^\dagger_{\bsl{R}+\bsl{\tau}_{\Se},z}  - \frac{1}{\sqrt{2}} c^\dagger_{\bsl{R}+\bsl{\tau}_{\Se},x}  + \sqrt{\frac{1}{6}} c^\dagger_{\bsl{R}+\bsl{\tau}_{\Se},y} ) + y_2 c^\dagger_{\bsl{R}+\bsl{a}_2, d_{z^2}} + y_3 (\frac{\sqrt{3}}{2} c^\dagger_{\bsl{R}, d_{xy}} - \frac{1}{2} c^\dagger_{\bsl{R}, d_{x^2-y^2}}) \\
& \quad \quad +  y_3 (\frac{\sqrt{3}}{2} c^\dagger_{\bsl{R}+\bsl{a}_1+\bsl{a}_2, d_{xy}} - \frac{1}{2} c^\dagger_{\bsl{R}+\bsl{a}_1+\bsl{a}_2, d_{x^2-y^2}}) + ... \ ,
}
where
\eq{
\label{eq:values_x_1_x_2_lower_3-band_basis}
y_1 = 0.8615\ ,\ y_2 = -0.2144\ , \ y_3 = -0.2904\ ,
}
``$...$ "labels the terms that are smaller than $0.1$.
We note that if we neglect the small terms in ``$...$ " in \eqref{eq:lower_3-band_model_basis}, the Wannier functions are not strictly orthonormal anymore (but still almost so).

To provide more precise approximated forms of the Wannier functions, we use the following format:
\eq{
(\widetilde{c}^\dagger_{\bsl{R}_0+\bsl{\tau}_{\Se},1}, \widetilde{c}^\dagger_{\bsl{R}_0+\bsl{\tau}_{\Se},2}, \widetilde{c}^\dagger_{\bsl{R}_0+\bsl{\tau}_{\Se},3} ) = \sum_{\bsl{R},\bsl{\tau}} c^\dagger_{\bsl{R}+\bsl{\tau}+\bsl{R}_0+\bsl{\tau}_{\Se}} \widetilde{\xi}_{\Se,\bsl{R}+\bsl{\tau}}\ ,
}
where $c^\dagger_{\bsl{R}+\bsl{\tau}}$ is defined in \eqref{eq:c_dagger_basis}.
From the symmetry reps in \eqref{eq:lower_3band_sym_rep}, we obtain the constraints on $\xi_{\bsl{R}+\bsl{\tau}}$, which read
\eqa{
\label{eq:lower_3band_basis_xi_tilde_sym}
& U_{g} \widetilde{\xi}_{\Se,\bsl{R}+\bsl{\tau}} = \widetilde{\xi}_{\Se,g(\bsl{R}+\bsl{\tau})} R^\dagger U_{g} R\\
& \widetilde{\xi}_{\Se,\bsl{R}+\bsl{\tau}}^* = \widetilde{\xi}_{\Se,\bsl{R}+\bsl{\tau}}\\
& \sum_{\bsl{R},\bsl{\tau}} \widetilde{\xi}_{\Se,\bsl{R}+\bsl{\tau}}^\dagger \widetilde{\xi}_{\Se,\bsl{R}+\bsl{\tau}} = 1\ ,
}
where $U_{g=C_3,m_x}$ are in \eqref{eq:3band_sym_rep_Ug}. The last equality is the normalization condition of the Wannier orbitals.
To the NNN terms, the numerical data suggests
\eq{
\widetilde{\xi}_{\Se,\bsl{ 0}} =\left(
\begin{array}{ccc}
 0.5009 & 0.5009 & 0.5009 \\
 0 & 0.6070 & -0.6070 \\
 -0.7009 & 0.3504 & 0.3504 \\
\end{array}
\right)\ ,
}
\eq{
\widetilde{\xi}_{\Se,-\bsl{\tau}_{\Se}} = \left(
\begin{array}{ccc}
 -0.2133 & -0.0584 & -0.0584 \\
 0 & 0.2286 & -0.2286 \\
 -0.0142 & 0.1805 & 0.1805 \\
\end{array}
\right) \ ,
}
\eq{
\widetilde{\xi}_{\Se,- C_3 \bsl{\tau}_{\Se}} = \widetilde{\xi}_{\Se,\bsl{a}_1+\bsl{a}_2-\bsl{\tau}_{\Se}} = \left(
\begin{array}{ccc}
 -0.0584 & -0.2133 & -0.0584 \\
 -0.0420 & 0.0123 & -0.2706 \\
 -0.2882 & 0.0071 & 0.1077 \\
\end{array}
\right) \ ,
}
\eq{
\widetilde{\xi}_{\Se,- C_3^2 \bsl{\tau}_{\Se}} = \widetilde{\xi}_{\Se,\bsl{a}_2-\bsl{\tau}_{\Se}} = \left(
\begin{array}{ccc}
 -0.0584 & -0.0584 & -0.2133 \\
 0.0420 & 0.2706 & -0.0123 \\
 -0.2882 & 0.1077 & 0.0071 \\
\end{array}
\right)\ ,
}
\eq{
\widetilde{\xi}_{\Se,\bsl{a}_1} =
\left(
\begin{array}{ccc}
 0.0049 & -0.0228 & 0.0244 \\
 0.0067 & -0.0241 & 0.0297 \\
 0.0378 & 0.0027 & -0.0006 \\
\end{array}
\right)\ ,
}
\eq{
\widetilde{\xi}_{\Se,C_3\bsl{a}_1} = \widetilde{\xi}_{\Se,\bsl{a}_2} = \left(
\begin{array}{ccc}
 0.0244 & 0.0049 & -0.0228 \\
 -0.0144 & -0.0361 & 0.0097 \\
 0.0260 & -0.0131 & -0.0222 \\
\end{array}
\right)\ ,
}
\eq{
\widetilde{\xi}_{\Se,C_3^2 \bsl{a}_{1}} = \widetilde{\xi}_{\Se,-\bsl{a}_1-\bsl{a}_2} = 
\left(
\begin{array}{ccc}
 -0.0228 & 0.0244 & 0.0049 \\
 0.0143 & -0.0153 & 0.0293 \\
 0.0195 & -0.0254 & -0.0247 \\
\end{array}
\right)
\ ,
}
\eq{
\widetilde{\xi}_{\Se,m_x\bsl{a}_1} = \widetilde{\xi}_{\Se,-\bsl{a}_1} =\left(
\begin{array}{ccc}
 0.0049 & 0.0244 & -0.0228 \\
 -0.0067 & -0.0297 & 0.0241 \\
 0.0378 & -0.0006 & 0.0027 \\
\end{array}
\right)\ ,
}
\eq{
\widetilde{\xi}_{\Se,C_3 m_x\bsl{a}_1} = \widetilde{\xi}_{\Se,-C_3\bsl{a}_1} = \widetilde{\xi}_{\Se,-\bsl{a}_2} = 
\left(
\begin{array}{ccc}
 -0.0228 & 0.0049 & 0.0244 \\
 -0.0143 & -0.0293 & 0.0153 \\
 0.0195 & -0.0247 & -0.0254 \\
\end{array}
\right) \ ,
}
and
\eq{
\widetilde{\xi}_{\Se,C_3^2 m_x\bsl{a}_1} = \widetilde{\xi}_{\Se,-C_3^2 \bsl{a}_{1}} = \widetilde{\xi}_{\Se,\bsl{a}_1+\bsl{a}_2} = \left(
\begin{array}{ccc}
 0.0244 & -0.0228 & 0.0049 \\
 0.0144 & -0.0097 & 0.0361 \\
 0.0260 & -0.0222 & -0.0131 \\
\end{array}
\right)\ .
}

\subsection{Other Materials}
\label{app:other_materials}

\begin{figure}[h]
    \centering
    \includegraphics[width=\columnwidth]{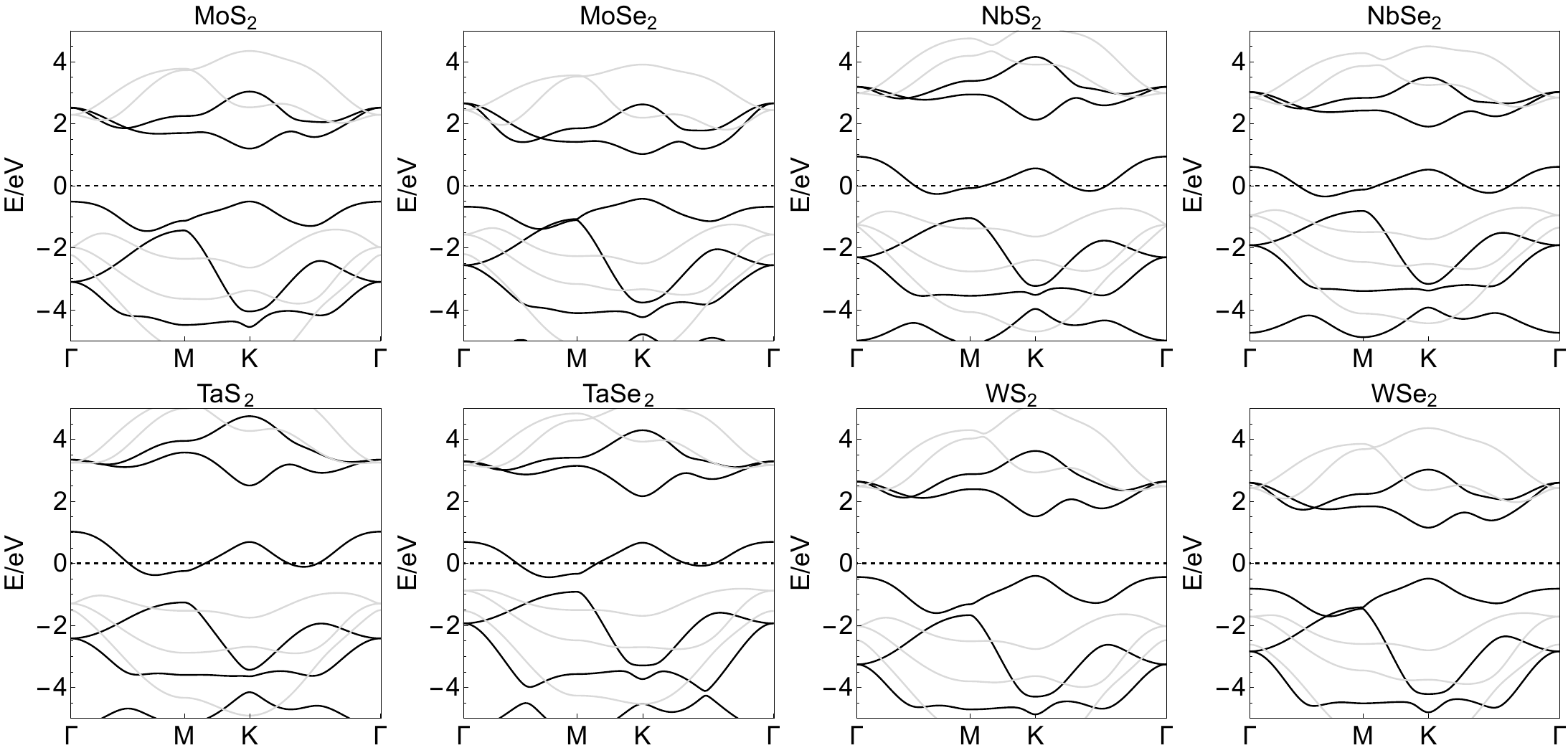}
    \caption{The electron band structures from the DFT calculation for 8 TMD materials. Only the closest 11 or 10 bands to the Fermi level are plotted. 
    The black lines are $m_z$ even, while the gray lines are $m_z$ odd.
    Note that there is a band near the Fermi energy that is isolated in the $m_z$-even sector for 1H-MoS$_2$, NbS$_2$, NbSe$_2$, TaS$_2$, TaSe$_2$, and WS$_2$.
    The 1 isolated $m_z$-even band near the Fermi energy is obstructed atomic---A$_1$@1c. Depending on the number of valence electrons, this isolated flat band can be either fully filled or half-filled. 
    }
    \label{fig:el-band_8TMD}
\end{figure}

\begin{table}[h]
    \centering
    $
    \begin{array}{|c|c|c|c|c|}
    \hline
 & t_{w}(\bsl{a}_1) & t_{w}(\bsl{a}_2) & t_{w}(\bsl{a}_3)  & \frac{1}{N}\sum_{\bsl{k}}\left| \bra{0} w_{compact,\bsl{k}} w^\dagger_{\bsl{k}} \ket{0}
\right|^2 \\
  \hline
\text{MoS}_ 2 & 0.0148 & 0.0949 & 0.0202 & 0.9069 \\
 \text{NbS}_ 2 & 0.0645 & 0.0980 & 0.0153 & 0.9309 \\
 \text{NbSe}_ 2 & 0.0178 & 0.0955 & 0.0157 & 0.9379 \\
 \text{TaS}_ 2 & 0.0730 & 0.1165 & -0.0021 & 0.9090 \\
 \text{TaSe}_ 2 & 0.0389 & 0.1076 & -0.0128 & 0.9346 \\
 \text{WS}_ 2 & 0.0263 & 0.1134 & 0.0053 & 0.9048 \\
 \hline
\end{array}
$
    \caption{ The middle three columns correspond to the hopping amplitudes for the 1 band model of MoS$_2$, NbS$_2$, NbSe$_2$, TaS$_2$, TaSe$_2$, and WS$_2$. 
    The 1 band model is defined in \eqnref{eq:one-band_model}. 
    The last column shows the probability overlap between the approximated compact Wannier function and the DFT wannier function (\eqnref{eq:wannier_overlap_prob}) for the isolated 1 band. 
    }
    \label{tab:1-band_hoppings}
\end{table}

\begin{table}[h]
    \centering
    $
    \begin{array}{|c|c|c|c|c||c|c|c|}
    \hline
 & f_{w,onsite}(\bsl{a}_1) & f_{w,NN}(\bsl{a}_1) & f_{w,NNN}(\bsl{a}_1) & t_{w}(\bsl{a}_1) & f_{w,NN}(\bsl{a}_1 + 2\bsl{a}_2 ) & f_{w,NNN}(\bsl{a}_1 + 2\bsl{a}_2 ) & t_{w}(\bsl{a}_1+2\bsl{a}_2)  \\
  \hline
 \text{MoS}_ 2 & -0.1071 & 0.0632 & 0.0199 & -0.0240 & 0.0057 & \
0.0637 & 0.0694 \\
 \text{NbS}_ 2 & -0.0844 & 0.0764 & 0.0249 & 0.0169 & 0.0322 & 0.0516 \
& 0.0838 \\
 \text{NbSe}_ 2 & -0.0709 & 0.0651 & 0.0008 & -0.0050 & 0.0265 & \
0.0556 & 0.0820 \\
 \text{TaS}_ 2 & -0.0867 & 0.0818 & 0.0235 & 0.0186 & 0.0550 & 0.0507 \
& 0.1056 \\
 \text{TaSe}_ 2 & -0.0883 & 0.0767 & 0.0117 & 0.0001 & 0.0432 & \
0.0496 & 0.0928 \\
 \text{WS}_ 2 & -0.1072 & 0.0610 & 0.0246 & -0.0215 & 0.0320 & 0.0638 \
& 0.0957 \\
 \hline
\end{array}
$
    \caption{The estimated hopping amplitudes for the 1 band model of MoS$_2$, NbS$_2$, NbSe$_2$, TaS$_2$, TaSe$_2$, and WS$_2$.
    The estimation is derived from the 6-band model according to \eqnref{eq:t_1band_from_3band}.
    }
    \label{tab:1-band_hoppings_from_6band}
\end{table}

The obstructed atomic band not only occurs in NbSe$_2$, but also exists in other 2D TMD materials.
In \figref{fig:el-band_8TMD}, we plot the band structure for eight TMD materials.
Among them, 1H-MoS$_2$, NbS$_2$, NbSe$_2$, TaS$_2$, TaSe$_2$, and WS$_2$ have one isolated $m_z$-even band near or at the Fermi energy is obstructed atomic---A$_1$@1c, and their Wannier functions can be approximated by the most compact Wannier functions in the three band model with more than 90\% probability, as shown in \tabref{tab:1-band_hoppings}.
Interestingly, all of them have NN hoppings smaller than the NNN hoppings among the obstructed Wannier functions, especially for MoS$_2$, TaSe$_2$, NbSe$_2$ and WS$_2$ which have NN hoppings nearly one-order-of-magnitude smaller than the NNN hoppings. (See \tabref{tab:1-band_hoppings}.)
The fact that NN hoppings are smaller than the NNN hoppings can also be explained approximately as the cancellation between the atomic onsite terms and the atomic NN terms for the NN hoppings among the obstructed Wannier functions as shown in \tabref{tab:1-band_hoppings_from_6band},

\section{Perturbative Understanding from the 6-band Model}
\label{app:analytic}

In this section, we provide more details on the analytic understanding of the 3-band model and of the obstructed atomic band from the 6-band Model from a perturbative approach.

\subsection{New Perturbation Theory for the 6-Band Model and Effective 3-Band Model}
\label{app:new_PT_and_3band_Model}
We first discuss the perturbation theory approach for the 6-band model.
To do so, we separate the 3 Nb orbitals and the 3 Se orbitals explicitly, and rewrite 6-band model in the $\bsl{k}$-space Hamiltonian:
\eqa{
\label{eq:H_6_Nb_Se_S}
    & H_{6 \text{band}}= \sum_{\bsl{k}} (c_{\Nb, \bsl{k}}^\dagger c_{\Se, \bsl{k}}^\dagger) \left( \begin{array}{cc}
 H_{\Nb}(\bsl{k})  & S(\bsl{k}) \\
 S^\dagger(\bsl{k}) & H_{\Se}(\bsl{k}) \\
\end{array}
\right)\left(\begin{array}{cc}
c_{\Nb,\bsl{k}}  \\
 c_{\Se,\bsl{k}}  \\
\end{array}
\right)
} where our Fourrier transform is 
\eq{
c_{\bsl{R}+\bsl{\tau},\alpha_{\bsl{\tau}}}= \frac{1}{\sqrt{N}}\sum_{\bsl{k}} e^{\ii \bsl{k} \cdot \left( \bsl{R}+ \bsl{\tau}\right)} c_{\bsl{k},\bsl{\tau},\alpha_{\bsl{\tau}}}\ ,
}
$c_{\Nb, \bsl{k}}^\dagger=\left(c_{\bsl{k},\bsl{\tau}_{\Nb},d_{z^2}}^\dagger, c_{\bsl{k},\bsl{\tau}_{\Nb},d_{xy}}^\dagger, c_{\bsl{k},\bsl{\tau}_{\Nb},d_{x^2-y^2}}^\dagger\right)$, and $c_{\Se, \bsl{k}}^\dagger=\left(c_{\bsl{k},\bsl{\tau}_{\Se},p_z}^\dagger, c_{\bsl{k},\bsl{\tau}_{\Se},p_x}^\dagger, c_{\bsl{k},\bsl{\Se},p_y}^\dagger\right)$.
We now have the eigenvalue equation:
\begin{eqnarray}
    \left( \begin{array}{cc}
 H_{\Nb}(\bsl{k})  & S(\bsl{k}) \\
 S^\dagger(\bsl{k}) & H_{\Se}(\bsl{k}) \\
\end{array}
\right)\left(\begin{array}{cc}
\psi_{\Nb, \bsl{k} }  \\
 \psi_{\Se,\bsl{k}}  \\
\end{array}
\right)= E_{\bsl{k}}\left(\begin{array}{cc}
\psi_{\Nb, \bsl{k} }  \\
 \psi_{\Se,\bsl{k}}  \\
\end{array}
\right)\ ,
\end{eqnarray} 
where
$\psi_{\Nb,\bsl{k}}$ and $\psi_{\Se,\bsl{k}}$ are three-component column vectors.
The creation operator for an eigenstate at energy $E(k)$ reads
\begin{eqnarray}
    \gamma_{E_{\bsl{k}},\bsl{k}} = \psi_{\Nb, \bsl{k}}^\dagger c_{\Nb,\bsl{k}} + \psi_{\Se, \bsl{k}}^\dagger c_{\Se,\bsl{k}}
\end{eqnarray} where $\psi_{\Nb, \bsl{k}}^\dagger c_{\Nb,\bsl{k}}$ should be understood as row-column multiplication, and similarly for Se. 
Then, the eigenvalue equation can be split into two parts:
\eqa{
 \label{Perturbation1} 
    & \psi_{\Se,\bsl{k}}= \left[ E_{\bsl{k}}- H_{\Se}(\bsl{k}) \right]^{-1} S^\dagger(\bsl{k}) \psi_{\Nb,\bsl{k}} \\
    & \left[ H_{\Nb}(\bsl{k}) + S(\bsl{k})  (E_{\bsl{k}}- H_{\Se}(\bsl{k}) )^{-1} S^\dagger(\bsl{k}) \right] \psi_{\Nb,\bsl{k}}= E_{\bsl{k}} \psi_{Nb,\bsl{k}}\\
    & \text{if $E_{\bsl{k}}- H_{\Se}(\bsl{k})$ is invertible.}
} 
The second equation gives an ``effective" Nb Hamiltonian (which so far is not linear in $E_{k}$).  The creation operator for the energy eigenstate can now be cast in the form:
\begin{eqnarray}
 \label{eq:Perturbation_Gamma} 
    \gamma_{E_{\bsl{k}},\bsl{k}} = \psi_{\Nb, \bsl{k}}^\dagger ( c_{\Nb,\bsl{k}} + S(\bsl{k})  (E_{\bsl{k}}- H_{\Se}(\bsl{k}) )^{-1} c_{\Se,\bsl{k}}) 
\end{eqnarray}

\begin{figure}[t]
    \centering
    \includegraphics[width=0.4\columnwidth]{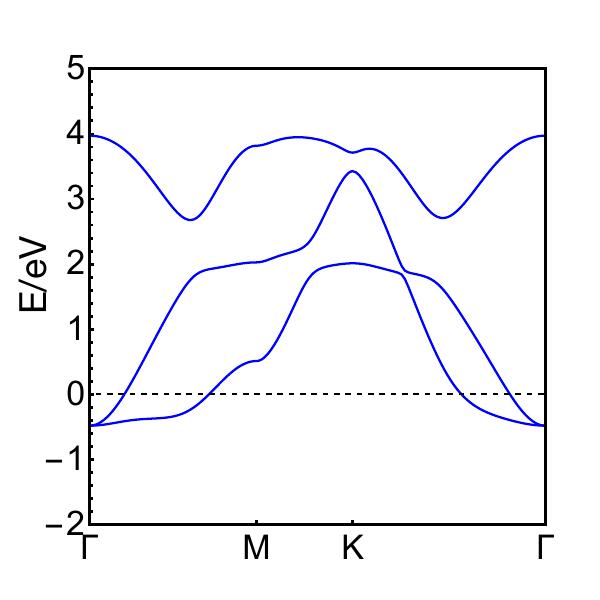}
    \caption{
    The eigenvalues of the DFT-precise $E_{k} - H_{\Se}(\bsl{k})$ along the high-symmetry lines. Here $E_{\bsl{k}}$ being the energy dispersion of the obstructed atomic band.
    }
    \label{fig:failure_conventional_perturbation_theory}
\end{figure}

\begin{figure}[t]
    \centering
    \includegraphics[width=\columnwidth]{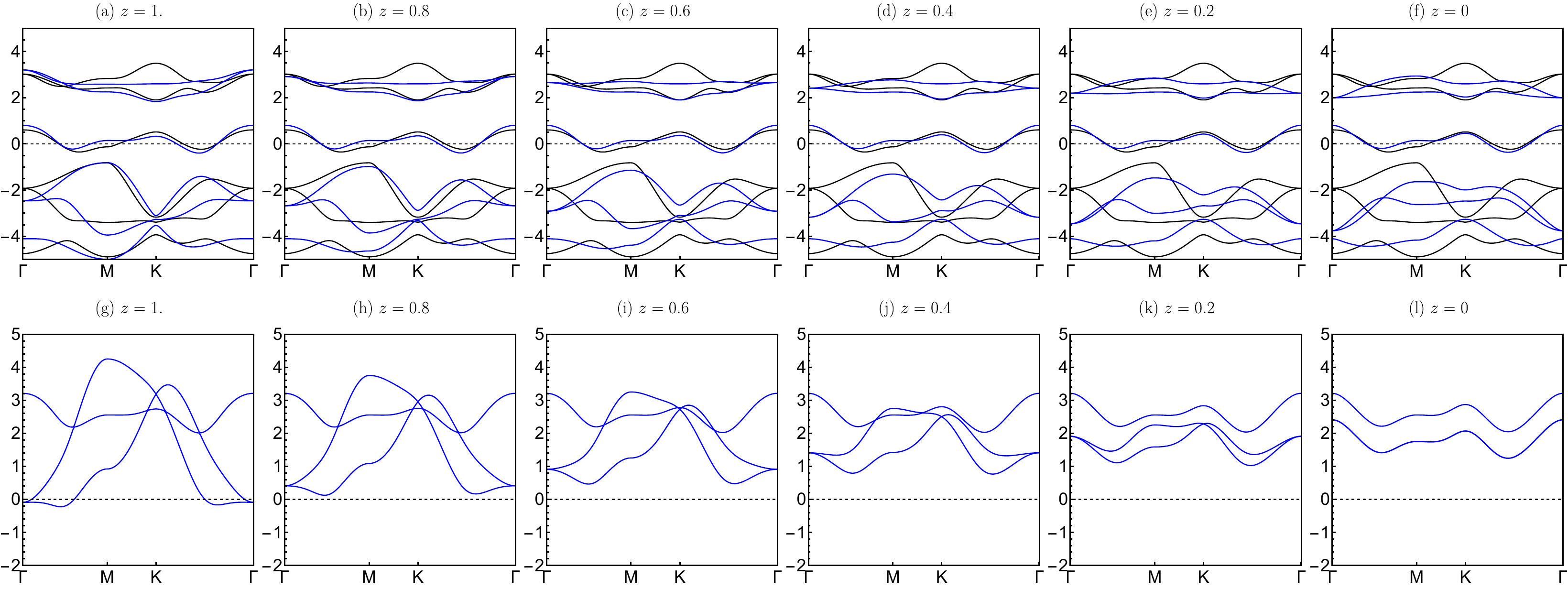}
    \caption{
    (a-f) The band structure of the simplified (or approximated) Hamiltonian in \eqnref{eq:SeSeHoppinga16Band_Sim} are plotted in blue for various values of $z$, while the DFT bands are plotted in black.
    (g-l) $E_{k} - H_{\Se}(\bsl{k})$ of the simplified (or approximated) Hamiltonian in \eqnref{eq:SeSeHoppinga16Band_Sim} for $E_{\bsl{k}}$ being the energy dispersion of the obstructed atomic band.    }
    \label{fig:DFT6bandSimPlots}
\end{figure}

So far, we have not made any approximations.
However, at this point we encounter an issue: the DFT-precise Hamiltonian does not satisfy the invertibility requirement in \eqref{Perturbation1}, as the DFT-precise $E_{\bsl{k}} - H_{\Se}(\bsl{k})$ has zero eigenvalues for $E_{\bsl{k}}$ being the energy dispersion of the obstructed atomic band, as shown in \figref{fig:failure_conventional_perturbation_theory}.
Therefore, we have to use an approximated version of the Hamiltonian in order to use the above formalisms.
To do so, we first neglect all the terms beyond and including the 3NN hoppings in the 6-band model, \ie, we neglect all terms other than on-site terms, Nb-Nb and Se-Se $\bsl{a}_1$ (and symmetry-related) hopping terms, and $\Nb-\Se$ nearest neighbor hopping at the distance $\bsl{\tau}_{\text{Se}}$ (and symmetry-related) hopping terms.
We further neglect most of the Se-Se hopping terms at distance $\bsl{a}_1$ in \eqref{SeSeHoppinga16Band} with the exception of the $p_x-p_x$ hopping which is large. Most other Se-Se hoppings are about one order of magnitude smaller and should not influence the upper 3 Nb bands which we aim to approximate.
Eventually, in order to access its effect, we allow a scaling factor for the $p_x-p_x$ term of the Se-Se hopping terms at distance $\bsl{a}_1$, resulting in the following simplified 6-band Hamiltonian
\eq{
\label{eq:HTB6band_Sim}
H_{6,Sim} = \sum_{\bsl{R}\bsl{R}',\bsl{\tau}\bsl{\tau}'}^{|\bsl{R}+\bsl{\tau}-\bsl{R}'-\bsl{\tau}'|\leq |\bsl{a}_1|}c^\dagger_{\bsl{R}+\bsl{\tau}}  t_{\bsl{\tau}\bsl{\tau}'}^{sim}(\bsl{R}+\bsl{\tau}-\bsl{R}'-\bsl{\tau}') c_{\bsl{R}'+\bsl{\tau}'}\ ,
}
where the onsite terms are the same as \eqref{onsiteSeNb6Band}, the NN hoppings are the same as \eqref{tNNSeNb6Band}, the $|\bsl{a}_1|$-hoppings among Nb atoms are the same as \eqref{SeSeHoppinga16Band}, and the $|\bsl{a}_1|$-hoppings among Se atoms are given by 
\eqa{
\label{eq:SeSeHoppinga16Band_Sim}
& t_{\bsl{\tau}_{\Se}\bsl{\tau}_{\Se}}^{sim}(\bsl{a}_1)= \left(
\begin{array}{ccc}
 t_{NNN,z,z} &  t_{NNN,z,x}  & t_{NNN,z,y} \\
- t_{NNN,z,x} & t_{NNN,x,x} & t_{NNN,x,y} \\
t_{NNN,z,y} & -t_{NNN,x, y} & t_{NNN,y,y} \\
\end{array}
\right) = \left(
\begin{array}{ccc}
 0 & 0 & 0 \\
 0 & 0.8330 z & 0 \\
 0 & 0 & 0 \\
\end{array}
\right)\ .
}
As shown in \figref{fig:DFT6bandSimPlots}(a-f), the simplified (or approximated) Hamiltonian in \eqref{eq:SeSeHoppinga16Band_Sim} maintains the shape of the 1 band near the Fermi energy and has the highest two bands roughly at the right energy, although it does not match the lower 3 bands well.
Importantly, as $z$ goes from the realistic value $1$ to $0$, the 1 band near the Fermi energy does not change much, but the $E_{k} - H_{\Se}(\bsl{k})$ becomes invertible for $E_{\bsl{k}}$ being the energy dispersion of the obstructed atomic band as $z$ becomes small. (See \figref{fig:DFT6bandSimPlots}(g-l).)
Therefore, we will choose the \eqnref{eq:HTB6band_Sim} with $z=0$, which we call the Se-onsite NNN 6-band model $H_{6,\text{Se-Onsite,NNN}}$, for our perturbative analysis unless specified otherwise; more explicilty, 
\eq{
\label{eq:HTB6band_Se-Onsite_NNN}
H_{6,\text{Se-Onsite,NNN}} = \left. H_{6,Sim} \right|_{z\rightarrow 0}\ .
}

With the simplified Hamiltonian, we are now allowed to use \eqref{Perturbation1} and \eqref{eq:Perturbation_Gamma}.
Eventually, we want to have a perturbative understanding of the obstructed atomic band \eqref{eq:Perturbation_Gamma} at the Fermi level.
For that purpose, we need to approximate $\left[ E_{\bsl{k}}- H_{\Se}(\bsl{k}) \right]^{-1}$ by a simple form and solve for $\psi_{\text{Nb},\bsl{k}}$ in \eqref{eq:Perturbation_Gamma}.
The two tasks can be resolved simultaneously by developing an effective 3-band model for top three bands.
We can use the conventional perturbation theory to write down the effective 3-band model, \emph{if} (i) the top three bands are separated from the lower three bands by a gap that is much larger than the matrix elements of $S(\bsl{k})$ (in absolute values) and (ii) the top three bands have similar energies,
However, it is not at all the case for the Se-onsite NNN 6-band model $H_{6,\text{Se-Onsite,NNN}}$---(i) the gap between the top three bands and the lowest three bands can be as small as $1.605$eV along the high-symmetry line, while the matrix elements of $S(\bsl{k})$ has absolute values as large as $2.784$eV along the high-symmetry line, and (ii) the band at the Fermi level has a gap from the highest two bands about $1.5$eV. 
Therefore, we cannot use the conventional perturbation theory.

To resolve this issue, we observe that in $H_{6,\text{Se-Onsite,NNN}}$, the band at the Fermi level is quasi-flat with dispersion ranging from roughly $-0.5$eV to $0.5$eV. 
We will then consider the band as a zero-energy band and the dispersion as a perturbation.
We also observe that in $H_{6,\text{Se-Onsite,NNN}}$, the two highest bands also have minimal dispersion, and their energy is around $E_2= -E_z= 2.410 eV$. The dispersion away from this energy is $\pm 0.5 eV$, and thus we treat the two bands as two exactly flat bands at this energy ($E_2= -E_z= 2.410 eV$).
Now we are ready to present a new perturbation theory valid for flat bands at two energies. 
With the previous two approximations, we try to approximate
    \begin{eqnarray}
        f(E) = \frac{1}{E- H_{\Se}(\bsl{k})} \approx a + b E 
    \end{eqnarray} 
    subject to the constraints $f(0)= -H_{\Se, \bsl{k}}^{-1}$ and $f({-E_z})= (-E_z- H_{\Se}( \bsl{k}))^{-1}$. We hence have $a=-H_{\Se}^{-1}(\bsl{k})$ and $b=\frac{1}{E_z} (-H_{\Se}^{-1}( \bsl{k}) - (-E_z- H_{\Se}^{-1}( \bsl{k})) $, resulting in 
\begin{eqnarray}
\label{eq:new_perturbation_theory}
        \frac{1}{E- H_{\Se}(\bsl{k})} \approx -H_{\Se}^{-1}(\bsl{k}) + \frac{1}{E_z} \left[ -H_{\Se}^{-1}(\bsl{k}) - (-E_z- H_{\Se}(\bsl{k}))^{-1} \right] E 
    \end{eqnarray} 
Notice that this is an unconventional perturbation theory, as the slope of the $E$ linear term around $0$ would have been $-H_{\Se}^{-1}(\bsl{k})$ in conventional perturbation theory. 
It also requires knowledge about the energies of the bands.
Nevertheless, it will give the effective 3-band model that captures well the energies of the top $3$ bands, as discussed in the following.

\begin{figure}[t]
    \centering
    \includegraphics[width=\columnwidth]{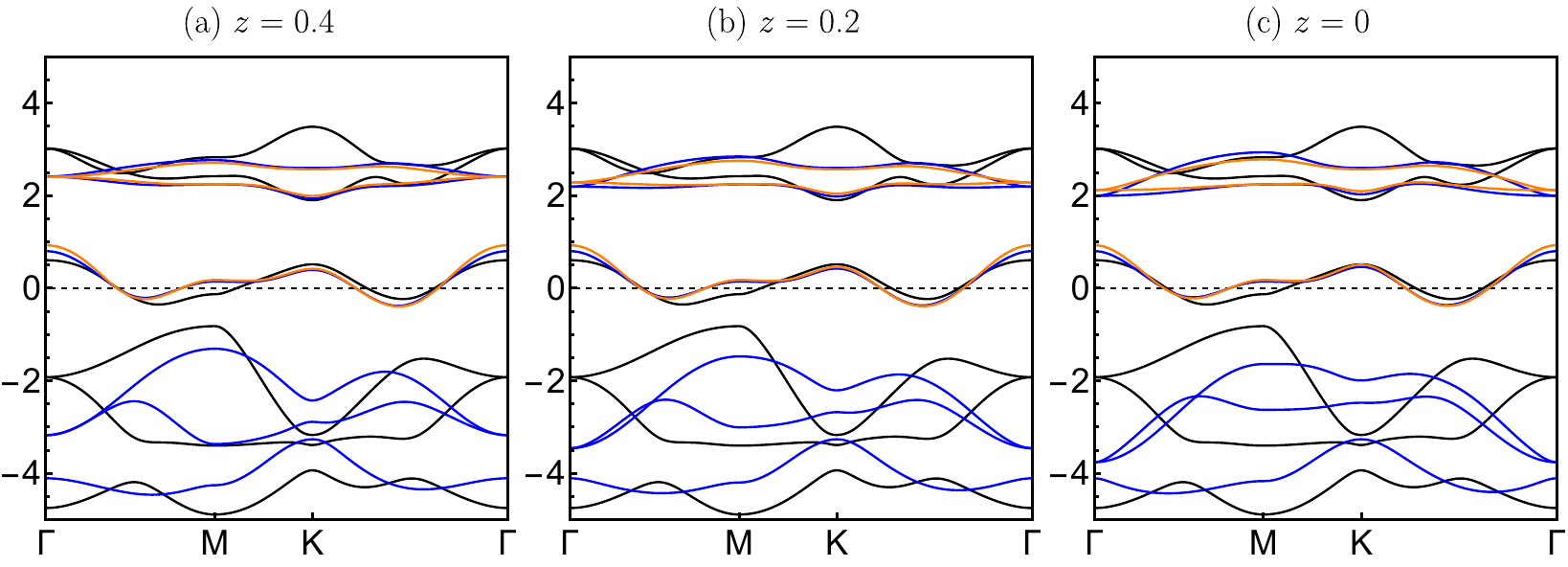}
    \caption{
    (a-c) The DFT band structure is plotted in black, and the band structure of the simplified (or approximated) 6-band Hamiltonian in \eqnref{eq:HTB6band_Sim} is plotted in blue for various values of $z$, and the band structure of the effective 3-band model in \eqref{eq:3band_eff_full} is plotted in orange.
    }
    \label{fig:gridplots3bandmodel}
\end{figure}

With the new perturbation \eqref{eq:new_perturbation_theory}, we now massage the second equation of \eqref{Perturbation1} to be:
\begin{eqnarray}
    (H_{\Nb}(\bsl{k}) - S(\bsl{k})   H_{\Se}^{-1}(\bsl{k}) S^\dagger(\bsl{k}) ) \psi_{\Nb,\bsl{k}}= E_{\bsl{k}}\left[ 1 + S(\bsl{k})\frac{1}{E_z} (H_{\Se}^{-1}(\bsl{k})  + (-E_z- H_{\Se}(\bsl{k}))^{-1})  S^\dagger(\bsl{k}) \right] \psi_{\Nb,\bsl{k}} \label{Perturbation2}
\end{eqnarray}
The term on the RHS of the eigenvalue equation is crucial in making the top 2 bands in the $3$-band model resemble the top $2$-bands in the $6$-band model. However, it is not necessary for the lower bands. In its absence, the top two bands would have dispersion more than an order of magnitude the correct one. The Hermitian  matrix $1 + S(\bsl{k})\frac{1}{E_z} (H_{\Se}^{-1}(\bsl{k})  + (-E_z- H_{\Se}(\bsl{k}))^{-1})  S^\dagger(\bsl{k}) $  has positive eigenvalues and can be diagonalized:
\begin{eqnarray}
    1 + S(\bsl{k})\frac{1}{E_z} (H_{\Se}^{-1}(\bsl{k})  + (-E_z- H_{\Se}(\bsl{k}))^{-1})  S^\dagger(\bsl{k}) = U_{\bsl{k}} D_{\bsl{k}} U^\dagger_{\bsl{k}} 
\end{eqnarray}
Now we can give eigenequation for a $3$-band effective Hamiltonian derived from the second equation of \eqref{Perturbation1}, which reads to be
\begin{eqnarray}
\label{eq:3band_eff_full}
  \frac{1}{\sqrt{D_{\bsl{k}}}}  U_{\bsl{k}}^\dagger  (H_{Nb}(\bsl{k})- S(\bsl{k})   H_{Se}^{-1}(\bsl{k}) S^\dagger(\bsl{k}) ) U_{\bsl{k}} \frac{1}{\sqrt{D_{\bsl{k}}}} \psi'_{Nb,\bsl{k}}= E_{\bsl{k}}   \psi'_{Nb,\bsl{k}},\;\;\;\; \psi'_{Nb,\bsl{k}}= \sqrt{D_{\bsl{k}}} U_{\bsl{k}}^\dagger \psi_{\Nb,\bsl{k}} \label{Perturbation3}
\end{eqnarray} 
where $\sqrt{D_k}$ is the square root of a diagonal, positive definite matrix. 
As shown in \figref{fig:gridplots3bandmodel}(c), we see that the effective 3-band model in \eqref{eq:3band_eff_full} performs a remarkable job in capturing the top 3 bands of $H_{6,\text{Se-Onsite,NNN}}$.
The match is still remarkable even if we increase the value of $z$ in \eqref{eq:HTB6band_Sim} to $0.4$ and $0.2$ from 0, as shown in \figref{fig:gridplots3bandmodel}(a-b). 
The agreement is not as good for the lower 3 bands, which originate on Se, but is rather excellent for the upper $3$ bands.  
From \eqref{eq:3band_eff_full}, we can also solve for the $\psi_{\Nb,\bsl{k}}$, which combined with 
\eq{
\label{eq:psi_Se_approx}
\psi_{\Se,\bsl{k}} \approx \left[ - H_{\Se}(\bsl{k}) \right]^{-1} S^\dagger(\bsl{k})\psi_{\Nb,\bsl{k}}
}
derived from the first equation of \eqref{Perturbation1}, gives the approximate form of the creation operator of the obstructed atomic band:
\eq{
\gamma_{approx,w,\bsl{k}}^\dagger = N_{\bsl{k}} \left[ c^\dagger_{\Nb,\bsl{k}} +   c^\dagger_{\Se,\bsl{k}}  (- H_{\Se}(\bsl{k}))^{-1} S^\dagger(\bsl{k}) \right] \psi_{Nb,\bsl{k}} = (c^\dagger_{\Nb,\bsl{k}}, c^\dagger_{\Se,\bsl{k}} ) \psi_{w,\bsl{k},approx}\ ,
}
where $H_{\Se}(\bsl{k})$ and $S(\bsl{k})$ are taken from the Se-onsite NNN 6-band model $H_{6,\text{Se-Onsite,NNN}}$, $N_{\bsl{k}}$ is the normalization factor, $c^\dagger_{\Nb,\bsl{k}}$ and $c^\dagger_{\Se,\bsl{k}}$ are defined in \eqref{eq:c_k_6band}, and 
\eq{
\label{eq:approx_psi_w}
\psi_{w,\bsl{k},approx} = N_{\bsl{k}}\mat{ \psi_{\Nb, \bsl{k}} \ ,
(- H_{\Se}(\bsl{k}) )^{-1} S^\dagger(\bsl{k}) \psi_{\Nb, \bsl{k}}}
}
The $\gamma_{approx,w,\bsl{k}}^\dagger$ has remarkably high probability overlap with the DFT-precise Bloch state $w_{\bsl{k}}^\dagger$ for the single band at the Fermi level, \ie,
\eq{
\frac{1}{N}\sum_{\bsl{k}} \left| \bra{0} \gamma_{approx,w,\bsl{k}} w_{\bsl{k}}^\dagger \ket{0} \right|^2 =  0.9638\ ,
}
where $\ket{0}$ is the vacuum state. This shows the power of our perturbation theory for the band around the Fermi level. 

\subsection{Perturbative Analysis for the Obstructed Atomic Band}

\label{app:1-band_approx}

In \appref{app:new_PT_and_3band_Model}, we used the new perturbation theory to derive an effective 3-band model that captures the dispersion of the top three bands.
In this section, we will provide a perturbative understanding of the obstructed atomic band at the Fermi energy.
Unless specified otherwise, we will still use the Se-onsite NNN 6-band model, $H_{6,\text{Se-Onsite,NNN}}$ in \eqref{eq:HTB6band_Se-Onsite_NNN}, to keep $E_{\bsl{k}}-H_{\Se}(\bsl{k})$ invertiable for $E_{\bsl{k}}$ being the energy of the obstructed atomic band.
At the end of \appref{app:new_PT_and_3band_Model}, we see that if we use the new perturbation theory defined in \eqref{eq:new_perturbation_theory}, we can obtain an approximated wavefunction that has 0.964 probability overlap with the obstructed atomic band at the Fermi level.
In this section, we will show that even if we neglect the second term in \eqnref{eq:new_perturbation_theory} and further simplify $H_{6,\text{Se-Onsite,NNN}}$, we can still obtain a good approximated wavefunction for the obstructed atomic band at the Fermi level.

As we neglect the second term in \eqnref{eq:new_perturbation_theory}, we have the following effective three-band Hamiltonian 
\eq{
\label{eq:3band_eff}
H_{\Nb,eff}(\bsl{k}) = H_{\Nb}(\bsl{k}) - S(\bsl{k})  (H_{\Se}(\bsl{k}) )^{-1} S^\dagger(\bsl{k}) \ ,
}
where $H_{\Nb}(\bsl{k})$, $H_{\Se}(\bsl{k})$ and $S(\bsl{k})$ are defined in \eqref{eq:H_6_Nb_Se_S} by replacing $H_6$ to $H_{6,\text{Se-Onsite,NNN}}$.
Then, the approximated wavefunction for the obstructed atomic band is given by \eqnref{eq:approx_psi_w} with $\psi_{\Nb,\bsl{k}}$ derived from
\begin{eqnarray}
\label{Perturbation5}
    & H_{\Nb,eff}(\bsl{k})  \psi_{\Nb,\bsl{k}}= E_{\bsl{k}} \psi_{Nb,\bsl{k}} \nonumber \ ,
\end{eqnarray} 
where $E_{\bsl{k}}$ is the lowest band of $H_{\Nb,eff}(\bsl{k})$.

Since we use $H_{6,\text{Se-Onsite,NNN}}$ which Se atoms only have onsite terms, we have
\begin{eqnarray}
    & H_{\Se}(\bsl{k})= t_{\bsl{\tau}_{\Se}\bsl{\tau}_{\Se}}(\bsl{0}) = \left(
\begin{array}{ccc}
 E_{z} & 0 & 0 \\
 0 & E_{x} & 0 \\
 0 & 0 & E_{x} \\
\end{array}
\right)\ .
\end{eqnarray} 
With this, $ H_{\text{Nb, eff}}(\bsl{k})$ reads:
\begin{eqnarray}
    &  H_{\text{Nb, eff}}(\bsl{k}) = H_{\Nb}(\bsl{k}) - S(\bsl{k})  ( t_{\bsl{\tau}_{\Se}\bsl{\tau}_{\Se}}(\bsl{0}) )^{-1} S^\dagger(\bsl{k}) \ .  \label{eq:Nb_eff}
\end{eqnarray}
We then separate $H_{\text{Nb, eff}}(\bsl{k})$ into the on-site term and the Nb-Nb $|\bsl{a}_1|$ hopping terms:
\begin{eqnarray}
    &  H_{\text{Nb, eff}}(\bsl{k}) = t_{\bsl{\tau}_{\Nb}\bsl{\tau}_{\Nb}}(\bsl{0})   - [ S(\bsl{k})  ( t_{\bsl{\tau}_{\Se}\bsl{\tau}_{\Se}}(\bsl{0}) )^{-1} S^\dagger(\bsl{k}) ]_{\text{on site}}     + \nonumber \\  &+ (H_{\Nb}(\bsl{k})- t_{\bsl{\tau}_{\Nb}\bsl{\tau}_{\Nb}}(\bsl{0}))  - [S(\bsl{k})  ( t_{\bsl{\tau}_{\Se}\bsl{\tau}_{\Se}}(\bsl{0}) )^{-1} S^\dagger(\bsl{k})  - [ S(\bsl{k})  ( t_{\bsl{\tau}_{\Se}\bsl{\tau}_{\Se}}(\bsl{0}) )^{-1} S^\dagger(\bsl{k}) ]_{\text{on site}}  ]
\end{eqnarray} 
where the first and second lines are the on-site term and the Nb-Nb $|\bsl{a}_1|$ hopping terms, respectively, as the renormalized Hamiltonian $S(\bsl{k})  ( t_{\bsl{\tau}_{\Se}\bsl{\tau}_{\Se}}(\bsl{0}) )^{-1} S^\dagger(\bsl{k}) $ involves only those two kinds of terms. 
If we Fourier transform back to real space, we find the new effective the on-site term and the Nb-Nb $|\bsl{a}_1|$ hopping terms:

\begin{eqnarray}
\label{eq:Nb_eff_real_hoppings}
    &t_{\bsl{\tau}_{\Nb}\bsl{\tau}_{\Nb} \text{eff}}(\bsl{0})  = t_{\bsl{\tau}_{\Nb}\bsl{\tau}_{\Nb}}(\bsl{0})   - \sum_{m=1}^3 t^\dagger_{\bsl{\tau}_{\Se}\bsl{\tau}_{\Nb}}(C_3^{m-1} \bsl{\tau}_{\Se})    ( t_{\bsl{\tau}_{\Se}\bsl{\tau}_{\Se}}(\bsl{0}) )^{-1} t_{\bsl{\tau}_{\Se}\bsl{\tau}_{\Nb}}(C_3^{m-1} \bsl{\tau}_{\Se})     \nonumber \\  & t_{\bsl{\tau}_{\Nb}\bsl{\tau}_{\Nb} \text{eff}}(\bsl{a}_1) = t_{\bsl{\tau}_{\Nb}\bsl{\tau}_{\Nb}}(\bsl{a}_1)  -t^\dagger_{\bsl{\tau}_{\Se}\bsl{\tau}_{\Nb}}(C_3^{} \bsl{\tau}_{\Se})    ( t_{\bsl{\tau}_{\Se}\bsl{\tau}_{\Se}}(\bsl{0}) )^{-1} t_{\bsl{\tau}_{\Se}\bsl{\tau}_{\Nb}}(C_3^{2} \bsl{\tau}_{\Se})    
\end{eqnarray}
Explicitly, we have
\eqa{
& t_{\bsl{\tau}_{\Nb}\bsl{\tau}_{\Nb} \text{ eff}}(\bsl{0}) =\left(
\begin{array}{ccc}
 E_{d_{z^2} \text{eff}}& 0 & 0 \\
 0 & E_{d_{xy} \text{eff}}  & 0 \\
 0 & 0 & E_{d_{xy} \text{eff}} \\
\end{array}
\right) \\
 & E_{d_{z^2}\text{ eff}}= E_{d_{z^2}}- 3\frac{t^2_{NN, y, d_{z^2} } E_z + t^2_{NN, z, d_{z^2} } E_x  }{ E_{z} E_{x}}  \\ 
 & E_{d_{xy} \text{eff}}= E_{d_{xy}}- \frac{3}{2}\frac{( t^2_{NN, x, d_{xy}}+  t^2_{NN, y, d_{x^2-y^2} }) E_z + t^2_{NN, z, d_{x^2-y^2} } E_x  }{ E_{z} E_{x}}
}
and
\eqa{ \label{Pertubration9}
&t_{\bsl{\tau}_{\Nb}\bsl{\tau}_{\Nb} \text{eff}}(\bsl{a}_{1})=      \left(
\begin{array}{ccc}
 t_{NNN,d_{z^2},d_{z^2} \text{eff}} &  t_{NNN,d_{z^2},d_{xy}\text{eff}}  & t_{NNN,d_{z^2},d_{x^2-y^2}\text{eff}} \\
- t_{NNN,d_{z^2},d_{xy}\text{eff}} & t_{NNN,d_{xy},d_{xy} \text{eff}} & t_{NNN,d_{xy}, d_{x^2-y^2}\text{eff}} \\
t_{NNN,d_{z^2},d_{x^2-y^2}\text{eff}} & -t_{NNN,d_{xy}, d_{x^2-y^2}\text{eff}} & t_{NNN,d_{x^2-y^2},d_{x^2-y^2}\text{eff}} \\
\end{array}
\right)  \\ & t_{NNN,d_{z^2},d_{z^2} \text{eff}}= t_{NNN,d_{z^2},d_{z^2}}-  (\frac{ t_{NN,z,d_{z^2}}^2}{E_z}-\frac{t_{NN,y,d_{z^2}}^2}{2 E_x})    \\  & t_{NNN,d_{z^2},d_{xy}\text{eff}} = t_{NNN,d_{z^2},d_{xy}} - \frac{\sqrt{3} (2 E_x  t_{NN,z,d_{x^2-y^2}}  t_{NN,z,d_{z^2}}-E_z t_{NN,y,d_{z^2}} (t_{NN,x,d_{xy}}+ t_{NN,y,d_{x^2-y^2}}))}{4 E_x E_z}   \\ &t_{NNN,d_{z^2}, d_{x^2-y^2}\text{eff}}= t_{NNN,d_{z^2}, d_{x^2-y^2}}- \frac{1}{4} \left(\frac{t_{NN,y,d_{z^2}} ( t_{NN,y,d_{x^2-y^2}}-3 t_{NN,x,d_{xy}})}{E_x}-\frac{2  t_{NN,z,d_{x^2-y^2}}  t_{NN,z,d_{z^2}}}{E_z}\right)   \\ &  t_{NNN,d_{xy},d_{xy} \text{eff}} =  t_{NNN,d_{xy},d_{xy}} -  (-\frac{t_{NN,x,d_{xy}}^2-6 t_{NN,x,d_{xy}} t_{NN,y,d_{x^2-y^2}}-3  t_{NN,y,d_{x^2-y^2}}^2}{8 E_x}-\frac{3  t_{NN,z,d_{x^2-y^2}}^2}{4 E_z})  \\ &  t_{NNN,d_{xy}, d_{x^2-y^2}\text{eff}} =  t_{NNN,d_{xy}, d_{x^2-y^2}} - \frac{\sqrt{3} \left(2 E_x  t_{NN,z,d_{x^2-y^2}}^2-E_z (t_{NN,x,d_{xy}}- t_{NN,y,d_{x^2-y^2}})^2\right)}{8 E_x E_z}  \\ & t_{NNN,d_{x^2-y^2},d_{x^2-y^2}\text{eff}}= t_{NNN,d_{x^2-y^2},d_{x^2-y^2}}-\frac{1}{8} \left(\frac{3 t_{NN,x,d_{xy}}^2+6 t_{NN,x,d_{xy}} t_{NN,y,d_{x^2-y^2}}- t_{NN,y,d_{x^2-y^2}}^2}{E_x}+\frac{2  t_{NN,z,d_{x^2-y^2}}^2}{E_z}\right)\ .
}

We have focused on the on-site term and the Nb-Nb $|\bsl{a}_1|$ hopping terms. The perturbation presented here and in the following, can however work for more complicated Hamiltonians including longer-range hoppings, etc. 
We now describe the perturbation theory used to solve the Hamiltonian $H_{\text{Nb, eff}}(\bsl{k})$ in \eqref{eq:Nb_eff} further.

To motivate the perturbation theory we use, we show the numerical values of the hopping terms in \eqref{eq:Nb_eff_real_hoppings} in the rotated basis:
\eqa{
\label{eq:terms_3band_eff_values}
& R t_{\bsl{\tau}_{\Nb}\bsl{\tau}_{\Nb} \text{eff}}(\bsl{0}) R^\dagger = \left(
\begin{array}{ccc}
 2.7909 & -0.1275 & -0.1275 \\
 -0.1275 & 2.7909 & -0.1275 \\
 -0.1275 & -0.1275 & 2.7909 \\
\end{array}
\right) \\
& R t_{\bsl{\tau}_{\Nb}\bsl{\tau}_{\Nb} \text{eff}}(\bsl{a}_1) R^\dagger = \left(
\begin{array}{ccc}
 0.7047 & -0.1774 & 0.1275 \\
 0.1275 & -0.1585 & 0.3590 \\
 -0.1774 & -1.2323 & -0.1585 \\
\end{array}
\right) \ ,
}
where the unit is eV, and the values are derived from $H_{6,\text{Se-onsite, NNN}}$ in \eqref{eq:HTB6band_Se-Onsite_NNN}.
From the numerical values, we can see that besides the average onsite energy $2.7909$eV, the largest value is the $(3,2)$ elements of $R t_{\bsl{\tau}_{\Nb}\bsl{\tau}_{\Nb} \text{eff}}(\bsl{a}_1) R^\dagger$.
The zeroth-order Hamiltonian for our perturbation theory will be chosen to include only the onsite energy and the 23 element of $R t_{\bsl{\tau}_{\Nb}\bsl{\tau}_{\Nb} \text{eff}}(\bsl{a}_1) R^\dagger$ (and its symmetry-related partners).
As discussed in \appref{app:1band_model}, such a zeroth-order Hamiltonian describes a compact obstructed atomic orbital at 1c position, whose eigenstates are created by
\begin{eqnarray}
\label{eq:compact_wannier_U}
   &  (c^\dagger_{\bsl{R},0}, c^\dagger_{\bsl{R},+}, c^\dagger_{\bsl{R},-})  = (c^\dagger_{\bsl{R} + \bsl{a}_1 + \bsl{a}_2, 1}, c^\dagger_{\bsl{R},2},c^\dagger_{\bsl{R} + \bsl{a}_1 ,3})U, 
\end{eqnarray} 
where
\eq{
(c^\dagger_{ \bsl{R}, 1}, c^\dagger_{\bsl{R},2},c^\dagger_{\bsl{R} ,3}) = (c^\dagger_{\bsl{R},d_{z^2}}, c^\dagger_{\bsl{R}, d_{xy}}, c^\dagger_{\bsl{R}, d_{x^2-y^2}}) R\ ,
}
$c^\dagger_{\bsl{R},d_{z^2}}, c^\dagger_{\bsl{R}, d_{xy}}, c^\dagger_{\bsl{R}, d_{x^2-y^2}}$ are defined in \eqref{eq:6band_basis}, $R$ is defined in \eqref{eq:R_rotation}, and 
\eq{
U=  \frac{1}{\sqrt{3}}\left( \begin{array}{ccc}
 1 & e^{i \frac{4\pi}{3}} & e^{-i \frac{4\pi}{3}} \\
 1 & 1 & 1 \\
 1 & e^{i \frac{2\pi}{3}} & e^{-i \frac{2\pi}{3}} \\
\end{array}\right) \ .
}
We now write down the zeroth-order Hamiltonian
\eqa{
&\sum_{\bsl{R}}  (c^\dagger_{ \bsl{R} + \bsl{a}_1 + \bsl{a}_2,1}, c^\dagger_{ \bsl{R},2},c^\dagger_{ \bsl{R} + \bsl{a}_1 ,3})   \left( \begin{array}{ccc}
 \varepsilon_0 & t & t \\
 t & \varepsilon_0 & t \\
 t & t & \varepsilon_0\\
\end{array}\right)     \left( \begin{array}{c}
 c_{\bsl{R} + \bsl{a}_1 + \bsl{a}_2,1} \\
 c_{ \bsl{R},2} \\
 c_{ \bsl{R} + \bsl{a}_1,3 } \\
\end{array}\right)\\
&  =  \sum_{\bsl{R}}(c^\dagger_{\bsl{R},0}, c^\dagger_{\bsl{R},+}, c^\dagger_{\bsl{R},-}) \left( \begin{array}{ccc}
 \varepsilon_0 + 2 t& 0 & 0 \\
 0 & \varepsilon_0-t & 0 \\
 0 & 0 & \varepsilon_0-t\\
\end{array}\right)  \left( \begin{array}{c}
 c_{\bsl{R},0}  \\
 c_{\bsl{R},+} \\
 c_{\bsl{R},-} \\
\end{array}\right)    \\ 
& =  \sum_{\bsl{k}} (c^\dagger_{Nb, \bsl{k},d_{z^2}}, c^\dagger_{Nb, \bsl{k},d_{xy}},c^\dagger_{Nb, \bsl{k} ,d_{x^2-y^2}})   H_{eff,0}(\bsl{k})     \left( \begin{array}{c}
 c_{Nb, \bsl{k},d_{z^2}} \\
 c_{Nb, \bsl{k},d_{xy}} \\
 c_{Nb, \bsl{k},d_{x^2-y^2} } \\
\end{array}\right)\ ,
} 
where
\eq{
\label{eq:H_eff_0}
H_{eff,0}(\bsl{k}) = R  \mat{ e^{-\ii \bsl{k}\cdot(\bsl{a}_1+\bsl{a}_2)} & & \\ & 1 & \\ & & e^{-\ii \bsl{k}\cdot\bsl{a}_1}} \left( \begin{array}{ccc}
 \varepsilon_0 & t & t \\
 t & \varepsilon_0 & t \\
 t & t & \varepsilon_0\\
\end{array}\right)  \mat{ e^{\ii \bsl{k}\cdot(\bsl{a}_1+\bsl{a}_2)} & & \\ & 1 & \\ & & e^{\ii \bsl{k}\cdot\bsl{a}_1}}  R^\dagger \ ,
}
$E_0$ is the diagonal element of $R t_{\bsl{\tau}_{\Nb}\bsl{\tau}_{\Nb} \text{eff}}(\bsl{0}) R^\dagger$, and $t$ would be $\left[R t_{\bsl{\tau}_{\Nb}\bsl{\tau}_{\Nb} \text{eff}}(\bsl{a}_1) R^\dagger \right]_{32}$. (See the expressions of $t_{\bsl{\tau}_{\Nb}\bsl{\tau}_{\Nb} \text{eff}}(\bsl{0})$ and $t_{\bsl{\tau}_{\Nb}\bsl{\tau}_{\Nb} \text{eff}}(\bsl{0})$ in \eqnref{eq:Nb_eff_real_hoppings}.) 

As we can see, $H_{eff,0}(\bsl{k})$ can be solved exactly, with the eigenstates created by 
\eq{
\label{OAIkbasis}
(c^\dagger_{\bsl{k},0}, c^\dagger_{\bsl{k},+}, c^\dagger_{\bsl{k},-}) = (c^\dagger_{Nb, \bsl{k},d_{z^2}}, c^\dagger_{Nb, \bsl{k},d_{xy}},c^\dagger_{Nb, \bsl{k} ,d_{x^2-y^2}}) R U_w(\bsl{k})\ ,
}
where
\eq{
U_w(\bsl{k}) = \mat{ e^{-\ii \bsl{k}\cdot(\bsl{a}_1+\bsl{a}_2)} & & \\ & 1 & \\ & & e^{-\ii \bsl{k}\cdot\bsl{a}_1}}U\ .
}
The expressions of the eigenstates of the zeroth order $H_{eff,0}(\bsl{k})$ allow us to derive the higher order terms explicitly.

\begin{figure}[t]
    \centering
    \includegraphics[width=\columnwidth]{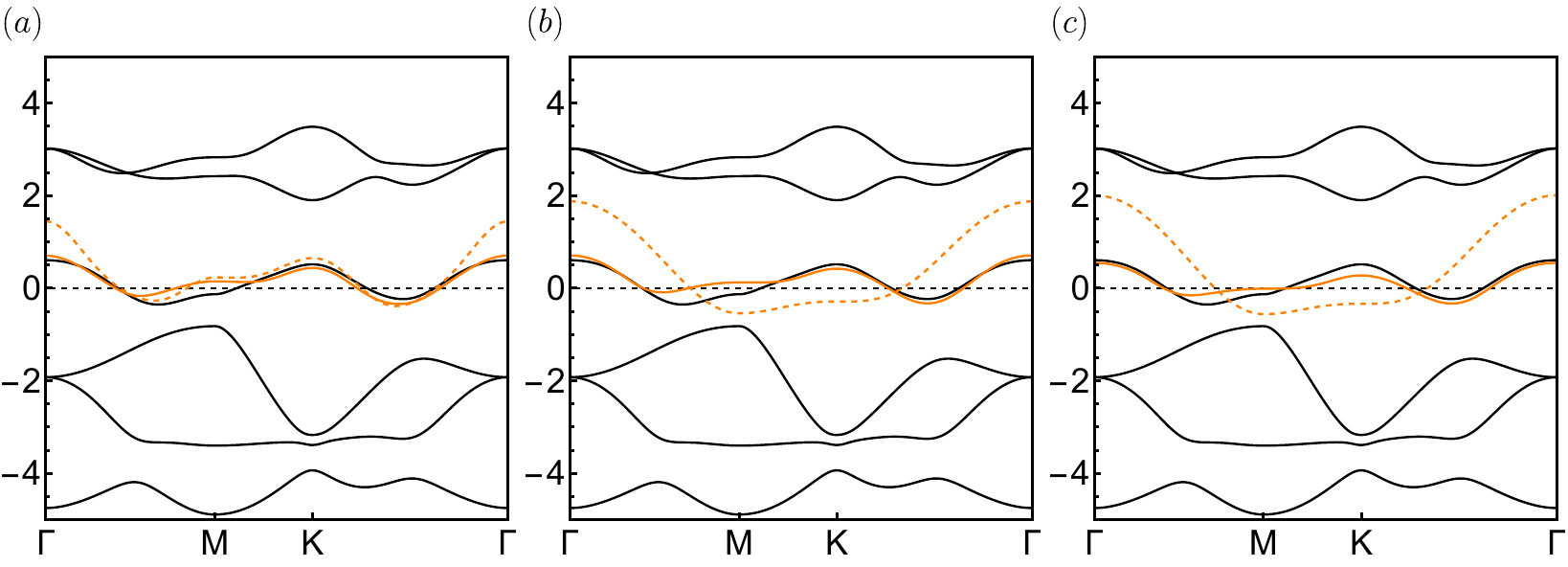}
    \caption{
    The band structures of DFT are plotted in black.
    The band structure given by \eqref{eq:approx_E_w_k} and \eqnref{eq:exp_value} for $H_{6,\text{Se-onste,NNN}}$ in \eqref{eq:HTB6band_Se-Onsite_NNN} is plotted as orange dashed and solid lines, respectively, in (a).
    The band structure given by \eqref{eq:E_w_k_sim1} and \eqnref{eq:exp_value} for the simplification in \eqref{eq:sim1_approx1} and \eqref{eq:sim1_approx2} is plotted as orange dashed and solid lines, respectively, in (b).
    The band structure given by \eqref{eq:E_w_k_sim2} and \eqnref{eq:exp_value} for the simplification in \eqref{eq:sim2_approx1} and \eqref{eq:sim2_approx2} is plotted as orange dashed and solid lines, respectively, in (c).
    The great match between the orange and black lines evidencees that our OAI states are well captured by our approximation. 
    }
    \label{fig:perturbation_1band}
\end{figure}

The 3-band effective Hamiltonian can now be split into 
\eq{
H_{\Nb,eff}(\bsl{k}) = H_{eff,0}(\bsl{k}) + H_{eff,1}(\bsl{k}) \ ,
}
where
\eq{
H_{eff,1}(\bsl{k}) = H_{\Nb,eff}(\bsl{k}) - H_{eff,0}(\bsl{k}) \ .
}
To perform perturbation theory, we rotate $H_{\Nb,eff}(\bsl{k})$ to the eigenbasis of $H_{eff,0}(\bsl{k})$:
\eq{
\overline{H}_{\Nb,eff}(\bsl{k}) =U_w^\dagger(\bsl{k})   R^\dagger H_{\Nb,eff}(\bsl{k}) R U_w(\bsl{k}) = \overline{H}_{eff,0}(\bsl{k}) + \overline{H}_{eff,1}(\bsl{k})\ ,
}
where
\eq{
\overline{H}_{eff,0}(\bsl{k}) 
= 
\mat{ 
 E_0(\bsl{k}) & & \\
& E_+(\bsl{k}) & \\
& & E_-(\bsl{k})
}\ .
}
Then, the approximated dispersion for the obstructed atomic band reads
\eq{
\label{eq:approx_E_w_k}
E_{w,\bsl{k}} = E_0(\bsl{k}) + \left[ \overline{H}_{eff,1}(\bsl{k}) \right]_{11} + \frac{ \left| [\overline{H}_{eff,1 }( \bsl{k})]_{21} \right|^2}{E_0(\bsl{k})-E_+(\bsl{k})} + \frac{ \left| [\overline{H}_{eff,1 }( \bsl{k})]_{31} \right|^2}{E_0(\bsl{k})-E_-(\bsl{k})}\ ,
}
and the corresponding eigenstate is created by
\eqa{
c^\dagger_{eff,0,\bsl{k}} & = (c^\dagger_{\bsl{k},0}, c^\dagger_{\bsl{k},+}, c^\dagger_{\bsl{k},-}) \left( \begin{array}{c}
1  \\
 \frac{1}{E_0(\bsl{k})-E_+(\bsl{k})} [\overline{H}_{eff,1 }( \bsl{k})]_{21}\\
\frac{1}{E_0(\bsl{k})-E_-(\bsl{k})} [\overline{H}_{eff,1 }( \bsl{k})]_{31}  \\
\end{array}
\right) \\
& = (c^\dagger_{Nb, \bsl{k},d_{z^2}}, c^\dagger_{Nb, \bsl{k},d_{xy}},c^\dagger_{Nb, \bsl{k} ,d_{x^2-y^2}}) R U_w(\bsl{k}) 
\left( \begin{array}{c}
1  \\
 \frac{1}{E_0(\bsl{k})-E_+(\bsl{k})} [\overline{H}_{eff,1 }( \bsl{k})]_{21}\\
\frac{1}{E_0(\bsl{k})-E_-(\bsl{k})} [\overline{H}_{eff,1 }( \bsl{k})]_{31}  \\
\end{array}
\right)\ ,
}
meaning that the approximate expression of $\psi_{\Nb}(\bsl{k})$ derived from the perturbation theory reads
\eq{
\label{eq:approx_psi_NB}
\psi_{\Nb}(\bsl{k}) = R U_w(\bsl{k}) \left( \begin{array}{c}
1  \\
 \frac{1}{E_0(\bsl{k})-E_+(\bsl{k})} [\overline{H}_{eff,1 }( \bsl{k})]_{21}\\
\frac{1}{E_0(\bsl{k})-E_-(\bsl{k})} [\overline{H}_{eff,1 }( \bsl{k})]_{31}  \\
\end{array}
\right) \ .
}
By substituting \eqnref{eq:approx_psi_NB} into \eqref{eq:approx_psi_w}, we can obtain the approximated expression of the creation operator of the obstructed atomic band.
With the parameter values for $H_{6,\text{Se-onste,NNN}}$ in \eqref{eq:HTB6band_Se-Onsite_NNN}, we find that our perturbation procedure gives an approximated state of the obstructed atomic band that has a remarkable probability overlap with the DFT-precise one:
\eq{
\frac{1}{N}\sum_{\bsl{k}} \left| \bra{0} \gamma_{approx,w,\bsl{k}} w_{\bsl{k}}^\dagger \ket{0} \right|^2 =  0.9538\ .
}
However, as shown by the orange dashed line in \figref{fig:perturbation_1band}(a), the approximated dispersion given by \eqnref{eq:approx_E_w_k} does not perfectly match the obstructed atomic band near Fermi energy of $H_{6,\text{Se-onste,NNN}}$ in \eqref{eq:HTB6band_Se-Onsite_NNN} and that of the DFT precise Hamiltonian.
On the other hand, if we compare the expectation value
\eq{
\label{eq:exp_value}
\bra{0} \gamma_{approx,w,\bsl{k}} H_{6,\text{Se-onste,NNN}} \gamma_{approx,w,\bsl{k}}^\dagger \ket{0} 
}
to the dispersion of the obstructed atomic band of $H_{6,\text{Se-onste,NNN}}$ in \eqref{eq:HTB6band_Se-Onsite_NNN} and that of the DFT precise Hamiltonian, we can see the good match in \figref{fig:perturbation_1band}(a) between the two, as the orange and black lines.

In the above analysis, we still numerically evaluate $\psi_{\Nb}(\bsl{k})$ and $\psi_{\Se}(\bsl{k})$.
In the following, we will further simply $H_{6,\text{Se-onste,NNN}}$ in order to obtain an analytical expression for $\psi_{\Nb}(\bsl{k})$ and $\psi_{\Se}(\bsl{k})$.

\subsubsection{Simplification 1}

We first consider the following approximations (in eV):
\eqa{ 
\label{eq:sim1_approx1}
 & t_{\bsl{\tau}_{\Se}\bsl{\tau}_{\Se}}(\bsl{0}) = \left(
\begin{array}{ccc}
 E_z & 0 & 0 \\
 0 & E_x & 0 \\
 0 & 0 & E_x \\
\end{array}
\right)=  \left(
\begin{array}{ccc}
 -\frac{12}{5} & 0 & 0 \\
 0 & -\frac{8}{5} & 0 \\
 0 & 0 & -\frac{8}{5} \\
\end{array}
\right) \\
&  t_{\bsl{\tau}_{\Se}\bsl{\tau}_{\Nb}} (\bsl{\tau_{\Se}})=    \left(
\begin{array}{ccc}
t_B & 0 & -\sqrt{2}t_B \\
 0 & 2t_A & 0 \\
 \sqrt{2}t_A & 0 &t_A \\
\end{array}
\right)\text{ with $t_A = -0.6300$eV and $t_B = 0.7700$}\ .
}
The values exhibit at most about 10\% deviation from the corresponding values in \eqref{onsiteSeNb6Band} and \eqnref{tNNSeNb6Band}.
Based on the values of $t_{\bsl{\tau}_{\Nb}\bsl{\tau}_{\Nb} \text{ eff}}(\bsl{0})$ and $t_{\bsl{\tau}_{\Nb}\bsl{\tau}_{\Nb} \text{ eff}}(\bsl{a}_1)$ for $H_{6,\text{Se-onsite,NNN}}$ (\eqref{eq:terms_3band_eff_values}), we approximate $t_{\bsl{\tau}_{\Nb}\bsl{\tau}_{\Nb} \text{ eff}}(\bsl{0})$ and $t_{\bsl{\tau}_{\Nb}\bsl{\tau}_{\Nb} \text{ eff}}(\bsl{a}_1)$ by
\eqa{
\label{eq:sim1_approx2}
& R^\dagger t_{\bsl{\tau}_{\Nb}\bsl{\tau}_{\Nb} \text{ eff}}(\bsl{0}) R = \frac{11}{4} I_{3\times 3} \\
& R^\dagger t_{\bsl{\tau}_{\Nb}\bsl{\tau}_{\Nb} \text{eff}}(\bsl{a}_1) R =   \left(
\begin{array}{ccc}
 5/(4\sqrt{3}) & 0 & 0 \\
 0 & 0 & 0 \\
 0 & -5/4 & 0 \\
\end{array}
\right)\ ,
} 
where the errors are not large than $0.4$eV.
With the simplification \eqnref{eq:sim1_approx1} and \eqnref{eq:sim1_approx2}, we are now ready to provide analytical expressions of $E_{w,\bsl{k}}$, $\psi_{\Nb,\bsl{k}}$ and $\psi_{\Se,\bsl{k}}$.

First, we note that the simplification on the effective 3-band model leads to 
\eqa{
\label{eq:H_bar_eff_sim_1}
& \overline{H}_{eff,0}(\bsl{k}) = \left(
\begin{array}{ccc}
 \frac{1}{4} & 0 & 0 \\
 0 & 4 & 0 \\
 0 & 0 & 4 \\
\end{array}
\right) \\
& \left[ \overline{H}_{eff,1}(\bsl{k}) \right]_{11} = \frac{5 \left(2 \cos \left(\frac{k_x a}{2}\right) \cos \left(\frac{\sqrt{3} k_y a}{2}\right)+\cos (k_x a)\right)}{6 \sqrt{3}} \\
& \left[ \overline{H}_{eff,1}(\bsl{k}) \right]_{21} = \frac{5 \left(-\sqrt[3]{-1} \cos \left(\frac{1}{2} \left(k_x a+\sqrt{3} k_y a\right)\right)+\cos \left(\frac{1}{2} \left(k_x a-\sqrt{3} k_y a\right)\right)+(-1)^{2/3} \cos (k_x a)\right)}{6 \sqrt{3}} \\
& \left[ \overline{H}_{eff,1}(\bsl{k}) \right]_{31} = \frac{5 \left((-1)^{2/3} \cos \left(\frac{1}{2} \left(k_x a+\sqrt{3} k_y a\right)\right)+\cos \left(\frac{1}{2} \left(k_x a-\sqrt{3} k_y a\right)\right)-\sqrt[3]{-1} \cos (k_x a)\right)}{6 \sqrt{3}} \ .
}
Then, combined with \eqnref{eq:approx_E_w_k}, we obtain the analytic expression of the approximated dispersion as
\eqa{
\label{eq:E_w_k_sim1}
& E_{w,\bsl{k}} = \frac{1}{324} \left[20 \cos \left(\frac{k_x a}{2}\right) \left(4 \cos (k_x a)+9 \sqrt{3}\right) \cos \left(\frac{\sqrt{3} k_y a}{2}\right)+20 (1-2 \cos (k_x a)) \cos \left(\sqrt{3} k_y a\right)+90 \sqrt{3} \cos (k_x a) \right.\\
& \qquad \left. +20 \cos (k_x a)-20 \cos (2 k_x a)+21\right] \ .
}
The analytical expression of the $E_{w,\bsl{k}}$ now does not match the DFT band structure well, as shown by the orange dashed line in \figref{fig:perturbation_1band}(b).
From \eqref{eq:approx_psi_NB} and \eqref{eq:psi_Se_approx}, we obtain the analytical expression of $\psi_{\Nb}$ and $\psi_{\Se}$ as 
\eqa{
\psi_{\Nb,1}(\bsl{k}) & =
 \frac{1}{81} e^{-\frac{1}{2} i \left(k_x a+\sqrt{3} k_y a\right)} \left[2 \cos \left(\frac{k_x a}{2}\right) \left(2 \sqrt{3} \cos \left(\frac{\sqrt{3} k_y a}{2}\right)+e^{\frac{1}{2} i \sqrt{3} k_y a} \left(2 \sqrt{3} \cos (k_x a)+27\right)\right) \right. \\
 & \qquad \left.+2 \sqrt{3} e^{i \sqrt{3} k_y a} (\cos (k_x a)-2)-8 \sqrt{3} \cos (k_x a)+2 \sqrt{3}+27\right] \\
\psi_{\Nb,2}(\bsl{k}) & =\frac{1}{27} i \sqrt{2} e^{-\frac{1}{2} (i k_x a)} \sin \left(\frac{k_x a}{2}\right) \left(-2 \cos \left(\frac{k_x a}{2}\right) \left(\cos \left(\frac{\sqrt{3} k_y a}{2}\right)-3 i \sin \left(\frac{\sqrt{3} k_y a}{2}\right)\right)+2 \cos (k_x a)+9 \sqrt{3}\right) \\
\psi_{\Nb,3}(\bsl{k}) &  = \frac{1}{81} \sqrt{2} e^{-i \left(k_x a+\sqrt{3} k_y a\right)} \left[ e^{\frac{1}{2} i \left(k_x a+\sqrt{3} k_y a\right)} \left(2 \sqrt{3} \cos (k_x a)+\sqrt{3}-27\right)  -\sqrt{3} \left(1+e^{i k_x a}\right) \right. \\
& \qquad \left. +e^{\frac{1}{2} i \left(k_x a+2 \sqrt{3} k_y a\right)} \cos \left(\frac{k_x a}{2}\right) \left(2 \sqrt{3} \cos (k_x a)-2 \sqrt{3}+27\right)+\sqrt{3} e^{\frac{1}{2} i \left(k_x a+3 \sqrt{3} k_y a\right)} (\cos (k_x a)-2)\right] 
}
and
\eqa{
\psi_{\Se,1}(\bsl{k}) & =
 \frac{5}{324} t_B e^{-\frac{1}{6} i \left(3 k_x a+8 \sqrt{3} k_y a\right)} \left[ 8 \sqrt{3} e^{i \sqrt{3} k_y a} \sin \left(\frac{k_x a}{2}\right) \sin (k_x a)+2 e^{\frac{3}{2} i \sqrt{3} k_y a} \cos (k_x a) \left(2 \sqrt{3} \cos (k_x a)+27\right) \right.\\
& \qquad  \left. +e^{\frac{1}{2} i \sqrt{3} k_y a} \left(27-4 \sqrt{3} \cos (k_x a)\right)+2 \sqrt{3} e^{2 i \sqrt{3} k_y a} \left(\cos \left(\frac{3 k_x a}{2}\right)-2 \cos \left(\frac{k_x a}{2}\right)\right)+2 \sqrt{3} \cos \left(\frac{k_x a}{2}\right) \right] \\
 \psi_{\Se,2}(\bsl{k}) & = \frac{5}{144 \sqrt{2}} \left(-1+e^{i k_x a}\right) t_A e^{-\frac{1}{6} i \left(9 k_x a+5 \sqrt{3} k_y a\right)} \left[ \left(1+e^{i k_x a}\right) e^{i \sqrt{3} k_y a}+5 e^{\frac{1}{2} i \left(k_x a+3 \sqrt{3} k_y a\right)} -3 \left(1+e^{i k_x a}\right)\right.\\
 & \qquad \left. +e^{\frac{1}{2} i \left(k_x a+\sqrt{3} k_y a\right)} \left(4 \cos (k_x a)+9 \sqrt{3}-5\right)\right] \\
\psi_{\Se,3}(\bsl{k}) &  = \frac{5}{432 \sqrt{2}} t_A e^{-\frac{1}{6} i \left(6 k_x a+5 \sqrt{3} k_y a\right)} \left[ \left(-\sqrt{3}\right) \left(1+e^{i k_x a}\right) e^{\frac{3}{2} i \sqrt{3} k_y a}+2 e^{\frac{1}{2} i \left(k_x a+2 \sqrt{3} k_y a\right)} \left(\sqrt{3} \cos (k_x a)-7 \sqrt{3}+27\right) \right. \\
& \qquad \left. +2 e^{\frac{1}{2} i \left(k_x a+\sqrt{3} k_y a\right)} \cos \left(\frac{k_x a}{2}\right) \left(16 \sqrt{3} \cos (k_x a)-\sqrt{3}-27\right)-2 \sqrt{3} e^{\frac{i k_x a}{2}} (7 \cos (k_x a)+1) \right]\ .
}
Substituting the analytical expressions into \eqref{eq:approx_psi_w}, we obtain the approximate $\gamma_{approx,w,\bsl{k}}^\dagger$, which has 0.9281 probability overlap with the DFT-precise Bloch state $w_{\bsl{k}}^\dagger$ for the single band at the Fermi level, \ie,
\eq{
\frac{1}{N}\sum_{\bsl{k}} \left| \bra{0} \gamma_{approx,w,\bsl{k}} w_{\bsl{k}}^\dagger \ket{0} \right|^2 =  0.9281\ ,
}
where $\ket{0}$ is the vacuum state.
The expectation value
$
\bra{0} \gamma_{approx,w,\bsl{k}} H_{6,\text{Se-onste,NNN}} \gamma_{approx,w,\bsl{k}}^\dagger \ket{0} 
$ given by the approximate Bloch state has great match with the dispersion of the obstructed atomic band of $H_{6,\text{Se-onste,NNN}}$ in \eqref{eq:HTB6band_Se-Onsite_NNN} and with that of the DFT precise Hamiltonian, as shown in \figref{fig:perturbation_1band}(b).

\subsubsection{Simplification 2}

We now further simplify (more simplification than \eqref{eq:sim1_approx1})  the Hamiltonian by considering the following approximations (in eV):
\eqa{ 
\label{eq:sim2_approx1}
 & t_{\bsl{\tau}_{\Se}\bsl{\tau}_{\Se}}(\bsl{0}) = \left(
\begin{array}{ccc}
 E_{\Se} & 0 & 0 \\
 0 & E_{\Se} & 0 \\
 0 & 0 & E_{\Se} \\
\end{array}
\right)=  - 1.8755 \mathds{1}_3  \\
&  t_{\bsl{\tau}_{\Se}\bsl{\tau}_{\Nb}} (\bsl{\tau_{\Se}})=   t_{\Se\Nb} \left(
\begin{array}{ccc}
 1 & 0 & 0 \\
 0 & -\frac{\sqrt{3}}{2} & \frac{\sqrt{3}}{2} \\
 0 & -\frac{1}{2} & -\frac{1}{2} \\
\end{array}
\right) R^\dagger \text{ with $t_{\Se\Nb} =- 2 E_{\Se}/3$}\ ,
}
which at most has about 30\% away from the corresponding values in \eqref{onsiteSeNb6Band} and \eqnref{tNNSeNb6Band}.
We futher approximate $t_{\bsl{\tau}_{\Nb}\bsl{\tau}_{\Nb} \text{ eff}}(\bsl{0})$ and $t_{\bsl{\tau}_{\Nb}\bsl{\tau}_{\Nb} \text{ eff}}(\bsl{a}_1)$ by
\eqa{
\label{eq:sim2_approx2}
& R^\dagger t_{\bsl{\tau}_{\Nb}\bsl{\tau}_{\Nb} \text{ eff}}(\bsl{0}) R = (E_{\Nb}+2 t_{\Se\Nb}) I_{3\times 3} \\
& R^\dagger t_{\bsl{\tau}_{\Nb}\bsl{\tau}_{\Nb} \text{eff}}(\bsl{a}_1) R =   t_{\Se\Nb} \left(
\begin{array}{ccc}
 5/8 & 0 & 0 \\
 0 & 0 & 0 \\
 0 & -6/5 & 0 \\
\end{array}
\right)\ ,
} 
where $E_{\Nb} = 0.5313$, the errors are not larger than $0.5$eV compared to \eqref{eq:terms_3band_eff_values}, and the derivation from  \eqref{eq:sim1_approx1} is no more than $0.3$eV.
With the simplification \eqnref{eq:sim2_approx1} and \eqnref{eq:sim2_approx2}, we are ready to provide analytical expressions of $E_{w,\bsl{k}}$, $\psi_{\Nb,\bsl{k}}$ and $\psi_{\Se,\bsl{k}}$.

First, we note that the simplification on the effective 3-band model leads to 
\eqa{
& \overline{H}_{eff,0}(\bsl{k}) =\left(
\begin{array}{ccc}
 E_{\Nb}-\frac{2 t_{\Se\Nb}}{5} & 0 & 0 \\
 0 & E_{\Nb}+\frac{16 t_{\Se\Nb}}{5} & 0 \\
 0 & 0 & E_{\Nb}+\frac{16 t_{\Se\Nb}}{5} \\
\end{array}
\right) \\
& \left[ \overline{H}_{eff,1}(\bsl{k}) \right]_{11} = \frac{5}{12} t_{\Se\Nb} \left(2 \cos \left(\frac{k_xa}{2}\right) \cos \left(\frac{\sqrt{3} k_ya}{2}\right)+\cos (k_xa)\right) \\
& \left[ \overline{H}_{eff,1}(\bsl{k}) \right]_{21} = \frac{5}{12} t_{\Se\Nb} \left(-\sqrt[3]{-1} \cos \left(\frac{1}{2} \left(k_xa+\sqrt{3} k_ya\right)\right)+\cos \left(\frac{1}{2} \left(k_xa-\sqrt{3} k_ya\right)\right)+(-1)^{2/3} \cos (k_xa)\right) \\
& \left[ \overline{H}_{eff,1}(\bsl{k}) \right]_{31} = \frac{5}{12} t_{\Se\Nb} \left((-1)^{2/3} \cos \left(\frac{1}{2} \left(k_xa+\sqrt{3} k_ya\right)\right)+\cos \left(\frac{1}{2} \left(k_xa-\sqrt{3} k_ya\right)\right)-\sqrt[3]{-1} \cos (k_xa)\right) \ .
}
Compared to those in \eqref{eq:H_bar_eff_sim_1}, $\left[ \overline{H}_{eff,1}(\bsl{k}) \right]_{11}$, $\left[ \overline{H}_{eff,1}(\bsl{k}) \right]_{21}$ and $\left[ \overline{H}_{eff,1}(\bsl{k}) \right]_{31}$ are scaled by a constant favtor.
Then, combined with \eqnref{eq:approx_E_w_k}, we obtain the analytic expression of the approximated dispersion as
\eqa{
\label{eq:E_w_k_sim2}
E_{w,\bsl{k}} &  = E_{\Nb}+\frac{5 t_{\Se\Nb}}{2592} \left[4 \cos \left(\frac{k_xa}{2}\right) (25 \cos (k_xa)+108) \cos \left(\frac{\sqrt{3} k_ya}{2}\right)+25 (1-2 \cos (k_xa)) \cos \left(\sqrt{3} k_ya\right)\right.\\
& \quad \left. +241 \cos (k_xa)-25 \cos (2 k_xa)\right]-\frac{2353 t_{\Se\Nb}}{4320}\ .
}
The analytical expression of the $E_{w,\bsl{k}}$ now does not match the DFT band structure well, as shown by the orange dashed line in \figref{fig:perturbation_1band}(b).
However, the dispersion derived from the approximate wavefunction will match well with the DFT one.
From \eqref{eq:approx_psi_NB} and \eqref{eq:psi_Se_approx}, we obtain the analytical expression of $\psi_{\Nb}$ and $\psi_{\Se}$ as 
\eqa{
\psi_{\Nb,1}(\bsl{k}) & =
 \frac{25}{648} \left(-2 e^{-\frac{1}{2} i \left(k_xa+\sqrt{3} k_ya\right)}+e^{-i k_xa}+1\right) \cos (k_xa)+\frac{1}{648} e^{-\frac{1}{2} i \left(k_xa+\sqrt{3} k_ya\right)} \left[482 \cos \left(\frac{k_xa}{2}\right) \cos \left(\frac{\sqrt{3} k_ya}{2}\right) \right. \\
 & \quad \left. +25 e^{i \sqrt{3} k_ya} (\cos (k_xa)-2)+432 i \cos \left(\frac{k_xa}{2}\right) \sin \left(\frac{\sqrt{3} k_ya}{2}\right)-50 \cos (k_xa)+241\right] \\
\psi_{\Nb,2}(\bsl{k}) & =\frac{1}{108 \sqrt{6}}i e^{-\frac{1}{2} (i k_xa)} \sin \left(\frac{k_xa}{2}\right) \left[ -25 \cos \left(\frac{k_xa}{2}\right) \left(\cos \left(\frac{\sqrt{3} k_ya}{2}\right)-3 i \sin \left(\frac{\sqrt{3} k_ya}{2}\right)\right)+25 \cos (k_xa)+216 \right]\\
\psi_{\Nb,3}(\bsl{k}) &  = \frac{1}{648 \sqrt{2}}e^{-i \left(k_xa+\sqrt{3} k_ya\right)} \left[ \left(1+e^{i k_xa}\right) e^{i \sqrt{3} k_ya} (25 \cos (k_xa)+191)+25 e^{\frac{1}{2} i \left(k_xa+3 \sqrt{3} k_ya\right)} (\cos (k_xa)-2) \right. \\
& \quad \left. +e^{\frac{1}{2} i \left(k_xa+\sqrt{3} k_ya\right)} (50 \cos (k_xa)-407)-25 \left(1+e^{i k_xa}\right) \right]
}
and
\eqa{
\psi_{\Se,1}(\bsl{k}) & =
 \frac{1}{324 \sqrt{3}}e^{-\frac{1}{6} i \left(3 k_xa+8 \sqrt{3} k_ya\right)} \left[ 100 e^{i \sqrt{3} k_ya} \sin \left(\frac{k_xa}{2}\right) \sin (k_xa)+25 e^{2 i \sqrt{3} k_ya} \left(\cos \left(\frac{3 k_xa}{2}\right)-2 \cos \left(\frac{k_xa}{2}\right)\right) \right. \\
 & \quad \left. -2 e^{\frac{1}{2} i \sqrt{3} k_ya} (25 \cos (k_xa)-108)+2 e^{\frac{3}{2} i \sqrt{3} k_ya} \cos (k_xa) (25 \cos (k_xa)+216)+25 \cos \left(\frac{k_xa}{2}\right) \right]\\
 \psi_{\Se,2}(\bsl{k}) & = \frac{1}{1296}e^{-\frac{1}{6} i \left(9 k_xa+5 \sqrt{3} k_ya\right)} \left[-25 \left(-1+e^{i k_xa}\right) \left(\left(1+e^{i k_xa}\right) e^{i \sqrt{3} k_ya}+3 e^{\frac{1}{2} i \left(k_xa+3 \sqrt{3} k_ya\right)}-2 \left(1+e^{i k_xa}\right)\right) \right. \\
 & \quad \left. -2 i e^{\frac{1}{2} i \left(2 k_xa+\sqrt{3} k_ya\right)} \sin \left(\frac{k_xa}{2}\right) (50 \cos (k_xa)+357)\right]\\
\psi_{\Se,3}(\bsl{k}) &  =\frac{1}{1296 \sqrt{3}}e^{-\frac{1}{6} i \left(6 k_xa+5 \sqrt{3} k_ya\right)} \left[ 25 \left(1+e^{i k_xa}\right) e^{\frac{3}{2} i \sqrt{3} k_ya}-\left(1+e^{i k_xa}\right) e^{\frac{1}{2} i \sqrt{3} k_ya} (250 \cos (k_xa)-457) \right.\\
&  \quad \left. -2 e^{\frac{1}{2} i \left(k_xa+2 \sqrt{3} k_ya\right)} (25 \cos (k_xa)+332)+50 e^{\frac{i k_xa}{2}} (4 \cos (k_xa)+1)\right]\ .
}
Substituting the analytical expressions into \eqref{eq:approx_psi_w}, we obtain the approximate $\gamma_{approx,w,\bsl{k}}^\dagger$, which has 0.8805 probability overlap with the DFT Bloch state $w_{\bsl{k}}^\dagger$ for the single band at the Fermi level, \ie,
\eq{
\frac{1}{N}\sum_{\bsl{k}} \left| \bra{0} \gamma_{approx,w,\bsl{k}} w_{\bsl{k}}^\dagger \ket{0} \right|^2 =  0.8805\ ,
}
where $\ket{0}$ is the vacuum state.
The expectation value
$
\bra{0} \gamma_{approx,w,\bsl{k}} H_{6,\text{Se-onste,NNN}} \gamma_{approx,w,\bsl{k}}^\dagger \ket{0} 
$ given by the approximate Bloch state now has reasonable match with the dispersion of the obstructed atomic band of $H_{6,\text{Se-onste,NNN}}$ in \eqref{eq:HTB6band_Se-Onsite_NNN} and with that of the DFT precise Hamiltonian, as shown in \figref{fig:perturbation_1band}(b).

\subsection{Conditions for a Compact Wannier Basis }

We note that the basis of the $3$-band model \eqref{eq:3-band_model_basis} and \eqref{eq:lower_3-band_model_basis} is both very localized (nearest neighbor Wannier) as well as containing much fewer elements than the full symmetry allows. 

We now find the conditions of the $6$-band model such that the $3$-band models of Nb and Se have a orthonormal compact basis.
(We will always assume the basis to be orthonormal from here on. )
We can write the most general nearest neighbor Wannier state as

\begin{equation}
    \widetilde{c}^\dagger_{\bsl{R},\alpha} = x_1 {c}^\dagger_{\bsl{R},i} R_{i\alpha} 
    +  c^\dagger_{\bsl{R}+\bsl{\tau}_{m}, i} R^m_{i \alpha} 
\end{equation} for Nb and 

\begin{equation}
    \widetilde{c}^\dagger_{\bsl{R} + \bsl{\tau}_{Se} ,\alpha} = y_1 {c}^\dagger_{\bsl{R} + \bsl{\tau}_{Se},i} R_{i\alpha} 
    +  c^\dagger_{\bsl{R}+\bsl{\tau}_{Se} - \bsl{\tau}_{m}, i} V^m_{i \alpha} 
\end{equation} for Se (upper index $m$ does not mean power). Note the use of $i$ for the original basis and the use of $\alpha$ for the rotated basis (by $R$). 
 Double index means summation and $i= 1,2,3 \equiv d_{z^2} , d_{xy}, d_{x^2-y^2}$ if the atom is Nb and $i= 1,2,3 \equiv p_{z} , p_{x}, p_{y}$ if the atom is Se (whether the atom is one or another is implied in their positions). We also have $\bsl{\tau}_m = (C_{3})^{m-1}\bsl{\tau}_{Se}$, $m=1,2,3$ while $R^m$ and $V^m$ are two sets of $3$ matrices, which by spatial symmetry, satisfy (they are real due to time-reversal):

\begin{eqnarray}
&
U_{m_x} R^1 \tilde{U}_{m_x}^\dagger = R^1,\;\; \;\;\;U_{m_x} V^1 \tilde{U}_{m_x}^\dagger = V^1
\nonumber \\ &R^m = (U_{C3})^{m-1} R^1 (\tilde{U}_{C{3}}^\dagger)^{m-1},\;\;\;\; V^m = (U_{C3})^{m-1} V^1 (\tilde{U}_{C{3}}^\dagger)^{m-1}
\end{eqnarray}
where $\tilde{U}_{C_3}, \tilde{U}_{m_x}$ are the symmetry matrices in the $c_\alpha$ basis:
\begin{eqnarray}
    &\tilde{U}_{C_3}=  \mat{ 0 & 0 & 1\\ 1 & 0 & 0 \\ 0 & 1 & 0 } \\
& \tilde{U}_{m_x}= \mat{ 1 & 0 & 0\\ 0 & 0 & 1 \\ 0 & 1 & 0 }
\end{eqnarray} And $U_{C_3}$  $U_{m_x}$ are the original basis matrices given in \eqref{eq:3band_sym_rep_Ug}. Hence the $m=2,3$ matrices can be obtained from the $m=1$ ones. Using $m_{x}$ we find the general form of $R^1$:
\begin{eqnarray}
    & R^1= \mat{ r_{11} & r_{12} & r_{12}\\ 0 & r_{22} & -r_{22} \\ r_{31} & r_{33} & r_{33} }
\end{eqnarray}
and similar for $V^1$. Notice that it is dependent on $5$ parameters whereas the special forms of the Wanniers in \eqref{eq:3-band_model_basis} and \eqref{eq:lower_3-band_model_basis} are given by the $  r_{31} = 0, r_{12} = 0, r_{22} = \sqrt{3} r_{33},  r_{11}=-2 r_{33}= x_2$, and $  rr_{31} = 0, rr_{12} = 0, rr_{22} = \sqrt{3} rr_{33},   rr_{11}=y_2,  rr_{33}= -y_3/2$, values which are much more restrictive than the symmetry-allowed freedom. We try to explain this constraint from the $6$-band model. Normalization gives 
\begin{eqnarray}
    &x_1^2+ \sum_{i,m} (R_{i\alpha}^m)^2 =1,\;\;\;\; y_1^2+ \sum_{i,m} (V_{i\alpha}^m)^2 =1
\end{eqnarray} We accept orthonormality between Wannier center in different unit cells to be only approximately satisfied (to be checked in retrospect). If the orthonormality were exact, it will impose the following constraint:
\begin{eqnarray}
    x_1 \sum_i R_{i\alpha}V^1_{i\beta} + y_1 \sum_i R_{i\alpha}^1 R_{i \beta} =0 \label{wannierorthogonality1}
\end{eqnarray}
We now form the projectors into the Nb and Se compact Wannier subspace:
\begin{eqnarray}
    P_{Nb} = \sum_{\bsl{R},\alpha}  \widetilde{c}^\dagger_{\bsl{R},\alpha} \ket{0}\bra{0} \widetilde{c}_{\bsl{R},\alpha},\;\;\; P_{Se }= \sum_{\bsl{R},\alpha} \widetilde{c}^\dagger_{\bsl{R} + \bsl{\tau}_{Se} ,\alpha}\ket{0}\bra{0} \widetilde{c}_{\bsl{R} + \bsl{\tau}_{Se} ,\alpha}\ .
\end{eqnarray} 
They are true projectors that satisfy $P_{Nb}+P_{Se}=1$ if all orthogonality conditions are achieved (which is not exactly the case for our wavefunctions, but almost). Notice the insertion $\ket{0}\bra{0}$ which suggests these are single-particle projectors.

We now know that:
\begin{eqnarray}
   H_{3}= P_{Nb} H_6P_{Nb}, \;\;\;  H_{lower-3}= P_{Se} H_6P_{Se}
\end{eqnarray} By our proofs (\secref{sec:proposition}) then $ H_{3}, H_{lower-3}$ give the exact correct spectrum of the upper and lower $3$ bands \emph{iff} 
\begin{equation}
    P_{Nb} H_6 P_{Se}=0. 
\end{equation} We now compute the LHS.
\eqa{
    H_6 &  = \sum_{\bsl{R}_2,\bsl{R}_3, i, j } \left\{ c^\dagger_{\bsl{R}_2, i }\left[t_{\Nb\Nb}(\bsl{R}_2- \bsl{R}_3)\right]_{ij}c_{\bsl{R}_3, j }+ c^\dagger_{\bsl{R}_2+ \bsl{\tau}_{Se}, i }\left[t_{\Se\Se}(\bsl{R}_2- \bsl{R}_3)\right]_{ij}c_{\bsl{R}_3 + \bsl{\tau}_{Se}, j } \right.\\ 
    & \left. + c^\dagger_{\bsl{R}_2+ \bsl{\tau}_{Se}, i }\left[t_{\Se\Nb}(\bsl{R}_2+ \bsl{\tau}_{Se}- \bsl{R}_3)\right]_{ij}c_{\bsl{R}_3, j }+ c^\dagger_{\bsl{R}_2,  i }\left[t_{\Nb\Se}(\bsl{R}_2- \bsl{R}_3 - \bsl{\tau}_{Se})\right]_{ij}c_{\bsl{R}_3 + \bsl{\tau}_{Se}, j } \right\}
}
The position of the atom denotes the type of the atom.

Using the relations, obtained from using the Wannier expressions:
\eqa{
    & P_{Nb}c^\dagger_{\bsl{R}_2 i } = x_1 \sum_{\alpha} \tilde{c}^\dagger_{\bsl{R}_2 \alpha } \ket{0}\bra{0} R^\star_{i\alpha}, \;\;\; P_{Nb}c^\dagger_{\bsl{R}_2 + \bsl{\tau}_{Se} i } = \sum_{\alpha,m} \tilde{c}^\dagger_{\bsl{R}_2 +\bsl{\tau}_{Se}  - \bsl{\tau}_m \alpha } \ket{0}\bra{0} R^{m\star}_{i\alpha} \\ 
    & { c^\dagger_{\bsl{R}_3 j } P_{Se} = \sum_{\gamma, m} V^{m}_{j\gamma} \ket{0}\bra{0}  \tilde{c}_{\bsl{R}_3 +\bsl{\tau}_m  \gamma},\;\;\; c^\dagger_{\bsl{R}_3+\bsl{\tau}_{Se} j } P_{Se} = y_1 \sum_{\gamma} R_{j\gamma} \ket{0}\bra{0} \tilde{c}_{\bsl{R}_3 +\bsl{\tau}_{Se}  \gamma} }
} We hence find (double index means summation, so $i, j, \alpha, \gamma, m, n$ are summed over where they appear) 
\eqa{
    P_{Nb} H_6 P_{Se} &  = \sum_{\bsl{R}_2,\bsl{R}_3} \tilde{c}^\dagger_{\bsl{R}_2 \alpha }\ket{0} x_1 R_{i\alpha}^\star \left[t_{\Nb\Nb}(\bsl{R}_2- \bsl{R}_3)\right]_{ij} V_{j \gamma}^m\bra{0}\tilde{c}_{\bsl{R}_3 + \bsl{\tau}_m \gamma } \\
    & \quad + \tilde{c}^\dagger_{\bsl{R}_2+ \bsl{\tau}_{Se}- \bsl{\tau}_m \alpha } \ket{0} R^{m\star}_{i \alpha} \left[t_{\Se\Se}(\bsl{R}_2- \bsl{R}_3)\right]_{ij} y_1 R_{j\gamma} \bra{0}\tilde{c}_{\bsl{R}_3 + \bsl{\tau}_{Se} \gamma}\\ 
    & 
    \quad  + \tilde{c}^\dagger_{\bsl{R}_2+ \bsl{\tau}_{Se} - \bsl{\tau}_m  \alpha  }\ket{0} R^{m\star}_{i\alpha} \left[t_{\Se\Nb}(\bsl{R}_2- \bsl{R}_3 + \bsl{\tau}_{Se})\right]_{ij}  V^n_{j \gamma} \bra{0}\tilde{c}_{\bsl{R}_3 + \bsl{\tau}_n \gamma } \\
    & 
    \quad  +     \tilde{c}^\dagger_{\bsl{R}_2  \alpha }\ket{0} x_1 R_{i\alpha}^\star  \left[t_{\Nb\Se}(\bsl{R}_2- \bsl{R}_3- \bsl{\tau}_{Se})\right]_{ij} y_1 R_{j\gamma} \bra{0}\tilde{c}_{\bsl{R}_3 + \bsl{\tau}_{Se} \gamma }   \\ 
    & = \sum_{\bsl{R}_2,\bsl{R}_3} \tilde{c}^\dagger_{\bsl{R}_2 \alpha }\ket{0} \bra{0}\tilde{c}_{\bsl{R}_3 + \bsl{\tau}_{Se} \gamma } \cdot  (x_1 R_{i\alpha}^\star t^{Nb; ij}_{\bsl{R}_2- \bsl{R}_3 +\bsl{\tau}_m - \bsl{\tau}_{Se}} V_{j \gamma}^m + y_1 R^{m\star}_{i \alpha} t^{Se; ij}_{\bsl{R}_2- \bsl{R}_3+\bsl{\tau}_m - \bsl{\tau}_{Se}} R_{j\gamma} \\ 
    & \quad  +  R^{m\star}_{i\alpha} t^{Se-Nb; ij}_{\bsl{R}_2- \bsl{R}_3 +\bsl{\tau}_m +\bsl{\tau}_n-\bsl{\tau}_{Se}} V^n_{j \gamma}  + x_1 y_1R_{i\alpha}^\star t^{Nb-Se; ij}_{\bsl{R}_2- \bsl{R}_3- \bsl{\tau}_{Se}} R_{j\gamma} ) \equiv 0 
}
These are the conditions to be satisfied by the compact Wannier states. We assume a $H_6$ model with only Nb-Se NN and Nb-Nb {NNN} and Se-Se {NNN} hopping; all others are zero. We then only have the following nontrivial options for conditions (all others are trivially zero; double index means summation, and the conditions should be satisfied $\forall \alpha, \gamma$)

\begin{eqnarray}\label{P1H6P2Eq1}
    &\bsl{R}_2=\bsl{R}_3=\bsl{R}:\; x_1 R_{i\alpha}^\star t^{Nb; ij}_{ \bsl{\tau}_m - \bsl{\tau}_{Se}} V_{j \gamma}^m + y_1 R^{m\star}_{i \alpha} t^{Se; ij}_{\bsl{\tau}_m - \bsl{\tau}_{Se}} R_{j\gamma} +  R^{m\star}_{i\alpha} t^{Se-Nb; ij}_{\bsl{\tau}_m +\bsl{\tau}_n-\bsl{\tau}_{Se}} V^n_{j \gamma}  + x_1 y_1R_{i\alpha}^\star t^{Nb-Se; ij}_{- \bsl{\tau}_{Se}} R_{j\gamma} =0\nonumber \\ 
    & \bsl{R}_2= \bsl{R} _3 + \bsl{a}_1: \;\;\; x_1 R_{i\alpha}^\star t^{Nb; ij}_{ \bsl{a}_1+\bsl{\tau}_m - \bsl{\tau}_{Se}} V_{j \gamma}^m + y_1 R^{m\star}_{i \alpha} t^{Se; ij}_{\bsl{a}_1+\bsl{\tau}_m - \bsl{\tau}_{Se}} R_{j\gamma} + \nonumber \\ 
    & +  R^{m\star}_{i\alpha} t^{Se-Nb; ij}_{\bsl{a}_1+\bsl{\tau}_m +\bsl{\tau}_n-\bsl{\tau}_{Se}} V^n_{j \gamma}  + x_1 y_1R_{i\alpha}^\star t^{Nb-Se; ij}_{\bsl{a}_1- \bsl{\tau}_{Se}} R_{j\gamma} =0\nonumber \\ 
    & \bsl{R}_2= \bsl{R} _3 +2 \bsl{a}_1+ \bsl{a}_2: \;\;\; x_1 R_{i\alpha}^\star t^{Nb; ij}_{ 2\bsl{a}_1+\bsl{a}_2+\bsl{\tau}_m - \bsl{\tau}_{Se}} V_{j \gamma}^m + y_1 R^{m\star}_{i \alpha} t^{Se; ij}_{2\bsl{a}_1+\bsl{a}_2+ \bsl{\tau}_m - \bsl{\tau}_{Se}} R_{j\gamma} + \nonumber \\ & +  R^{m\star}_{i\alpha} t^{Se-Nb; ij}_{2\bsl{a}_1+\bsl{a}_2+\bsl{\tau}_m +\bsl{\tau}_n-\bsl{\tau}_{Se}} V^n_{j \gamma}  + x_1 y_1R_{i\alpha}^\star t^{Nb-Se; ij}_{2\bsl{a}_1 + \bsl{a}_2- \bsl{\tau}_{Se}} R_{j\gamma} =0
\end{eqnarray}

Numerically, if we only include up to Nb-Se NN hoppings, we have for the LHS
\eqa{
& \bsl{R}_2=\bsl{R}_3=\bsl{R}: \left(
\begin{array}{ccc}
 -0.0313746 & -0.0437002 & -0.0437002 \\
 0.0945302 & 0.143265 & -0.133272 \\
 0.0945302 & -0.133272 & 0.143265 \\
\end{array}
\right) \\
 & \bsl{R}_2= \bsl{R} _3 + \bsl{a}_1: \left(
\begin{array}{ccc}
 -0.0647369 & -0.108607 & 0.0535254 \\
 -0.108607 & -0.0647369 & 0.0535254 \\
 0.024617 & 0.024617 & 0.102663 \\
\end{array}
\right) \\
&  \bsl{R}_2= \bsl{R} _3 +2 \bsl{a}_1+ \bsl{a}_2: 
\left(
\begin{array}{ccc}
 0.0875213 & -0.0251414 & -0.0256658 \\
 -0.0395956 & -0.0467516 & 0.0791911 \\
 -0.0256658 & 0.0502827 & 0.0513315 \\
\end{array}
\right)\ .
}

If we only include up to Nb-Nb/Se-Se NNN hoppings, we have  for the LHS
\eqa{
& \bsl{R}_2=\bsl{R}_3=\bsl{R}:\left(
\begin{array}{ccc}
 -0.296818 & 0.236501 & 0.236501 \\
 -0.00122464 & 0.0994799 & -0.0497001 \\
 -0.00122464 & -0.0497001 & 0.0994799 \\
\end{array}
\right)\\
 & \bsl{R}_2= \bsl{R} _3 + \bsl{a}_1:\left(
\begin{array}{ccc}
 -0.253689 & -0.00237398 & 0.0909641 \\
 -0.00237398 & -0.253689 & 0.0909641 \\
 0.0522668 & 0.0522668 & -0.189565 \\
\end{array}
\right) \\
&  \bsl{R}_2= \bsl{R} _3 +2 \bsl{a}_1+ \bsl{a}_2: 
\left(
\begin{array}{ccc}
 0.0325957 & -0.0596218 & -0.0151371 \\
 0.0105558 & -0.00862789 & -0.0241369 \\
 0.0904242 & 0.225571 & -0.0947826 \\
\end{array}
\right)\ .
}
We now write the first equation in \ref{P1H6P2Eq1}:
\begin{eqnarray}
    &\bsl{R}_2=\bsl{R}_3=\bsl{R}:\; 0=x_1 (R_{i\alpha}^\star t^{Nb; ij}_{ 0} V_{j \gamma}^1  + R_{i\alpha}^\star t^{Nb; ij}_{ -\bsl{a}_1 - \bsl{a}_2} V_{j \gamma}^2+ R_{i\alpha}^\star t^{Nb; ij}_{  - \bsl{a}_{2}} V_{j \gamma}^3) +\nonumber\\ &+  y_1 ( R^{1\star}_{i \alpha} t^{Se; ij}_{0} R_{j\gamma} +R^{2\star}_{i \alpha} t^{Se; ij}_{-\bsl{a}_1 - \bsl{a}_{2}} R_{j\gamma} +R^{3\star}_{i \alpha} t^{Se; ij}_{ - \bsl{a}_{2}} R_{j\gamma} )     +\nonumber \\ &+  R^{1\star}_{i\alpha} t^{Se-Nb; ij}_{ \bsl{\tau}_{1}} V^1_{j \gamma}+ R^{1\star}_{i\alpha} t^{Se-Nb; ij}_{\bsl{\tau}_2} V^2_{j \gamma}+ R^{1\star}_{i\alpha} t^{Se-Nb; ij}_{\bsl{\tau}_3} V^3_{j \gamma} \nonumber\\ 
    & + R^{2\star}_{i\alpha} t^{Se-Nb; ij}_{\bsl{\tau}_2 } V^1_{j \gamma}  + R^{2\star}_{i\alpha} t^{Se-Nb; ij}_{\bsl{\tau}_2 -\bsl{a}_1 - \bsl{a}_{2}} V^2_{j \gamma}  + R^{2\star}_{i\alpha} t^{Se-Nb; ij}_{\bsl{\tau}_2 - \bsl{a}_{2}} V^3_{j \gamma} \nonumber\\ 
    & + R^{3\star}_{i\alpha} t^{Se-Nb; ij}_{\bsl{\tau}_3} V^1_{j \gamma}   + R^{3\star}_{i\alpha} t^{Se-Nb; ij}_{\bsl{\tau}_3-\bsl{a}_1 - \bsl{a}_{2}} V^2_{j \gamma}  + R^{3\star}_{i\alpha} t^{Se-Nb; ij}_{\bsl{\tau}_3- \bsl{a}_{2}} V^3_{j \gamma}  \nonumber\\
    & + x_1 y_1R_{i\alpha}^\star t^{Nb-Se; ij}_{- \bsl{\tau}_{Se}} R_{j\gamma} =0 \nonumber \ ,
\end{eqnarray}
which is simplified to 
\begin{eqnarray}\label{P1H6P2Eq1_sim}
    &\bsl{R}_2=\bsl{R}_3=\bsl{R}:\; 0=x_1 (R_{i\alpha}^\star t^{Nb; ij}_{ 0} V_{j \gamma}^1  + R_{i\alpha}^\star t^{Nb; ij}_{ -\bsl{a}_1 - \bsl{a}_2} V_{j \gamma}^2+ R_{i\alpha}^\star t^{Nb; ij}_{  - \bsl{a}_{2}} V_{j \gamma}^3) +\nonumber\\ &+  y_1 ( R^{1\star}_{i \alpha} t^{Se; ij}_{0} R_{j\gamma} +R^{2\star}_{i \alpha} t^{Se; ij}_{-\bsl{a}_1 - \bsl{a}_{2}} R_{j\gamma} +R^{3\star}_{i \alpha} t^{Se; ij}_{ - \bsl{a}_{2}} R_{j\gamma} )     +\nonumber \\ &+  R^{1\star}_{i\alpha} t^{Se-Nb; ij}_{ \bsl{\tau}_{1}} V^1_{j \gamma}+ R^{1\star}_{i\alpha} t^{Se-Nb; ij}_{\bsl{\tau}_2} V^2_{j \gamma}+ R^{1\star}_{i\alpha} t^{Se-Nb; ij}_{\bsl{\tau}_3} V^3_{j \gamma} + R^{2\star}_{i\alpha} t^{Se-Nb; ij}_{\bsl{\tau}_2 } V^1_{j \gamma}  + R^{3\star}_{i\alpha} t^{Se-Nb; ij}_{\bsl{\tau}_3} V^1_{j \gamma}  \nonumber\\
    & + x_1 y_1R_{i\alpha}^\star t^{Nb-Se; ij}_{- \bsl{\tau}_{Se}} R_{j\gamma} =0\ ,
\end{eqnarray}
owing to the fact that we omit NNNN and longer hoppings---only $(m,n)\in \{ (1,1), (1,2), (1,3), (2,1), (3,1)\}$ gives nonzero contributions to the third term in \ref{P1H6P2Eq1} in this case.

For the second equation in \ref{P1H6P2Eq1} we find 
\begin{eqnarray}
    & \bsl{R}_2= \bsl{R} _3 + \bsl{a}_1: \;\;\;0= x_1 R_{i\alpha}^\star t^{Nb; ij}_{ \bsl{a}_1} V_{j \gamma}^1 +R_{i\alpha}^\star t^{Nb; ij}_{ -\bsl{a}_2} V_{j \gamma}^2  + \nonumber\\ &+ y_1 R^{1\star}_{i \alpha} t^{Se; ij}_{\bsl{a}_1} R_{j\gamma} + y_1 R^{2\star}_{i \alpha} t^{Se; ij}_{-\bsl{a}_2} R_{j\gamma}   \nonumber \\ 
    & +R_{i\alpha}^{2 \star} t^{Se-Nb; ij}_{\bsl{a}_1+\bsl{\tau}_2 } V^1_{j \gamma}+ R^{1\star}_{i\alpha} t^{Se-Nb; ij}_{\bsl{a}_1+\bsl{\tau}_2 } V^2_{j \gamma}
\end{eqnarray}

For no $Nb-Nb$ and $Se-Se$ hopping at $a_1$ distance, this becomes:
\begin{eqnarray}
       & \bsl{R}_2= \bsl{R} _3 + \bsl{a}_1: \;\;\;0= 
  R^{1\star}_{i\alpha} t^{Se-Nb; ij}_{\bsl{a}_1+\bsl{\tau}_2 } V^2_{j \gamma} +  R^{2\star}_{i\alpha} t^{Se-Nb; ij}_{\bsl{a}_1+\bsl{\tau}_2 } V^1_{j \gamma}  =  R^{1\star}_{i\alpha} t^{Se-Nb; ij}_{\bsl{\tau}_3 } V^2_{j \gamma} +  R^{2\star}_{i\alpha} t^{Se-Nb; ij}_{\bsl{\tau}_3 } V^1_{j \gamma} 
\end{eqnarray}

\subsection{A Simple Form of the  6-Band Model Hopping Matrices}

For an interesting fact, we present a simple form of the  6-Band Model hopping matrice.
With errors of about only $0.2\%$, the values of the NN Nb-Se hoppings  in \eqref{tNNSeNb6Band} are 

\begin{eqnarray}
    & t_{\bsl{\tau}_{\Se}\bsl{\tau}_{\Nb}}(\bsl{\tau}_{\Se}) =  \left(
\begin{array}{ccc}
 t_{NN,z,d_{z^2}} & 0 & t_{NN,z,d_{x^2-y^2}} \\
 0 & t_{NN,x,d_{xy}} & 0 \\
 t_{NN,y,d_{z^2}} & 0 & t_{NN,y,d_{x^2-y^2}} \\
\end{array}
\right) = \left(
\begin{array}{ccc}
 0.77754 & 0 & -0.973852 \\
 0 & -1.22464 & 0 \\
 -0.836268 & 0 & -0.631248 \\
\end{array}
\right) 
\end{eqnarray}
We also observe the following relations, to better than $1\%$ accuracy
\begin{eqnarray}
     &t_{NN,x,d_{xy}}= -\sqrt{\frac{3}{2}} ;\;\;\;  t_{NN,z,d_{z^2}} -t_{NN,z,d_{x^2-y^2}}=2\nonumber \\ & t_{NN,y,d_{x^2-y^2}} = t_{NN,y,d_{z^2}}-t_{NN,z,d_{x^2-y^2}} -  t_{NN,z,d_{z^2}}\nonumber \\ & t_{NN,y,d_{x^2-y^2}} = -\frac{1}{2t_{NN,z,d_{z^2}} } \nonumber \\ & t_{NN,z,d_{x^2-y^2}} \cdot  t_{NN,x,d_{xy}}  \cdot  t_{NN,y,d_{z^2}} = -1
\end{eqnarray}
Which allows us to obtain values for each of the components.

\begin{eqnarray}
    &  t_{\bsl{\tau}_{\Se} \bsl{\tau}_{\Nb}}=   \left( \begin{array}{ccc}
2- \sqrt{\frac{3}{2}} & 0 & t_{NN,z,d_{x^2-y^2}} \\
 0 &   - \sqrt{\frac{3}{2}} & 0 \\
 t_{NN,y,d_{z^2}} & 0 &  \frac{1}{10} \left(-\sqrt{6}-4\right)\\
\end{array}\right) 
\end{eqnarray}

\subsection{An Alternative Way to Understand the Approximated Wannier Functions of the 3-band Model}

If we rotate the Nb basis with $R$ in \eqref{eq:R_rotation}, \ie,
\eq{
\mat{ c^\dagger_{\bsl{R}+\bsl{\tau}_{\Nb},1} & c^\dagger_{\bsl{R}+\bsl{\tau}_{\Nb},2}  & c^\dagger_{\bsl{R}+\bsl{\tau}_{\Nb},3}  } = \mat{ c^\dagger_{\bsl{R}+\bsl{\tau}_{\Nb},d_{z^2}} & c^\dagger_{\bsl{R}+\bsl{\tau}_{\Nb},d_{xy}}  & c^\dagger_{\bsl{R}+\bsl{\tau}_{\Nb},d_{x^2-y^2}}  } R \ ,
}
then the NN hopping between Nb and Se becomes
\eq{
t_{\bsl{\tau}_{\Se}\bsl{\tau}_{\Nb}}(\bsl{\tau}_{\Se}) R =\left(
\begin{array}{ccc}
 1.24406 & 0.0513429 & 0.0513369 \\
0 & -0.865951 & 0.865951 \\
 0.0325924 & -0.740525 & -0.740527 \\
\end{array}
\right) \approx 
t_{\Se\Nb}\left(
\begin{array}{ccc}
 1 & 0 & 0 \\
 0 & -\frac{\sqrt{3}}{2} & \frac{\sqrt{3}}{2} \\
 0 & -\frac{1}{2} & -\frac{1}{2} \\
\end{array}
\right) = \left(
\begin{array}{ccc}
 1.24406 & 0. & 0. \\
 0. & -1.07739 & 1.07739 \\
 0. & -0.62203 & -0.62203 \\
\end{array}
\right)\label{tNbSeApproximation1}
}
where we choose
\eq{
t_{\Se\Nb} = 1.244 \eV\ .
}
With this observation, we can build a simple NN 6-band model, which only contains the onsite $\Nb$ energy, the onsite $\Se$ energy, and one NN hopping term:
\eqa{
\label{eq:6-band_simplified}
& R^\dagger t_{\bsl{\tau}_{\Nb}\bsl{\tau}_{\Nb}}(\bsl{0}) R = E_{\Nb} \mathds{1}_{3\times 3} \\
& t_{\bsl{\tau}_{\Se}\bsl{\tau}_{\Se}}(\bsl{0}) =  E_{\Se} \mathds{1}_{3\times 3} \\
& t_{\bsl{\tau}_{\Se}\bsl{\tau}_{\Nb}}(\bsl{\tau}_{\Se}) R =
t_{\Se\Nb}\left(
\begin{array}{ccc}
 1 & 0 & 0 \\
 0 & -\frac{\sqrt{3}}{2} & \frac{\sqrt{3}}{2} \\
 0 & -\frac{1}{2} & -\frac{1}{2} \\
\end{array}
\right) \\
& t_{\bsl{\tau}_{\Se}\bsl{\tau}_{\Nb}}(C_3\bsl{\tau}_{\Se}) R = U_{C_3} t_{\bsl{\tau}_{\Se}\bsl{\tau}_{\Nb}}(\bsl{\tau}_{\Se}) U_{C_3}^\dagger R =
t_{\Se\Nb}\left(
\begin{array}{ccc}
 0 & 1 & 0 \\
 0 & 0 & \frac{\sqrt{3}}{2} \\
 1 & 0 & -\frac{1}{2} \\
\end{array}
\right) \\
& t_{\bsl{\tau}_{\Se}\bsl{\tau}_{\Nb}}(C_3^2\bsl{\tau}_{\Se}) R =U_{C_3}^2 t_{\bsl{\tau}_{\Se}\bsl{\tau}_{\Nb}}(\bsl{\tau}_{\Se}) U_{C_3}^{-2} R=
t_{\Se\Nb}\left(
\begin{array}{ccc}
 0 & 0 & 1 \\
 0 & -\frac{\sqrt{3}}{2} & 0 \\
 1 & -\frac{1}{2} & 0 \\
\end{array}
\right)
}
where we choose 
\eqa{
& E_{\Nb} = \frac{E_{d_{z^2}}+2E_{d_{xy}}}{3} = 0.530349 \eV \\
& E_{\Se} = \frac{E_{z}+2E_{xy}}{3} = -1.877 \eV \ .
}
The model gives a bad band structure, but it contains two isolated sets of three bands, as shown in \figref{fig:lower_3band_DFT_el_6bandsim}(b).
In particular, the basis of the upper isolated set of three bands has $0.97$ probability overlap with the basis  (\eqref{eq:3-band_model_basis}) of the three-band model. Hence these bands can be used to obtain a simple 3-band model. 

However, direct diagonalization of simple NN 6-band model does not give compact Wannier functions. 
Nevertheless, we can use the simplified 6-band model (\eqref{eq:6-band_simplified}) to partially/perturbatively understand the approximated form of the basis (\eqref{eq:3-band_model_basis}) of the three-band model.
Since the three components of the basis are related by the $C_3$ symmetry, we only need to study $\widetilde{c}^\dagger_{\bsl{R},1}$.
As $t_{\Se\Nb}/(E_{\Nb}-E_{\Se}) \sim 1/2$, we treat $t_{\Se\Nb}/(E_{\Nb}-E_{\Se})$ as a perturbation.
To the zeroth order in $t_{\Se\Nb}/(E_{\Nb}-E_{\Se})$, we neglect the NN hopping in the simplified 6-band model (\eqref{eq:6-band_simplified}), and then we simply have $\widetilde{c}^\dagger_{\bsl{R},1} = c^\dagger_{\bsl{R}+\bsl{\tau}_{\Nb},1} = \left( c^\dagger_{\bsl{R},d_{z^2}} - \sqrt{2} c^\dagger_{\bsl{R},d_{x^2-y^2}} \right)/\sqrt{3}$.
To the first order in $t_{\Se\Nb}/(E_{\Nb}-E_{\Se})$, we only include $\Se$ orbitals that directly hop to $c^\dagger_{\bsl{R}+\bsl{\tau}_{\Nb},1}$, which, according to \eqref{eq:6-band_simplified}, are $c_{\bsl{R}+\bsl{\tau}_{\Se},z}$, $c_{\bsl{R}-\bsl{a}_1-\bsl{a}_2+\bsl{\tau}_{\Se},y}$ and $c_{\bsl{R}-\bsl{a}_2+\bsl{\tau}_{\Se},y}$.
So to the first order of $t_{\Se\Nb}/(E_{\Nb}-E_{\Se})$, we have a $4\times 4$ model in the basis of $\left(c^\dagger_{\bsl{R}+\bsl{\tau}_{\Nb},1} ,c^\dagger_{\bsl{R}+\bsl{\tau}_{\Se},z},c^\dagger_{\bsl{R}-\bsl{a}_1-\bsl{a}_2+\bsl{\tau}_{\Se},y},c^\dagger_{\bsl{R}-\bsl{a}_2+\bsl{\tau}_{\Se},y}\right)$, which reads
\eq{
\mat{
E_{\Nb} & t_{\Se\Nb} & t_{\Se\Nb} & t_{\Se\Nb} \\
 t_{\Se\Nb}  & E_{\Se} & 0&0 \\
 t_{\Se\Nb} &0 &  E_{\Se} &0 \\
 t_{\Se\Nb} &0 & 0 & E_{\Se} \ .
}
}
Diagonalizing the $4\times 4$ matrix, the eigenvector for the highest eigenvalue (since we are looking at the higher 3-bands in the 6-band model) reads
\eq{
\frac{\left(\sqrt{E_{\Nb}^2-2 E_{\Nb} E_{\Se}+E_{\Se}^2+12 t_{\Se\Nb}^2}+E_{\Nb}-E_{\Se},2 t_{\Se\Nb},2 t_{\Se\Nb},2 t_{\Se\Nb}\right)}{\sqrt{\left(\sqrt{(E_{\Nb}-E_{\Se})^2+12 t_{\Se\Nb}^2}+E_{\Nb}-E_{\Se}\right)^2+12 t_{\Se\Nb}^2}} = (0.854208,0.300183,0.300183,0.300183)\ ,
}
which is close to the approximated form (which would correspond to $(x1,x2,x2,x2)=(0.86,0.27,0.27,0.27)$ in the current basis) of $\widetilde{c}^\dagger_{\bsl{R},1}$ in \eqref{eq:3-band_model_basis}.

\subsection{A Proposition}
\label{sec:proposition}

We now prove a spectral decomposition that we have used in this paper. 

\textbf{Proposition:}
Suppose we have a $6\times 6$ Hamiltonian $H$ with $H=\sum_{n=1,2,3} \ket{u_{A,n}} \bra{u_{A,n}} E_{A,n} + \sum_{n=1,2,3} \ket{u_{B,n}} \bra{u_{B,n}} E_{B,n}$ with $E_{A,3} > E_{A,2}  > E_{A,1} > E_{B,3} > E_{B,2}  > E_{B,1}$.
Suppose we have two rank-3 projectors $P_1=\sum_{a=1,2,3}\ket{u_{1,a}} \bra{u_{1,a}}$ and $P_2=\sum_{a=1,2,3}\ket{u_{2,a}} \bra{u_{2,a}}$ such that (i) $\bra{u_{1,a}} H\ket{u_{1,a'}} $ has three eigenvalues that equal to $E_{A,1}$,$E_{A,2}$ and $E_{A,3}$, (ii) $\bra{u_{2,a}} H\ket{u_{2,a'}} $ has three eigenvalues that equal to $E_{B,1}$,$E_{B,2}$ and $E_{B,3}$, and (iii) $P_1 P_2 = 0$ and $P_1 + P_2 = 1$.
Then, we have 
\eq{
H = P_1 H P_1 + P_2 H P_2\ .
}

\textbf{Proof:}
Label $v_{1,n}$ as the eigenvector of $\bra{u_{1,a}} H\ket{u_{1,a'}} $ with eigenvalue $E_{A,n}$, \ie, 
\eq{
\sum_{a'} \bra{u_{1,a}} H\ket{u_{1,a'}} \left[v_{1,n}\right]_{a'} = E_{A,n} \left[v_{1,n}\right]_{a}\ .
}
Then, we know, 
\eq{
\sum_{aa'}  \left[v_{1,n}\right]_{a}^* \bra{u_{1,a}} H\ket{u_{1,a'}} \left[v_{1,n}\right]_{a'} = E_{A,n}\ .
}
Let us label 
\eq{
\ket{\widetilde{u}_{1,n}} = \sum_{a'}  \ket{u_{1,a'}} \left[v_{1,n}\right]_{a'}\ ,
}
and then we have
\eq{
\bra{\widetilde{u}_{1,n}}H\ket{\widetilde{u}_{1,n}} = E_{A,n}\ .
}

Now we show $\ket{\widetilde{u}_{1,n}} = \ket{u_{A,n}}$ up to a phase factor.
Since $E_{A,3}$ is the largest eigenvalue, we have $\ket{\widetilde{u}_{1,3}} = \ket{u_{A,3}}$ up to a phase factor, since 
\eq{
E_{A,3} = \bra{\widetilde{u}_{1,3}}H\ket{\widetilde{u}_{1,3}} = \sum_{n=1,2,3} \left|\braket{\widetilde{u}_{1,3}}{u_{A,n}}\right|^2 E_{A,n} + \sum_{n=1,2,3} \left|\braket{\widetilde{u}_{1,3}}{u_{B,n}}\right|^2 E_{B,n}
}
holds iff $\left|\braket{\widetilde{u}_{1,3}}{u_{A,n=1,2}}\right|^2 = \left|\braket{\widetilde{u}_{1,3}}{u_{B,n=1,2,3}}\right|^2 = 0$ and $\left|\braket{\widetilde{u}_{1,3}}{u_{A,n=3}}\right|^2=1$.
Then, for the second-largest eigenvalue $E_{A,2}$, we have
\eq{
E_{A,2} = \bra{\widetilde{u}_{1,2}}H\ket{\widetilde{u}_{1,2}} = \sum_{n=1,2,3} \left|\braket{\widetilde{u}_{1,2}}{u_{A,n}}\right|^2 E_{A,n} + \sum_{n=1,2,3} \left|\braket{\widetilde{u}_{1,3}}{u_{B,n}}\right|^2 E_{B,n}\ ,
}
and owing to the fact that $\ket{\widetilde{u}_{1,2}}$ is orthogonal to $\ket{\widetilde{u}_{1,3}}= \ket{u_{A,3}}$, we have 
\eq{
E_{A,2} = \bra{\widetilde{u}_{1,2}}H\ket{\widetilde{u}_{1,2}} = \sum_{n=1,2} \left|\braket{\widetilde{u}_{1,2}}{u_{A,n}}\right|^2 E_{A,n} + \sum_{n=1,2,3} \left|\braket{\widetilde{u}_{1,3}}{u_{B,n}}\right|^2 E_{B,n}
}
which holds iff $\ket{\widetilde{u}_{1,2}} = \ket{u_{A,2}}$ up to a phase factor.
Thus, we have $\ket{\widetilde{u}_{1,2}} = \ket{u_{A,2}}$ up to a phase factor.
By iterating the process, we have $\ket{\widetilde{u}_{1,1}} = \ket{u_{A,1}}$ up to a phase factor.
Then, we know 
\eq{
P_1 = P_A = \sum_{n=1,2,3} \ket{u_{A,n}} \bra{u_{A,n}}\ .
}
Combined with $P_2=1-P_1=1-P_A=P_B=\sum_{n=1,2,3} \ket{u_{B,n}} \bra{u_{B,n}} $, we have
\eq{
P_1 H P_2 = P_2 H P_1 = 0\ .
}
\textbf{End of Proof.}

\bibliography{bibfile1.bib}

\end{document}